\documentclass[a4paper,11pt,oneside]{amsart}

\usepackage[english]{babel}

\usepackage{fullpage}

\usepackage[utf8]{inputenc} 
\usepackage{lmodern} 
\usepackage[T1]{fontenc}

\usepackage{amssymb}
	\newtheorem{thm}{Theorem}[section]
	\newtheorem{prp}[thm]{Proposition}
	\newtheorem{lem}[thm]{Lemma}
	\newtheorem{cor}[thm]{Corollary}
	\newtheorem*{thm*}{Theorem}
	\newtheorem*{prp*}{Proposition}
	\theoremstyle{definition}
	\newtheorem*{def*}{Definition}
	\newtheorem*{ex*}{Example}
	\newtheorem*{rmk*}{Remark}
	\newtheorem*{asn*}{Assumption}
\usepackage{mathtools} 
	\numberwithin{equation}{section}

\usepackage[bb=dsserif,bbscaled=1.15]{mathalpha}

\usepackage{verbatim} 

\usepackage[dvipsnames]{xcolor} 

\usepackage{graphicx}

\usepackage{tikz}
	\usetikzlibrary{decorations.pathreplacing}
	\usetikzlibrary{decorations.pathmorphing} 
	\usetikzlibrary{matrix,arrows} 
	\usetikzlibrary{patterns} 

\usepackage{multirow} 

\usepackage{enumitem} 
	\setlist[itemize]{left= \parindent .. 2\parindent, label=\raisebox{0.14ex}{\scriptsize$\bullet$}} 

\usepackage[alphabetic,short-journals,initials]{amsrefs}

\usepackage{scalerel,stackengine} 
	\newcommand\pig[1]{\scalerel*[5.5pt]{\Big#1}{%
		\ensurestackMath{\addstackgap[1.5pt]{\big#1}}}}
	\newcommand\pigl[1]{\mathopen{\pig{#1}}}
	\newcommand\pigr[1]{\mathclose{\pig{#1}}}

\makeatletter
	\newcommand\niton{\mathrel{\m@th\mathpalette\canc@l\owns}}
	\newcommand\canc@l[2]{{\ooalign{$\hfil#1/\mkern1mu\hfil$\crcr$#1#2$}}}
	\makeatother

\DeclarePairedDelimiterX{\ketbra}[2]{\lvert}{\rvert}{#1\rangle \langle#2}
\DeclarePairedDelimiterX{\braket}[2]{\langle}{\rangle}{#1\vert#2}

\DeclarePairedDelimiterX{\cbraket}[2]{\langle\!\langle}{\rangle}{#1\vert#2}
\DeclarePairedDelimiterX{\bracket}[2]{\langle}{\rangle\!\rangle}{#1\vert#2}
\DeclarePairedDelimiterX{\cbracket}[2]{\langle\!\langle}{\rangle\!\rangle}{#1\vert#2}

\DeclareRobustCommand\longtwoheadrightarrow{\relbar\joinrel\twoheadrightarrow}

\usepackage{bm} 
	\newcommand{\vect}[1]{\bm{{#1}}}

\usepackage{braket, relsize, nccmath} 

\newcommand{\I}{\mathrm{i}}
\newcommand{\E}{\mathrm{e}}

\newcommand{\C}{\mathbb{C}}

\newcommand{\Z}{\mathbb{Z}}

\newcommand{\s}{\mathbf{s}}

\newcommand{\SLid}{1}

 \let\Im\undefined \DeclareMathOperator{\Im}{Im}

\DeclareMathOperator*{\ordprod}{\prod\limits^{\vbox to -.5ex{\kern-0.5ex\hbox{$\leftharpoonup$}\vss}}}
\DeclareMathOperator*{\ordprodopp}{\prod\limits^{\vbox to -.5ex{\kern-0.5ex\hbox{$\rightharpoonup$}\vss}}}

\newcommand{\elt}{B}

\makeatletter
\DeclareRobustCommand\widecheck[1]{{\mathpalette\@widecheck{#1}}}
\def\@widecheck#1#2{%
	\setbox\z@\hbox{\m@th$#1#2$}%
	\setbox\tw@\hbox{\m@th$#1%
		\widehat{%
			\vrule\@width\z@\@height\ht\z@
			\vrule\@height\z@\@width\wd\z@}$}%
	\dp\tw@-\ht\z@
	\@tempdima\ht\z@ \advance\@tempdima2\ht\tw@ \divide\@tempdima\thr@@
	\setbox\tw@\hbox{%
		\raise\@tempdima\hbox{\scalebox{1}[-1]{\lower\@tempdima\box
				\tw@}}}%
	{\ooalign{\box\tw@ \cr \box\z@}}}
\makeatother

\usepackage{marginnote}

\usepackage[
	ocgcolorlinks, 
	linkcolor=BlueViolet, citecolor=OliveGreen, urlcolor=RawSienna, filecolor=Sepia,
	hyperfootnotes=false,
	]{hyperref} 

\newcommand{\alg}{\mathcal{A}}
\newcommand{\DA}{\mathcal{D}}
\newcommand{\evq}{
	\overline{\mathrm{ev}}_{\mspace{-2mu}\elt}
	}

\begin{document}
	
	\title{Modular families of elliptic long-range spin chains \\ from freezing}
	
	\author{Rob Klabbers$^a$, Jules Lamers$^b$}
	\address{%
		{\vspace{0.1cm} $^{a}$\,Humboldt-Universität zu Berlin,~Zum Großen Windkanal 2, 12489 Berlin, Germany} \\
		\vspace{0.1cm} $^{b}$\,School of Mathematics and Statistics, University of Glasgow, University Place, Glasgow G12 8QQ, UK
	} 
	
\begin{abstract}
	We consider the construction of quantum-integrable spin chains with $q$-deformed long-range interactions by `freezing' integrable quantum many-body systems with spins. The input is a (quantum) spin-Ruijsenaars system along with an equilibrium configuration of the underlying spinless classical Ruijsenaars--Schneider system.
	For a distinguished choice of equilibrium, the resulting long-range spin chain has a real spectrum and admits a short-range limit, providing an integrable interpolation from nearest-neighbour to long-range interacting spins. 
	
	We focus on the elliptic case. We first define an action of the modular group on the spinless elliptic Ruijsenaars--Schneider system to show that, for a fixed elliptic parameter, it has a whole modular family of classical equilibrium configurations. These typically have constant but nonzero momenta. Then we use the setting of deformation quantisation to provide a uniform framework for freezing elliptic spin-Ruijsenaars systems at any classical equilibrium whilst preserving quantum integrability. As we showed in previous work, the results include the Heisenberg, Inozemtsev and Haldane--Shastry chains along with their \textsc{xxz}-like $q$-deformations (face type), or the antiperiodic Haldane--Shastry chain of Fukui--Kawakami, its elliptic generalisation of Sechin--Zotov, and their completely anisotropic $q$-deformations due to Matushko--Zotov (vertex type). Finally, we show how freezing fits in the setting of `hybrid' integrable systems.
\end{abstract}

\begin{flushright}
	\scriptsize{HU-EP-25/26\,\raisebox{1pt}{\textbullet}\,HU-Mathematik 2025-02}
\end{flushright}


\maketitle
\vspace{20pt}
\tableofcontents

\section{Introduction}

\subsection{Mathematical version}

In the theory of quantum-integrable systems, two central topics are Heisenberg (spin) chains and quantum many-body systems of Calogero--Sutherland and Ruijsenaars type. These two sides exploit representation theory of infinite-dimensional quantum groups and Cherednik algebras, respectively. While these two worlds often seem quite separated, there are links between the two. This paper is about the link provided by what we call \emph{long-range integrability}, centred around integrable spin chains with long-range interactions, quantum many-body systems with `spins', and their relation.

Calogero--Sutherland systems are quantum-mechanical dynamical systems defined by a Schr\"o-dinger (differential) operator belonging to a family of commuting operators, and Ruijsenaars systems are $q$-deformations thereof (difference operators). Both have rational, trigonometric, hyperbolic and elliptic versions, indicating the dependence on the coordinates (particle positions). The trigonometric case has close ties to double affine Hecke algebras. The elliptic version is the most general\,---\,and subtle. In the setting with `spins', eigenfunctions are generalised to vector-valued functions, and the commuting operators become matrix valued. In this paper we are interested in the most general amongst these quantum many-body systems: the elliptic spin-Ruijsenaars system, defined by matrix-valued difference operators with elliptic coefficients. For quantum integrability, they are built from (quantum) $R$-matrices, which are again elliptic. This leads to two distinct variants: 
	\begin{itemize}
		\item \textbf{Vertex-type:} the elliptic spin-Ruijsenaars system due to Matushko and Zotov \cite{MZ_23a} based on Baxter--Belavin $R$-matrices; 
		\item \textbf{Face-type:} the elliptic spin-Ruijsenaars system that we recently proposed \cite{klabbers2024deformed} featuring dynamical $R$-matrices of Felder's elliptic quantum group. 
	\end{itemize}

A key notion in long-range integrability is `freezing'. This physically motivated procedure forms the link between integrable quantum many-body systems with spins and long-range spin chains, and holds the key to the integrability of the latter. 
The goal of this paper is to make this precise for elliptic spin-Ruijsenaars systems. The result of freezing the two above variants are
\begin{itemize}
	\item \textbf{Vertex-type:} the elliptic long-range spin chain of Matushko and Zotov \cite{MZ_23b},
	\item \textbf{Face-type:} the $q$-deformed Inozemtsev chain \cite{klabbers2024deformed}. 
\end{itemize}
Like the spin-Ruijsenaars systems, these spin chains contain rational, trigonometric and hyperbolic special cases, see \cite{klabbers2025landscapes} for an overview. At the spin-chain level, there is one further limit, to (variants of) Heisenberg chains, see again \cite{klabbers2025landscapes}. For the $q$-deformed Inozemtsev chain, this includes in particular the usual Heisenberg `\textsc{xxx}' chain from traditional (`nearest-neighbour') quantum integrability. Thus, the elliptic long-range spin chains connect long-range integrability to its well-known nearest-neighbour counterpart. However, convergence of the `nearest-neighbour limit' requires a carefully normalised elliptic long-range spin chain. In some cases, including the vertex-type example \cite{klabbers2025landscapes}, the correct normalisation can be arranged by incorporating appropriate additive and multiplicative constants, but such an ad hoc regularisation is not readily available for the $q$-deformed Inozemtsev chain. In this work we remedy this by providing a conceptual, rigorous and uniform way of producing $q$-deformed elliptic long-range spin chains by freezing, building on \cite{mikhailovCommutativePoissonAlgebras2024,chalykh2024integrability}.

\subsection{Theoretical-physical version} 

Integrable spin systems with long-range interactions provide theoretical laboratories to study physical phenomena such as fractional statistics and long-range order \cite{Hal_91b,thouless1969long,haukeSpreadCorrelationsLongRange2013}. Unlocking this potential, however, requires understanding the exact underlying mathematical structures. The best-understood member of the long-range family is the Haldane--Shastry chain \cite{haldane1988exact,shastry1988exact}. Its integrability descends from that of a related quantum many-body system, the trigonometric Calogero--Sutherland system with spins, through a procedure called \emph{freezing} \cite{Pol_93}. This enables one to leverage the rich algebraic structure of that quantum many-body system \cite{bernard1993yang}, see also \cite{lamers2024fermionic} and especially \cite{chalykh2024integrability}. Freezing relies on  `charge-spin separation', i.e.\ a decoupling of the dynamical degrees of freedom (coordinates and momenta), which become classical, from the spins, which remain fully quantum mechanical. As suggested in \cite{liashyk2024classical}, this situation is analogous to the Born--Oppenheimer approximation, in which electrons moving around atomic nuclei are viewed as quantum-mechanical particles moving in a classical background given by the (much more massive) nuclei. Similarly, the Haldane--Shastry chain comprises quantum-mechanical spins interacting in a background governed by the classical trigonometric Calogero--Moser--Sutherland system. This situation extends to the $q$-deformed level, connecting the spin-Ruijsenaars--Macdonald system with the \textsc{xxz}-type generalisation of the Haldane--Shastry chain \cite{Ugl_95u,Lam_18,lamers2022spin}.

For \emph{elliptic} long-range spin chains, a recent surge of activity has started to fill in several long-standing gaps in the understanding of their integrability. For the Inozemtsev chain \cite{Inozemtsev:1989yq}, which interpolates between the Heisenberg \textsc{xxx} and Haldane--Shastry chains, it has long been known that its exact eigenfunctions are built from those of the scalar elliptic Calogero--Sutherland system \cite{Inozemtsev_1995}. Our more recent reformulation in terms of physically motivated quantities \cite{klabbers2022coordinate} connects this solution directly to the exact (Bethe-ansatz/Jack) wave functions of the limiting spin chains, paving the way for a direct comparison of their integrable structures. A set of conserved quantities was proposed in \cite{inozemtsev1996invariants}, but their mutual commutativity, and hence the integrability of Inozemtsev chain, remained an open problem for a long time. In \cite{chalykh2024integrability}, Chalykh addressed it by constructing a hierarchy of commuting higher hamiltonians using elliptic Dunkl operators \cite{buchstaberEllipticDunklOperators1994} and freezing. 

Thanks to all this progress, freezing is now rather well understood.
In this paper, we leverage this by zooming in on freezing at the $q$-deformed elliptic level. Two long-range spin chains and underlying quantum many-body system with spins have recently been uncovered. 
\begin{itemize} 
	\item \textbf{Vertex-type.} In \cite{MZ_23a,MZ_23b}, Matushko and Zotov (MZ) constructed a fully an\-iso\-tro\-pic elliptic spin-Ruijsenaars model based on the Baxter--Belavin \textit{R}-matrix \cite{baxter1972one,belavinDynamicalSymmetryIntegrable1981}. The corresponding spin chain $q$-deforms Sechin and Zotov's elliptic generalisation \cite{sechin2018r} of the (antiperiodic) Fukui--Kawakami chain \cite{fukui1996exact}, cf.~\cite{klabbers2025landscapes}. 
	\item \textbf{Face-type.} In \cite{klabbers2024deformed}, we defined a partially (an)isotropic, i.e.\ \textsc{xxz}-like, elliptic spin-Ruijsenaars system based on Felder's dynamical \textit{R}-matrix \cite{felder1994elliptic}. The associated spin chain $q$-deforms the Inozemtsev chain and moreover is an elliptic generalisation of the $q$-deformed Haldane--Shastry chain. 
\end{itemize}	
For a detailed analysis of the (almost entirely disjoint) landscapes in which these two families of integrable long-range spin chains live, see \cite{klabbers2025landscapes}, summarised in Figures 2--3 therein. 

In \cite{MZ_23a}, the formalism for freezing from the trigonometric case \cite{TH_95,Ugl_95u,lamers2022spin} was used to obtain a spin chain from their elliptic (vertex-type) spin-Ruijsenaars system. This involves expanding around a classical equilibrium configuration of the (elliptic) scalar Ruijsenaars system for which all momenta vanish, $p_i^\star = 0$, where we use the superscript `$\star$' for a classical equilibrium. This expansion coincides with a strong-coupling expansion of the shift operators (multiplicative momenta), $\Gamma_i = \E^{\epsilon \mspace{2mu}\hat{p}_i} = 1 + \epsilon \, \hat{p}_i + O\bigl(\epsilon^2\bigr)$, in a parameter $\epsilon \propto 1/g$; this is the way it was presented in \cite{Ugl_95u,lamers2022spin}. However, compared to the trigonometric case, a new feature at the elliptic level is the existence of a large number of classical equilibria, almost all of which have non-vanishing momenta, as we shall see. It is precisely such equilibria that give rise to spin chains 
with a well-defined (i.e.\ convergent) short-range limit \cite{klabbers2024deformed,klabbers2025landscapes}. This requires a modification of the freezing procedure from \cite{MZ_23a} in order to account for the contribution from the nontrivial classical limit $\Gamma_i \to \E^{\epsilon \mspace{2mu} p_i^\star}$, which can be done following \cite{chalykh2024integrability}. Surprisingly, the resulting `MZ$'$ chain' \cite{klabbers2025landscapes} turns out to differ from the original MZ chain \cite{MZ_23a} simply by a shift over a multiple of the identity~\cite{klabbers2025landscapes}. In contrast, in the (face-type) dynamical case of \cite{klabbers2024deformed} the two ways of freezing produce $q$-deformed Inozemtsev chains that differ more drastically\,---\,do not even commute\,---\,with only the version considered in \cite{klabbers2024deformed,klabbers2025landscapes} admitting a short-range limit. The upshot is that both the MZ$'$ chain and the $q$-deformed Inozemtsev chain enable an analytic comparison between long- and short-range regimes via interpolation. However, it remains to be proven that these interpolating spin chains are indeed integrable. In the vertex case, this provides a direct and more general alternative to the combination of the proof of integrability for the MZ chain~\cite{MZ_23a} and the simple difference with the MZ$'$ chain established in \cite{klabbers2025landscapes}. In the face setting, it requires a separate proof. In this paper we provide a technical but crucial step to narrow this gap. Besides proving integrability, we expect this to be very useful for the study of the spectrum. Indeed, for the ($q$-deformed and ordinary) Haldane--Shastry chain, the eigenvalues, eigenvectors, and (nonabelian) symmetries are understood via the connection to quantum many-body systems provided by freezing.

Our main aim is to develop a framework for freezing elliptic spin-Ruijsenaars systems around \emph{any} equilibrium of their spinless classical limit.

\subsection{Overview of our results}

To describe our main results, let us denote the matrix-valued elliptic Ruijsenaars operators by $\widetilde{D}_n$, $1\leqslant n \leqslant N$. Schematically, the first of them has the form $\widetilde{D}_1 = \sum_i A_i(\vect{x})\,P_{i}(\vect{x})^{-1} \, \Gamma_i \, P_{i}(\vect{x})$. Here $A_i(\vect{x})$ are (scalar) coefficients depending on the positions $x_j$ of the particles, $P_{i}(\vect{x})$ are matrices (acting on spins) built from the elliptic $R$-matrix at hand, and the difference operator $\Gamma_i$ acting on functions of $\vect{x}$ by shifting the $i$th argument as $x_i \mapsto x_i -\I\,\epsilon\,\hbar$. The higher $\widetilde{D}_n$ have a similar form. Then the spin-chain hamiltonians obtained by freezing are
\begin{equation} \label{eq:H_n_intro}
	H_{n,\elt} = \mathrm{ev}_\elt \pigl( \partial_\hbar\big|_{\hbar=0} \, \widetilde{D}_n \pigr) \, , \qquad 1\leqslant n \leqslant N
\end{equation}
where the `evaluation' $\mathrm{ev}_\elt$ denotes the restriction of the (positions) $x_i$ and (multiplicative momenta) to an equilibrium configuration, labelled by $\elt$, of the classical scalar elliptic Ruijsenaars--Schneider system. As we emphasise throughout this paper, despite the expansion in $\hbar$, the operators \eqref{eq:H_n_intro} remain quantum mechanical: in freezing, the motion becomes classical, but the spin part stays fully quantum mechanical.

Informally stated, our main results are as follows. The first two results concern the classical equilibria. 
\begin{thm*}[Theorem~\ref{thm:RS_modular}]
	The classical Ruijsenaars--Schneider system is essentially invariant under an action of the modular group\/ $\mathrm{PSL}(2,\mathbb{Z})$ coming from a symplectomorphism on the (complexified) classical phase space.
\end{thm*}
While such a modular invariance would be expected from an elliptic system, we are not aware of any proofs in the literature. We explicitly construct the symplectomorphism.
\begin{thm*}[Theorem~\ref{thm:equilibria_modular}]
	Under a mild assumption on the associated velocities, the preceding modular action sends the classical Ruijsenaars--Schneider system's equilibrium configurations to equilibrium configurations associated to the same (elliptic) lattice.
\end{thm*}
This allows us to exhibit a large family of classical equilibria of the elliptic Ruijsenaars--Schneider system, by starting from a well-known equilibrium configuration and acting by $\elt \in \mathrm{PSL}(2,\mathbb{Z})$. This is the origin of the $\elt$ in \eqref{eq:H_n_intro}. We do not have a proof that our construction gives all (discrete) classical equilibria.

Given these preliminaries, our main result is that freezing preserves quantum integrability:
	\begin{thm*}[Theorem~\ref{thm:HnB commute}]
		Consider an elliptic spin-Ruijsenaars system given by matrix-valued Ruijsenaars operators $\widetilde{D}_n$, $1\leqslant n\leqslant N$ that obey a certain `semiclassical spin separation' property and commute pairwise. Then, for any $\elt \in \mathrm{PSL}(2,\mathbb{Z})$, the `frozen' spin-chain hamiltonians \eqref{eq:H_n_intro} again commute pairwise.
	\end{thm*}
	Both the vertex- and face-type elliptic spin-Ruijsenaars systems are semiclassically spin separated. The commutativity of the $\widetilde{D}_n$ was proven for the vertex-type example in \cite{MZ_23a}, and for the resulting spin chains with $B = 1$ in \cite{MZ_23b}. For the face-type example, we announced the analogous result in \cite{klabbers2024deformed}. The proof for our face-type $\widetilde{D}_n$ will appear elsewhere; Theorem~\ref{thm:HnB commute} then implies the integrability of the $q$-deformed Inozemtsev chain. As we explain, the theorem also follows from a recent result of Chalykh~\cite{chalykh2024integrability}; our proof is closer to the rest of the literature on freezing, cf.~e.g.\ \cite{Ugl_95u,lamers2022spin,MZ_23b}. Thus, given an integrable elliptic spin-Ruijsenaars system, freezing yields a family of elliptic long-range spin chains indexed by the modular group $\mathrm{PSL}(2,\mathbb{Z})$.

Finally, we connect our work to recent papers of Mikhailov and Vanhaecke~\cite{mikhailovCommutativePoissonAlgebras2024} and Liashyk, Reshetikihin and Sechin~\cite{liashyk2024classical}, and add a Poisson-algebraic formulation of freezing: 
\begin{thm*}[Theorem~\ref{thm:comm diagram}]
	Freezing can be reintepreted as a morphism of Poisson modules over (possibly noncommutative) Poisson algebras, mapping an elliptic spin-Ruijsenaars system, via an intermediate `hybrid' system, to an elliptic long-range spin chain.
\end{thm*}

\subsection{Outline}

In \textsection\ref{sec:scalar} we recall some basic facts about the scalar elliptic Ruijsenaars system and its classical limit in the framework of deformation quantitsation, prove Theorems~\ref{thm:RS_modular}--\ref{thm:equilibria_modular}, and use this to construct a modular family of classical equilibrium configurations.

In \textsection\ref{sec:ell_sRuij} we review the matrix-valued Ruijsenaars operators in a framework allowing for a uniform treatment of the face and vertex versions.

In \textsection\ref{sec:freezing} we use deformation quantisation to freeze any 
such a spin-Ruijsenaars system, and establish Theorem~\ref{thm:HnB commute}.
	We obtain an explicit formula for the higher spin-chain hamiltonians, and give an overview of the applications to various known examples and limiting cases.

In \textsection\ref{sec:hybrid_systems} we reinterpret this freezing process in the formalism of \cite{mikhailovCommutativePoissonAlgebras2024} and the physical picture of `hybrid' systems \cite{liashyk2024classical}, which naturally arise in the process, and provide a Poisson-algebraic interpretation of freezing (Theorem~\ref{thm:comm diagram}).

We conclude in \textsection\ref{sec:discussion}. 

The appendix consists of two parts. \textsection\ref{app:elliptics} contains all necessary definitions and properties of elliptic functions and $R$-matrices. \textsection\ref{app:deformed_spin_perm} has more details about the deformed spin permutations from which the spin-Ruijsenaars operators are built. 

\section{Scalar case} \label{sec:scalar}

\subsection{Elliptic Ruijsenaars system} \label{sec:ell_Ruijs}

\noindent
The (quantum, elliptic) Ruijsenaars system, see \cite{ruijsenaarsEllipticIntegrableSystems2004} for an overview, describes $N$ interacting particles moving on a circle with coordinates $x_j$. We will simply denote by $\mathrm{Fun}(\vect{x})$ the space 
of suitable functions of $\vect{x} = (x_1,\dots,x_N)$.
\footnote{\ \label{fn:regular}
	We will be interested in formal and algebraic structures rather than (functional) analysis. For simplicity, we take $\mathrm{Fun}(\vect{x})$ to consist of meromorphic functions on which the action of finite products of $\Gamma_i$ is well defined. One may wish to impose appropriate quasiperiodicity properties, cf.~equation (26) of \cite{hasegawaLoperatorBelavin1994}.} 
Fix an arbitrary elliptic parameter~$\tau$ with \mbox{$\mathrm{Im}\,\tau>0$}. We take the (odd) Jacobi theta function to be 
\begin{equation} \label{eq:theta_arb_tau}
	\theta(x\,|\,\tau) \coloneqq \frac{\sin(\pi \, x)}{\pi} \prod_{n=1}^{\infty} \frac{ \sin \bigl( \pi(n \, \tau + x)\bigr) \sin \bigl( \pi( n \, \tau - x)\bigr)}{\sin^2 (\pi \, n \, \tau)} \, .
\end{equation}
More details on the elliptic functions that we use, which are all defined in terms of \eqref{eq:theta_arb_tau}, can be found in \textsection\ref{app:elliptics}.
Given $\eta\in\mathbb{C}$, define the rational functions
	\begin{equation} \label{eq:A_coeffs}
		A_{I}(\vect{x};\eta \, | \, \tau ) \coloneqq \prod_{i \in I \niton j}  \!\! \frac{\theta(x_i - x_j + \eta \, | \, \tau)}{\theta(x_i - x_j \, | \, \tau )} \, ,
	\end{equation}
	where the product runs over $i\in I$ and $j \in \{1,\dots,N\} \setminus I$.
Further given $\epsilon \in \mathbb{C}$ and $\hbar \in \mathbb{C}$, consider the shift operators 
$\Gamma_j$ for $1\leqslant j\leqslant N$ acting on $\mathrm{Fun}(\vect{x})$ as 
\begin{equation} \label{eq:Gamma_i}
	(\Gamma_j \, f)(\vect{x}) 
	\coloneqq f(x_1,\dots,x_j- \I \, \hbar \, \epsilon,\dots, x_N) \, .
\end{equation}
We suppress its dependence
on $\hbar$ and $\epsilon$. 
\begin{def*} 
	The \emph{Ruijsenaars operators} are the $N$ difference operators \cite{ruijsenaarsCompleteIntegrabilityRelativistic1987}
	\begin{equation} \label{eq:quantum_RS_ops}
		D_n(\vect{x};\hbar, \eta,\epsilon \,|\, \tau) \coloneqq \!\!\!\! \sum_{\substack{ I \subset \{1,\dots,N\} \\ |I|= n }} \!\!\!\!\!\!\! A_I(\vect{x};\eta \, | \, \tau ) \, \Gamma_I \, , 
		\quad 
		\Gamma_I \coloneqq \prod_{i \in I} \Gamma_i \, , \qquad 
		1\leqslant n \leqslant N \, ,
	\end{equation}
	where the sum is over $n$-element subsets of $\{1,\ldots,N\}$.
\end{def*}
When there is no cause for confusion we will often suppress the range of the sum, or the dependence of the coefficients on either or both $\eta$ and~$\tau$,
so that \eqref{eq:quantum_RS_ops} acquires the compact form $D_n = \sum_I A_I(\vect{x}) \, \Gamma_I$. 

There is a symmetry between $D_n$ with $n<N/2$ and their counterparts $D_{N-n}$ `beyond the equator':
\begin{lem} \label{lem:D_-n}
	Set $A_{-I}(\vect{x}) \coloneqq A_{I}(-\vect{x})$ and $\Gamma_{-I} \coloneqq \Gamma_I^{-1}$. Then the difference operators $D_{-n} \coloneqq D_N^{-1} \, D_{N-n}$ can be expressed as
	\begin{equation} \label{eq:D_-n}	
		\begin{aligned}
			D_{-n} = \sum_{\substack{ I \subset \{1,\dots,N\} \\ \# I = n }} \!\!\!\!\!\! A_{-I}(\vect{x};\eta) \, \Gamma_{-I} = D_n \, \Big|_{\substack{ \eta \,\mapsto\, -\eta \\ \epsilon \,\mapsto\, -\epsilon}} \, , \quad 1\leqslant n \leqslant N \, .
		\end{aligned}
	\end{equation}
\end{lem}
\begin{proof}[Proof.]
	The $D_{-n}$ are well defined since $D_N = \Gamma_1 \cdots \Gamma_N$ is the total coordinate-shift operator, which is invertible. For the first equality in \eqref{eq:D_-n}, note that $A_{\{1,\dots,N\}\setminus I}(\vect{x}) = A_{I}(-\vect{x})$ and $\Gamma_{\{1,\dots,N\}\setminus I} = D_N \, \Gamma_I^{-1}$, and observe that $D_N$ can be moved through the coefficients $A_{\pm I}(\vect{x})$ as those depend only on differences of coordinates. The second equality follows by observing that $A_{-I}(\vect{x};\eta) = A_{I}(\vect{x};-\eta)$ and $\Gamma_{-I} = \Gamma_{I}|_{\epsilon \mspace{2mu}\mapsto\mspace{1mu} -\epsilon}$.
\end{proof}
The formula $D_{N-n} = D_N \, D_{-n}$ motivates further setting $D_0 \coloneqq 1$.

Up to a conjugation (or `gauge transformation') \cite{hasegawaRuijsenaarsCommutative1997}, the difference operators \eqref{eq:quantum_RS_ops}--\eqref{eq:D_-n} define the quantum many-body system discovered by Ruijsenaars \cite{ruijsenaarsCompleteIntegrabilityRelativistic1987} with $\mathrm{Fun}(\vect{x})$ as the space of states. This Ruijsenaars system is (quantum) integrable in the following sense.
\begin{thm*}[Ruijsenaars \cite{ruijsenaarsCompleteIntegrabilityRelativistic1987}]
	The difference operators~\eqref{eq:quantum_RS_ops}--\eqref{eq:D_-n} commute pairwise,
	\begin{equation} \label{eq:commutativity_scalar_quantum}
		[D_n, D_m] = 0 \, , \qquad -N\leqslant n,m\leqslant N \, .
	\end{equation}
\end{thm*}
Later we will be interested in \emph{matrix-valued} generalisations of \eqref{eq:quantum_RS_ops}
maintaining this commutativity.

\subsection{Classical limit} \label{sec:deformation_quantisation_scalar}
The appropriate setting for taking the classical limit is provided by deformation quantisation \cite{BAYEN197861,BAYEN1978111}, cf.~\cite{kontsevichDeformationQuantizationPoisson2003} and \cite{etingofLecturesCalogeroMoserSystems2007}. For the Ruijsenaars system it works as follows.

Let $\mathrm{Fun}(\vect{x})[\Gamma_1^{\pm1},\dots, \Gamma_N^{\pm1}]$ denote the associative algebra obtained by equipping the vector space $\mathrm{Fun}(\vect{x}) \otimes \mathbb{C}[\Gamma_1^{\pm1},\dots, \Gamma_N^{\pm1}]$ with the product induced by [cf.~\eqref{eq:Gamma_i}]
\begin{equation} \label{eq:crossed prod}
	f(\vect{x}) \, \Gamma_i \,\cdot\, g(\vect{x}) \, \Gamma_j = f(\vect{x}) \, g(\vect{x}-\I \, \hbar \, \epsilon \,\vect{e}_{i}) \; \Gamma_i \, \Gamma_j
\end{equation}
where we omit the tensor products when writing elements, and $\vect{e}_i$ is the $i$th standard basis vector. Notably, $\mathrm{Fun}(\vect{x})[\Gamma_1^{\pm1},\dots, \Gamma_N^{\pm1}]$ contains the difference operators~\eqref{eq:quantum_RS_ops}--\eqref{eq:D_-n}.

Now change point of view by reinterpreting $\hbar$ as a formal parameter. By extending~\eqref{eq:crossed prod}  $\C[\mspace{-2mu}[\hbar]\mspace{-2mu}]$-bilinearly we obtain the associative algebra 
\begin{equation} \label{eq:A_hbar}
	\alg_{\hbar} \coloneqq \mathrm{Fun}(\vect{x})[\Gamma_1^{\pm1},\dots, \Gamma_N^{\pm1}] \otimes \C[\mspace{-2mu}[\hbar]\mspace{-2mu}] 
\end{equation}
of Laurent polynomials in the difference operators $\Gamma_j$ with coefficients that are meromorphic functions in $x_1,\ldots,x_N$ times a formal power series in $\hbar$. Again, we omit the tensor product when writing elements.

Note that the action~\eqref{eq:Gamma_i} of the shift operator can be expanded as a formal power series,\,%
\footnote{\ We emphasise that the corresponding expression $\Gamma_{\mspace{-2mu}j} \mspace{1mu} = \E^{-\I\,\hbar\,\epsilon \, \partial_{x_j}}$ does not make sense in $\alg_{\hbar}$ as the latter does not contain derivative operators, and, in particular, the $\hbar$ in the exponent belongs to $\Gamma_{\mspace{-2mu}j}$ rather than $\C[\mspace{-2mu}[\hbar]\mspace{-2mu}]$.}
\begin{equation} \label{eq:Gamma_f}
	(\Gamma_{\mspace{-2mu}j} \mspace{1mu} f)(\vect{x}) = \sum_{k\geqslant 0} \frac{1}{k!} (-\I \, \epsilon)^k \, \partial_j^{\,k} f(\vect{x}) \; \hbar^k \ \in \alg_\hbar \, .
\end{equation}
Viewing $f(\vect{x})$ as a multiplication operator, we obtain the following commutation relations in $\alg_\hbar$:
\begin{subequations} \label{eq:Ahbar_bracket}
	\begin{align} \label{eq:[f,Gamma]}
		[f(\vect{x}),\Gamma_{\mspace{-2mu}j}] & = \I \, \epsilon \, \partial_j f(\vect{x}) \, \Gamma_j \, \hbar + O(\hbar^2) \, .
		\intertext{In particular, taking $f(\vect{x})=x_i$ a coordinate function gives}
		\label{eq:[x,p]}
		[ x_i , \Gamma_{\mspace{-2mu}j} ] & = \I \,\hbar \, \epsilon \, \delta_{ij}\,\Gamma_{\mspace{-2mu}j} \, .
	\end{align}
\end{subequations}

Let us illustrate this setting further with a result that will be useful later on.
\begin{lem} \label{lem: scalar commutator hbar expansion}
	In $\alg_\hbar$, the commutativity $[D_n,D_m]=0$ from \eqref{eq:commutativity_scalar_quantum} reads
	\begin{equation} \label{eq:comm_Dn_expansion}
		\begin{aligned}
			0 = [D_n, D_m] = {} & {-}\I \, \epsilon \, \hbar \sum_{I\mspace{-2mu},\mspace{2mu} J} \Biggl( A_I(\vect{x};\eta) \sum_{i \in I} \partial_i A_J(\vect{x};\eta) - A_J(\vect{x};\eta) \sum_{j \in J} \partial_{j} A_I(\vect{x};\eta) \Biggr) \, \Gamma_I \, \Gamma_{J} \\
			& - \frac{\epsilon^2 \, \hbar^2}{2} \sum_{I \mspace{-2mu},\mspace{2mu} J} \Biggl(A_I(\vect{x};\eta) \!\!\sum_{i ,\mspace{2mu} i'\in I}\!\! \partial_i \, \partial_{i'} A_J(\vect{x};\eta) - A_J(\vect{x};\eta) \!\!\sum_{j ,\, j'\in J}\!\! \partial_j \, \partial_{j'} A_I(\vect{x};\eta) \Biggr) \, \Gamma_I \, \Gamma_J \\
			& + O\bigl(\hbar^3\mspace{1mu}\big) \, ,
		\end{aligned}
	\end{equation}
	where we suppressed the summation ranges $I, J \subseteq \{1,\dots,N\}$ with $|I| = n$ and $|J|= m$.
\end{lem}
\begin{proof}[Proof.]
	The Ruijsenaars operators $D_{\pm n}$ naturally live in the space $\alg_\hbar$. 
	Let us expand the product $D_n \, D_m$ as a formal power series in $\hbar$. For brevity we suppress the details of the summation ranges $I, J \subseteq \{1,\dots,N\}$ with $|I| = n$ and $|J|= m$, as well as the argument $\eta$ of the coefficients. Then 
	\begin{equation}
		\begin{aligned}
			D_n \, D_m & = \sum_{I \mspace{-2mu},\mspace{2mu} J} A_J(\vect{x}) 
			\, \Gamma_I \; A_J(\vect{x}) \, \Gamma_J \\
			& = \sum_{I \mspace{-2mu},\mspace{2mu} J} A_I(\vect{x}) \, A_J\biggl(\vect{x} -\I\, \epsilon \, \hbar {\textstyle \sum\limits_{i \in I}} \vect{e}_i \biggr) \, \Gamma_I \, \Gamma_{J} \\ 
			& = \sum_{I \mspace{-2mu},\mspace{2mu} J} A_I(\vect{x}) \, \Biggl( A_J(\vect{x}) -\I \, \epsilon \, \hbar \sum_{i \in I} \partial_i A_{J}(\vect{x}) - \frac{\epsilon^2 \, \hbar^2}{2} \! \sum_{i, \mspace{2mu}i'\in I} \!\! \partial_i  \, \partial_{i'} A_J(\vect{x}) + O\bigl(\hbar^3\bigr) \Biggr) \,
			\Gamma_I \, \Gamma_J \, . 
		\end{aligned}
	\end{equation}
	The sum at order $\hbar^2$ is $\sum_{i, \mspace{1mu}i'\in I} \partial_i \, \partial_{i'} = \bigl(\sum_{i \in I} \partial_i \bigr){}^2$ applied to $A_J(\vect{x})$.
	Since the coefficients $A_I$ and $A_J$ commute with each other, and the difference operators $\Gamma_I,\Gamma_J$ do so too, we arrive at \eqref{eq:comm_Dn_expansion}.
\end{proof}

\begin{def*}
	Considering the commutative algebra
	\begin{subequations} \label{eq:cl_def}
		\begin{gather} \label{eq:A_0}
			\alg_0 \coloneqq \mathrm{Fun}(\vect{x})[\gamma_1^{\pm1},\dots,\gamma_N^{\pm1}] \, ,
			\intertext{the \emph{classical limit} is defined as}
			\label{eq:cl_mapping}
			c_0 \colon \alg_\hbar \longtwoheadrightarrow \alg_0 \, , \qquad 
			f(\vect{x}) \longmapsto f(\vect{x}) \, , \quad 
			\Gamma_i \longmapsto \gamma_i \, , \quad \hbar \longmapsto 0 \, . 
		\end{gather}
	\end{subequations}
\end{def*}
This is the classical limit of \eqref{eq:A_hbar} in the following precise sense.
\begin{prp*}[e.g.\ \cite{etingofLecturesCalogeroMoserSystems2007}]
	There exists a $\mathbb{C}[\mspace{-2mu}[\hbar]\mspace{-2mu}]$-linear map 
	\begin{equation} \label{eq:cl_mapping_isom_hbar}
		c_\hbar \colon \alg_\hbar \overset{\sim}{\longrightarrow} \alg_0[\mspace{-2mu}[\hbar]\mspace{-2mu}] 
	\end{equation}
	that reduces 
	to an isomorphism $\mathcal{A}_{\hbar}/\hbar \mspace{2mu} \alg_\hbar \cong \alg_0$ of commutative algebras through which \eqref{eq:cl_mapping} factors.
\end{prp*}
\begin{proof}[Proof.]
	Since the commutator~\eqref{eq:Ahbar_bracket} is of order $\hbar$, the quotient $\mathcal{A}_{\hbar}/\hbar \mspace{2mu} \alg_\hbar$ is 
	commutative. It is isomorphic as a commutative algebra to $\alg_0$ via the $\mathbb{C}$-linear map
	\begin{equation} \label{eq:cl_mapping_isom}
		\mathcal{A}_{\hbar}/\hbar \mspace{2mu} \alg_\hbar \overset{\sim}{\longrightarrow} \alg_0 \, , \qquad
		f(\vect{x}) \longmapsto f(\vect{x}) \, , \quad 
		\Gamma_i \longmapsto \gamma_i \, . 
	\end{equation}
	This exhibits $\alg_\hbar$ as a (flat, formal) deformation of $\alg_0$. 
	We can now choose an identification \eqref{eq:cl_mapping_isom_hbar} of $\mathbb{C}[\mspace{-2mu}[\hbar]\mspace{-2mu}]$-modules 
	that reduces $\mathrm{mod}\ \hbar$ to \eqref{eq:cl_mapping_isom}. 
	Then the composition $\alg_\hbar \longtwoheadrightarrow \alg_\hbar/\hbar \mspace{2mu} \alg_\hbar \overset{\sim}{\longrightarrow} \alg_0$
	coincides with the classical limit $c_0 = c_\hbar|_{\hbar = 0}$ announced in \eqref{eq:cl_mapping}.
\end{proof}

\begin{rmk*}
	The physical intuition behind \eqref{eq:cl_mapping} is that $\gamma_j=\E^{\epsilon \, p_j}$ is the classical limit of $\Gamma_j = \E^{\epsilon \, \hat{p}_j}$, with $p_j$ the classical momentum and $\hat{p}_j= -\I\, \hbar\, \partial_{x_j}$ the quantum-mechanical momentum operator. The map~\eqref{eq:cl_mapping_isom_hbar} amounts to a choice of a (normal) ordering of quantum operators.
	The relation \eqref{eq:[x,p]} is an intermediate version of the Heisenberg commutation relations $[x_i,\hat{p}_j] = \I \,\hbar \, \delta_{ij}$ (additive notation) and the Weyl-algebra relations $\E^{\I \, x_i} \, \Gamma_{\mspace{-2mu}j} = \E^{\epsilon\,\hbar} \, \Gamma_{\mspace{-2mu}j} \, \E^{\I \, x_i}$ (fully multiplicative notation). 
\end{rmk*}

The isomorphism \eqref{eq:cl_mapping_isom_hbar} of $\mathbb{C}[\mspace{-2mu}[\hbar]\mspace{-2mu}]$-modules can be used
to transport the (noncommutative) product on $\alg_\hbar$ to $\alg_0[\mspace{-2mu}[\hbar]\mspace{-2mu}]$, equipping the latter with the (associative, $\mathbb{C}[\mspace{-2mu}[\hbar]\mspace{-2mu}]$-bilinear) Moyal star-product
\begin{equation} \label{eq:star}
	a \star b \coloneqq c_\hbar\pigl(c^{-1}_\hbar(a) \, c^{-1}_\hbar(b)\pigr) = \sum_{k\geqslant 0} m_k(a,b) \, \hbar^k \, , \qquad m_0(a,b) = a \, b \, .	
\end{equation}
Each coefficient function $m_k$, which in general are non-zero due to \eqref{eq:Gamma_f}, 
is itself a product on $\alg_0$, with $m_0$ the original (commutative) product.
At 
linear order in $\hbar$ we obtain a bracket on $\alg_0$ given by
\begin{equation} \label{eq:Pbracket_scalar}
	\{ f , g \} \coloneqq -\I \, \bigl( m_1(f,g) - m_1(g,f) \bigr) = -\I \, \hbar^{-1} \, c_\hbar\pigl( \bigl[c_\hbar^{-1}(f),c_\hbar^{-1}(g)\bigr] \pigr) \ \, \mathrm{mod} \, \hbar \, .
\end{equation}
Note that this bracket is independent of the choice of quantisation scheme $c_\hbar^{-1}$ since the nonvanishing commutators \eqref{eq:Ahbar_bracket} are already of order~$\hbar$. 
From \eqref{eq:Ahbar_bracket}
we recognise \eqref{eq:Pbracket_scalar} as a Poisson bracket, obeying the relations
\begin{subequations} \label{eq:PB}
	\begin{gather}
		\label{eq:PB x gamma}
		\{ x_i , x_j \} = \{ \gamma_i , \gamma_j \} = 0 \, , \quad 
		\{ x_i , \gamma_j \} = \epsilon \, \delta_{ij} \, \gamma_j \, .
		\intertext{In terms of additive momenta $p_j = \log(\gamma_j)/\epsilon$ this corresponds to canonical Poisson relations,}
		\label{eq:PB x p}
		\{ x_i , x_j \} = \{ p_i , p_j \} = 0 \, , \quad \{x_i, p_j\} = \delta_{ij} \, , 
	\end{gather}
\end{subequations}
of the classical phase space $M = T^* \, \mathbb{R}^N \cong \mathbb{R}^{2N}$ with coordinates $x_i$ and conjugate momenta $p_j$. 
In the setting of algebraically integrable systems, one works with the complexified phase space
\begin{equation} \label{eq:cpx phase space}
	M_{\mathbb{C}} = T^* \, \mathbb{C}^N \cong \mathbb{C}^{2N} \, .
\end{equation}
The associated Poisson algebra of suitable\textsuperscript{\ref{fn:regular} (p.\,\pageref{fn:regular})} functions on $M_\mathbb{C}$ contains $\mathcal{A}_0$ as a Poisson subalgebra. Note that the multiplicative momenta $\gamma_j$ are valued in $\mathbb{C}^\times$.

In order to consider the spin case later on, it will be useful to extend the usual notion of Poisson algebra to the noncommutative setting, i.e.\ to allow for the underlying algebra to be noncommutative. Following \cite{mikhailovCommutativePoissonAlgebras2024}, we therefore use the following: 	
\begin{def*}
	Given a (unital, associative) algebra $\mathcal{A}$, a \emph{Poisson bracket} on $\mathcal{A}$ is a skew-symmetric bilinear map $\{\cdot,\cdot\} : \mathcal{A} \times \mathcal{A}\to \mathcal{A}$ satisfying the Leibniz and Jacobi identities. The pair $(\mathcal{A},\{\cdot,\cdot\})$ is a \emph{Poisson algebra}. Note that $\mathcal{A}$ may be noncommutative; in this case we call the Poisson algebra $(\mathcal{A},\{\cdot,\cdot\})$ \emph{noncommutative}. 
\end{def*}

\begin{rmk*}
	The quantum space $\alg_\hbar$ equipped with the rescaled commutator $[\,\cdot\mspace{2mu},\cdot\,]_\hbar \coloneqq -\I \, \hbar^{-1} \, [\,\cdot\mspace{2mu},\cdot\,]$ is a noncommutative Poisson algebra, and its classical limit \eqref{eq:cl_mapping} a Poisson-algebra homomorphism (see \textsection\ref{sec:hybrid_systems}). For now we prefer to work with the ordinary commutator to keep all factors of $\hbar$ explicit.
\end{rmk*}

The classical limit \eqref{eq:cl_mapping} of the difference operators~\eqref{eq:quantum_RS_ops}--\eqref{eq:D_-n} yields the \emph{Ruijsenaars--Schneider} (RS) system defined by the functions
\begin{equation} \label{eq:class_RS_fns}
	D_{\pm n}^{\mspace{1mu}\text{cl}}(\vect{x},\vect{p};\eta \, | \, \tau ) \coloneqq  c_0\big(D_{\pm n} \big)= \!\!\!\sum_{\substack{ I \subset \{1,\dots,N\} \\ \# I = n }} \!\!\!\!\!\! A_{\pm I}(\vect{x};\eta \, | \, \tau ) \, \gamma_{\pm I} \ \, \in \alg_0 \, , \quad 1\leqslant n \leqslant N \, , \qquad \gamma_{\pm I} \coloneqq \prod_{i \in I} \! \gamma_i^ {\pm 1} \, , 
\end{equation}
with the same coefficients \eqref{eq:A_coeffs}. 
These functions depend on the parameters $\eta,\epsilon$ and $\tau$. 
\begin{rmk*}
	For the physical interpretation of \eqref{eq:class_RS_fns}, 
	form the combinations
	\begin{equation} \label{eq:RS_H_P}
		P^{\mspace{1mu}\text{cl}} \coloneqq \textcolor{lightgray}{\mu_0 \, c \,} \frac{D_1^{\mspace{1mu}\text{cl}} - D_{-1}^{\mspace{1mu}\text{cl}}}{2} \, , \qquad
		H^{\mspace{1mu}\text{cl}} \coloneqq \textcolor{lightgray}{\mu_0 \, c^2 \,}  \frac{D_1^{\mspace{1mu}\text{cl}} + D_{-1}^{\mspace{1mu}\text{cl}}}{2} \, .
	\end{equation}
	\textcolor{lightgray}{For the reader who likes physical units, here we let $\mu_0$ denote the rest mass, and $c$ the speed of light.}
	For a single particle we recognise
	\begin{equation} \label{eq:RS_N=1}
		N=1 : \qquad
		\begin{aligned}
			P^{\mspace{1mu}\text{cl}} = & \ \textcolor{lightgray}{\mu_0 \, c^{\hphantom{2}} } \sinh(\epsilon \, p) \, , \\
			H^{\mspace{1mu}\text{cl}} = & \ \textcolor{lightgray}{\mu_0 \, c^2} \cosh(\epsilon \, p) = \sqrt{(\textcolor{lightgray}{\mu_0 \, c^2}\, 1)^2 + (\textcolor{lightgray}{c} \, P^{\mspace{1mu}\text{cl}})^2} \, ,
		\end{aligned}
		\qquad \textcolor{lightgray}{\epsilon = \frac{1}{\mu_0\,c} \, ,}
	\end{equation}
	as the momentum and energy of a relativistic particle with 		$p\textcolor{lightgray}{/\mu_0}$ playing the role of rapidity. The functions \eqref{eq:RS_H_P} are $N$-particle generalisations of \eqref{eq:RS_N=1} belonging to the Liouville-integrable family \eqref{eq:class_RS_fns}.
\end{rmk*}

Note that only $N$ of the functions \eqref{eq:class_RS_fns} are functionally independent, since $D_{-n}^{\mspace{1mu}\text{cl}} = D_{N-n}^{\mspace{1mu}\text{cl}}/ D_{N}^{\mspace{1mu}\text{cl}}$, cf.~Lemma~\ref{lem:D_-n}.
Coming from the quantum level, it readily follows that this classical many-body system is Liouville integrable:
\begin{thm*}[Ruijsenaars--Schneider \cite{ruijsenaars1986new}]
	For the bracket \eqref{eq:PB x gamma}, the
	functions \eqref{eq:class_RS_fns} Poisson commute,
	\begin{equation} \label{eq:class integrab}
		\{ \mspace{-1mu} D_n^{\mspace{1mu}\mathrm{cl}}, D_m^{\mspace{1mu}\mathrm{cl}} \mspace{1mu}\} = 0 \, , \qquad -N \leqslant n,m\leqslant N \, .
	\end{equation}
\end{thm*}
\begin{proof}[Proof.]
	We use the commutativity \eqref{eq:commutativity_scalar_quantum} of the Ruijsenaars operators. As an element of $\alg_\hbar$, like any commutator, $[D_n,D_m]$ is of order $\hbar$, as \eqref{eq:comm_Dn_expansion} shows explicitly. 
	Applying the isomorphism $c_\hbar$ to \eqref{eq:commutativity_scalar_quantum} and dividing both sides by $\I \, \hbar$ gives $-\I \, \hbar^{-1} \, c_\hbar([D_m,D_m]) = 0$. Note that $c_\hbar(D_n) = D_n^\mathrm{cl} + O(\hbar) \in \alg_0[\mspace{-2mu}[\hbar]\mspace{-2mu}]$, where the `quantum corrections' in $\hbar \, \alg_0 [\mspace{-2mu}[\hbar]\mspace{-2mu}]$ arise from the choice of $c_\hbar$; for example, here they can be chosen to vanish. Hence  $c_\hbar^{-1}(D_n^\mathrm{cl}) = D_n + O(\hbar)$. 
	Working $\mathrm{mod}~\hbar$ we recognise the definition of the Poisson bracket~\eqref{eq:Pbracket_scalar} and arrive at \eqref{eq:class integrab}.
\end{proof}
\begin{rmk*}
	It is instructive to examine the explicit expansion \eqref{eq:comm_Dn_expansion} in more detail. At linear order in $\hbar$, after dividing by $\I\,\hbar$, applying $c_\hbar$ and using \eqref{eq:Pbracket_scalar} as in the preceding proof, we obtain an identity in $\alg_0[\mspace{-2mu}[\hbar]\mspace{-2mu}]$ for the classical Ruijsenaars--Schneider functions \eqref{eq:class_RS_fns}:
	\begin{equation} \label{eq:prf_PB}
		0 =  \{ D_n^{\mspace{1mu}\text{cl}} , D_m^{\mspace{1mu}\text{cl}} \} + O(\hbar) = -\epsilon \sum_{I \mspace{-2mu},\mspace{2mu} J} \Biggl( A_I \sum_{i \in I} \partial_i A_J - A_J \sum_{j \in J} \partial_{j} A_I \Biggr) \, \gamma_I \, \gamma_{J} \; + O(\hbar) \, .
	\end{equation}
	The first equality gives \eqref{eq:class integrab}, and the second equality matches a direct computation using the Poisson brackets~\eqref{eq:PB x gamma},
	\begin{align} \label{eq:Poisson_DnDm}
		\bigl\{ D_n^{\mspace{1mu}\text{cl}} , D_m^{\mspace{1mu}\text{cl}} \bigr\} & = \sum_{i ,\mspace{2mu} j=1}^N \Biggl( \frac{\partial D_n^{\mspace{1mu}\text{cl}}}{\partial x_i} \, \frac{\partial D_m^{\mspace{1mu}\text{cl}}}{\partial \gamma_j} \, \{ x_i, \gamma_j\}
		+ \frac{\partial D_n^{\mspace{1mu}\text{cl}}}{\partial \gamma_i} \, \frac{\partial D_m^{\mspace{1mu}\text{cl}}}{\partial x_j} \, \{ \gamma_i, x_j\} \Biggr) \nonumber \\
		& = \sum_{I \mspace{-2mu},\mspace{2mu} J} \Biggl( \, \sum_{i=1}^N \partial_i A_I \: \gamma_I  \sum_{j \in J} A_J \: \gamma_{J \setminus \{ j \}} \times \epsilon\, \gamma_j \, \delta_{ij} + \sum_{i \in I} A_I \: \gamma_{I\setminus \{i\}} \sum_{j =1}^N \partial_j A_J \: \gamma_J \times {-}\epsilon\, \gamma_i \, \delta_{ji}\Biggr) \\
		& = -\epsilon \sum_{I \mspace{-2mu},\mspace{2mu} J} \Biggl( A_I \sum_{i \in I} \partial_i A_J - A_J \sum_{j \in J} \partial_{j} A_I \Biggr) \, \gamma_I \, \gamma_{J} \, . \nonumber
	\end{align}
	These terms come from the $m_1$-commutator of the $\star$-product \eqref{eq:star}, cf.~\eqref{eq:Pbracket_scalar}. At higher orders in $\hbar$, even if one picks $c_\hbar$ such that $D_n \longmapsto D_n^\mathrm{cl}$, the commutator~\eqref{eq:comm_Dn_expansion} receives contributions encoded in the $\star$-product of $\alg_0[\mspace{-2mu}[\hbar]\mspace{-2mu}]$. For instance, the terms of order $\hbar^2$ in \eqref{eq:comm_Dn_expansion} give rise to a non-vanishing $m_2$-commutator. The vanishing
	\eqref{eq:comm_Dn_expansion} at each order of $\hbar$ yields non-trivial identities obeyed by the coefficients $A_I(\vect{x};\eta)$ that will be useful when we prove Proposition~\ref{prp:partial class lim} in \textsection\ref{sec:freezing}.
\end{rmk*}

\subsection{Modularity of the classical Ruijsenaars--Schneider system} \label{sec:modularity}

The aim of this section is to show that, as one might expect from an elliptic model, the elliptic Ruijsenaars--Schneider system is invariant, up to a rescaling, under an action of the group
	\begin{equation}
		\mathrm{SL}(2,\mathbb{Z}) = \bigl\langle S, T \,\big|\, S^4 = \SLid = (S \, T)^6 \mspace{1mu}\bigr\rangle \, .
	\end{equation}
We will be interested in families of hamiltonians associated to a \emph{fixed} (quasi-)period lattice $\Lambda_{\tau} \coloneqq \mathbb{Z} + \tau \, \mathbb{Z} \subset \mathbb{C}$, with modular parameter $\tau \in \mathbb{H} \coloneqq \{ z \in \mathbb{C} : \Im z > 0 \}$ in the upper half plane. Such a lattice defines an elliptic curve $\mathbb{C}/\Lambda_\tau$, and homothetic lattices yield isomorphic curves. 
\begin{lem}
	There is an action of $\mathrm{SL}(2,\mathbb{Z})$ on $\mathbb{C} \times \mathbb{H}$ that fixes the lattice $\Lambda_\tau \subset \mathbb{C}$.
\end{lem}
\begin{proof}[Proof.]
	It is well known (see e.g.\ \cite{silvermanAdvancedTopicsArithmetic1994}) that all homothetic lattices are related by an action of the modular group
	\begin{equation}
		\mathrm{PSL}(2,\mathbb{Z}) = \bigl\langle S, T \,\big|\, S^2 = \SLid = (S \, T)^3 \mspace{1mu} \bigr\rangle \, , 
	\end{equation}
	whose action on $\mathbb{H}$ is generated by the signed inversion $S\colon \tau \longmapsto -1/\tau$ and the translation $T\colon \tau \longmapsto \tau+1$.
	To keep the lattice \emph{invariant}, this action has to be combined with a rescaling of the ambient copy of $\mathbb{C} \supset \Lambda_\tau$, see Fig.~\ref{fg:modular_action}. Indeed, $\Lambda_\tau = c \, \Lambda_{\tau'}$ for some $\tau' \in \mathbb{H}$ and $c \in \mathbb{C}^\times$ precisely when $\tau = \elt \cdot \tau'$ for some $\elt \in \mathrm{PSL}(2,\mathbb{Z})$. Note that, given $z$ in this ambient copy of $\mathbb{C}$, applying the $S$-transformation twice gives
	\begin{equation} \label{eq:double_application}
		(z \,|\, \tau) \ \longmapsto \ \bigl(-z/\tau \,\big|\, {-}1/\tau\bigr) \eqqcolon \bigl(z' \,\big|\,  \tau'\mspace{1mu} \bigr) \ \longmapsto \ \bigl(-z'/\tau' \,\big|\, {-}1/\tau' \mspace{1mu}\bigr) = (-z \,|\, \tau) \, ,
	\end{equation}
	where the second application uses the updated modular parameter $\tau'$. Due to the sign on the right-hand side of \eqref{eq:double_application}, we are led to an action of the double cover $\mathrm{SL}(2,\mathbb{Z})$
	of the modular group $\mathrm{PSL}(2,\mathbb{Z}) = \mathrm{SL}(2,\mathbb{Z})/\langle \pm\SLid\rangle$.%
	\footnote{\ One can also see that $S$ now has order $4$ by keeping track of both lattice vectors (periods). To preserve the orientation, the order of the two vectors is swapped under the action $S\colon \Lambda_{1,\tau} \longmapsto -\tau \, \Lambda_{1,-1/\tau} = \Lambda_{-\tau,1}$ (cf.~Fig.~\ref{fg:modular_action}). Applying $S$ again thus gives $\Lambda_{-\tau,1} \longmapsto \Lambda_{-1,-\tau} = -\Lambda_{1,\tau}$, so $S^2=-\SLid$.} 
\end{proof}
This action is illustrated in Fig.~\ref{fg:modular_action}.

\begin{figure}[t]
	\centering
	\begin{tikzpicture}{scale=0.5}
		\node at (0,0) {
			\begin{tikzpicture}[scale=0.8]
				\node at (0+.2,1.9) {$\Lambda_\tau \coloneqq \Z + \tau \, \Z$};
				\clip (-1.5+.2,-1.5) rectangle (1.5+.3,1.6); 
				\foreach \i in {-2,...,2} \draw[dotted, darkgray] (\i-3/3,-3) -- (\i+5/3,5);
				\foreach \j in {-1,1} \draw[dotted, darkgray] (-2,1.2*\j) -- (5.5,1.2*\j);
				\draw[thick] (-2,0) -- (5,0);
				\draw[thick] (0,-3) -- (0,5);
				\foreach \i in {-4,...,4} \draw[ultra thin] (\i,-.05) -- (\i,.05);
				\foreach \j in {-2,...,4} \draw[ultra thin] (-.05,\j) -- (.05,\j);
				\node at (1,0) [yshift=-.25cm]{$1$};
				\node at (0,1) [xshift=-.15cm]{$\I$};
				\draw[-latex,very thick] (0,0) -- (1,0);
				\draw[-latex,very thick] (0,0) -- (.4,1.2) node[shift={(-.05cm,.14cm)}] {$\tau$};
			\end{tikzpicture}
		};
		
		\draw[->, thick] (2,0) -- node[above]{$T\colon \tau \longmapsto \tau +1$} (4,0); 
		
		\node at (6,0) {
			\begin{tikzpicture}[scale=0.8]
				\node at (0+.2,1.9) {$\Lambda_\tau = \Lambda_{\tau+1}$};
				\clip (-1.5+.2,-1.5) rectangle (1.5+.3,1.6); 
				\foreach \i in {-2,...,2} \draw[dotted, darkgray] (\i-3/3,-3) -- (\i+5/3,5);
				\foreach \j in {-1,1} \draw[dotted, darkgray] (-2,1.2*\j) -- (5.5,1.2*\j);
				\draw[thick] (-2,0) -- (5,0);
				\draw[thick] (0,-3) -- (0,5);
				\foreach \i in {-4,...,4} \draw[ultra thin] (\i,-.05) -- (\i,.05);
				\foreach \j in {-2,...,4} \draw[ultra thin] (-.05,\j) -- (.05,\j);
				\node at (1,0) [yshift=-.25cm]{$1$};
				\node at (0,1) [xshift=-.15cm]{$\I$};
				\draw[-latex,very thick] (0,0) -- (1,0);
				\draw[-latex,very thick] (0,0) -- (1.4,1.2) node[shift={(-.1cm,.15cm)}] {$\tau+1$};
			\end{tikzpicture}
		};
		
		\draw[->, thick] (0,-2) -- node[right]{$S\colon \tau \longmapsto -\frac{1}{\tau}$} (0,-3);
		
		\node at (0,-5) {
			\begin{tikzpicture}[scale=0.8]
				\node at (0+.2,1.9) {$\Lambda_\tau =-\tau \,\Lambda_{-1/\tau}$};
				\clip (-1.5+.2,-1.5) rectangle (1.5+.3,1.6); 
				\foreach \i in {-2,...,2} \draw[dotted, darkgray] (\i-3/3,-3) -- (\i+5/3,5);
				\foreach \j in {-1,1} \draw[dotted, darkgray] (-2,1.2*\j) -- (5.5,1.2*\j);
				\draw[-latex,thick,gray!75!white] (0,0) -- (-.4/1.6,1.2/1.6) node[shift={(-.42cm,.075cm)},gray] {$-1/\tau$};
				\draw[thick] (-2,0) -- (5,0);
				\draw[thick] (0,-3) -- (0,5);
				\foreach \i in {-4,...,4} \draw[ultra thin] (\i,-.05) -- (\i,.05);
				\foreach \j in {-2,...,4} \draw[ultra thin] (-.05,\j) -- (.05,\j);
				\node at (1,0) [yshift=-.25cm]{$1$};
				\node at (0,1) [shift={(.15cm,.05cm)}]{$\I$};
				\draw[|->,thin,gray,shorten < =.1cm,shorten > =.1cm] (-.4/1.6,1.2/1.6) to[out=15,in=105] (1,0);
				\draw[|->,thin,gray,shorten < =.45cm,shorten > =.1cm] (1,0) to[out=-95,in=-10] (-.4,-1.2);
				\draw[-latex,very thick] (0,0) -- (1,0);
				\draw[-latex,very thick] (0,0) -- (-.4,-1.2) node[shift={(-.2cm,-.14cm)}] {$-\tau$};
			\end{tikzpicture}
		};
	\end{tikzpicture}
	\caption{The generator $T$ of $\mathrm{SL}(2,\Z)$ simply shifts the modular parameter of the lattice $\Lambda_\tau \subset \mathbb{C}$ as  $\tau \longmapsto \tau+1$. The generator $S$ acts by a signed inversion, $\tau \longmapsto -1/\tau$, which requires a simultaneous coordinate rescaling $x\longmapsto -\tau \, x$ in order to fix the lattice, as indicated}
	\label{fg:modular_action}
\end{figure}

It is not directly obvious how the Ruijsenaars--Schneider functions~\eqref{eq:class_RS_fns} with modular parameter $-1/\tau$ are related to their cousins with parameter $\tau$. We will now construct a variant of the modular transformations that allows for a simple relation between the transformed and original Ruijsenaars--Schneider functions. 

Consider the complexified phase space $M_{\mathbb{C}} = T^* \, \mathbb{C}^N \cong \mathbb{C}^{2N}$ from \eqref{eq:cpx phase space}, where we parametrise the multiplicative classical momenta $\gamma_j = \E^{\epsilon \, p_j}$ in terms of their additive counterparts $p_j$. In the setting of dynamical systems, the coordinates $x_j$ live in the ambient $\mathbb{C}$, so we simultaneously rescale $x_i \longmapsto c \, x_i$ with $c\in \C^\times$. To preserve the canonical Poisson brackets~\eqref{eq:PB x p} we also rescale all $p_j \longmapsto p_j/c$, while the Poisson brackets \eqref{eq:PB x gamma} require rescaling the parameter $\epsilon$ as $\epsilon \longmapsto c\, \epsilon$.\,%
\footnote{\ Note that this rescaling preserves the multiplicative momenta $\gamma_j = \E^{\epsilon \, p_j}$ at the classical level, and also the operators \eqref{eq:Gamma_i} at the quantum level.}
It is then natural to rescale the remaining parameter $\eta \longmapsto c \, \eta$ as well, in order to preserve the form of the coefficients~\eqref{eq:A_coeffs} appearing in \eqref{eq:class_RS_fns}.
Altogether, we are thus led to  
\begin{lem} \label{lem:extended action}
	There is an action of\/ $\mathrm{SL}(2,\mathbb{Z})$ on $M_\mathbb{C} \times \mathbb{C}^2 \times \mathbb{H}$ given by
	\begin{equation} \label{eq:rescaling_symplectomorphism}
		(\vect{x},\vect{p} ; \eta,\epsilon \,|\, \tau) 
		\longmapsto  (c \, \vect{x},\vect{p}/c ; c\,\eta,c \,\epsilon \,|\, \elt \cdot \tau)\, , \qquad \elt \in \mathrm{SL}(2,\mathbb{Z}) \, , \quad c\in \mathbb{C}^\times \, ,
	\end{equation}
	which restricts to a symplectomorphism of 
	$M_\mathbb{C}$.
	Explicitly, this action is generated by
	\begin{equation} \label{eq:S_T_extended}
		\begin{aligned}
			S & : (\vect{x},\vect{p} ; \eta , \epsilon \,|\, \tau) 
			\longmapsto \bigl(-\vect{x}/\tau , -\tau\,\vect{p} ; -\eta/\tau , -\epsilon/\tau \,\big|\, {-}1/\tau \bigr) \, , \\
			T & : (\vect{x},\vect{p} ; \eta,\epsilon \,|\, \tau) 
			\longmapsto (\vect{x} , \vect{p} ; \eta, \epsilon \,|\, \tau + 1) \, . 
		\end{aligned}
	\end{equation}
\end{lem}

We push forward the $\mathrm{SL}(2,\Z)$-action on $M_\C \times \C^2 \times \mathbb{H}$ to functions $f \colon M_\C \times \C^2 \times \mathbb{H} \longrightarrow \C$.
On a theta function $\theta(x \,|\, \tau)$ where $x$ is any linear combination of the $x_i$ and $\eta$, these two generators yield nothing but the Jacobi imaginary transformation and a simple shift of $\tau$: 
\begin{equation} \label{eq:S_T_theta}
	\theta\bigl(-x/\tau \,\big|\, {-}1/\tau\bigr) = \I \, (-\I\,\tau)^{1/2} \, \E^{\I \mspace{2mu} \pi \mspace{2mu} x^2/\tau} \, \theta( x \,|\, \tau) \, , \qquad  
	\theta(x \,|\, \tau+1) = \E^{\I \mspace{2mu} \pi/4} \, \theta(x \,|\, \tau)\, .
\end{equation}
Because of the prefactors in \eqref{eq:S_T_theta} it is not immediately obvious what the effect is of the modular action on the dynamics of the Ruijsenaars--Schneider system.
As one may expect from an elliptic model, it is possible to tweak the action such that the dynamics are preserved:
	\begin{thm} \label{thm:RS_modular}
		The action of\/ $\mathrm{SL}(2,\mathbb{Z})$ on $M_\mathbb{C} \times \mathbb{C}^2 \times \mathbb{H}$ can be modified as
		\begin{equation} \label{eq:S_T_extended_with_shift}
			\begin{aligned}
				S & \colon (\vect{x},\vect{p} ; \eta , \epsilon \,|\, \tau) 
				\longmapsto \biggl(-\frac{\vect{x}}{\tau} , -\tau\,\vect{p} - \frac{2\pi \mspace{2mu}\I \, \eta}{\epsilon} \, \sum_{j=1}^N \bigl( |\vect{x}| - N\mspace{1mu}x_{j} \bigr) \, \vect{e}_j \, ; -\frac{\eta}{\tau} , -\frac{\epsilon}{\tau} \,\bigg|\, {-}\frac{1}{\tau} \biggr) \, , \\
				T & \colon (\vect{x},\vect{p}; \eta,\epsilon \,|\, \tau) 
				\longmapsto (\vect{x},\vect{p}; \eta, \epsilon \,|\, \tau + 1) \, ,
			\end{aligned}
		\end{equation}
		where $|\vect{x}| \coloneqq \sum_{i=1}^N x_{i}$ and $\vect{e}_j$ denotes the unit vector with a $1$ in its $j$th entry. Restricted to $M_\mathbb{C}$, this is a symplectomorphism. Moreover, on functions invariant under $(\vect{x};\eta) \longmapsto (-\vect{x};-\eta)$ it induces an action of the modular group $\mathrm{PSL}(2,\mathbb{Z})$, which acts trivially on the Ruijsenaars--Schneider functions in the sense that $T$ leaves the $D_n^{\mspace{1mu}\mathrm{cl}}$ invariant and there exists $c_n(\eta) \in \mathbb{C}^\times$ such that
		\begin{equation} \label{eq:S.Dn^cl}
			S \,\cdot D_n^{\mspace{1mu}\mathrm{cl}}(\vect{x}, \vect{p};\eta, \epsilon \, |\, \tau) = c_n(\eta) \, D_n^{\mspace{1mu}\mathrm{cl}}(\vect{x}, \vect{p}; \eta, \epsilon \, |\, \tau) \, .
		\end{equation}	
	\end{thm}

\begin{proof}[Proof.]
	A direct computation shows that \eqref{eq:S_T_extended_with_shift} again defines an action of $\mathrm{SL}(2,\Z)$. The action of $T$ is the same as in Lemma~\ref{lem:extended action}. For $S$, the extra shift of the momenta as
		\begin{equation} \label{eq:pj_shift}
			p_{j} \, \longmapsto \, p_{j} + \frac{2\pi \mspace{2mu}\I \, \eta}{\epsilon \, \tau} \sum_{k=1}^N (x_k - x_{j}) =  p_{j} + \frac{2\pi \mspace{2mu}\I \, \eta}{\epsilon \, \tau} \, \bigl( |\vect{x}| - N\mspace{1mu}x_{j} \bigr) \, . 
		\end{equation}	
		preserve the symplectic structure, as one readily verifies that $\{p_i, p_j\}=0$ is respected by \eqref{eq:pj_shift}. The shifts cancel after every second $S$-transformation to leave just an overall minus sign for the $x_j$ as well as $\eta$. In particular, on functions invariant under a simultaneously reflection of $(\vect{x};\eta)$ the formulas \eqref{eq:S_T_extended_with_shift} factor through an action of the modular group $\mathrm{PSL}(2,\mathbb{Z})$ by quotienting out $S^2 = -1$.
	
	Since the theta function is odd, the signs of $(\vect{x};\eta)$ cancel in the coefficients $A_I(\vect{x})$, yielding an $\mathrm{PSL}(2,\mathbb{Z})$-action on the Ruijsenaars--Schneider functions. It remains to examin how this modular action transofrms the Ruijsenaars--Schneider system. One readily verify that
	\begin{equation}
		\label{eq:T_action_on_DRS}
		T\cdot D_n^{\mspace{1mu}\text{cl}}(\vect{x}, \vect{p} ; \eta, \epsilon \, |\, \tau) = D_n^{\mspace{1mu}\text{cl}}(\vect{x}, \vect{p} ;\eta, \epsilon \, |\, \tau+1) =  D_n^{\mspace{1mu}\text{cl}}(\vect{x}, \vect{p} ; \eta, \epsilon \, |\, \tau)\, , 
	\end{equation} 
	as follows from \eqref{eq:S_T_theta} and the fact that the coefficients $A_I$ are homogeneous of degree $0$ in theta functions. Since the Poisson brackets are preserved under this action, so are the dynamics. 
	
	The effect of the $S$-transformation is more complicated. We start from the action in Lemma~\ref{lem:extended action}.
	Observe that the Jacobi imaginary transformation implies
	\begin{equation}
		\frac{\theta\bigl(-(x+\eta)/\tau \,\big|\, {-}1/\tau\bigr)}{\theta\bigl(-x/\tau \,\big|\, {-}1/\tau\bigr)} = \E^{\I\mspace{2mu} \pi \mspace{1mu} ( \eta^2 + 2 \, \eta \,x)/\tau} \, \frac{\theta(x+\eta \,|\, \tau)}{\theta(x \,|\, \tau)} \, .
	\end{equation} 
	Thus
	\begin{align} \label{eq:D_n_S}
		S \cdot D_n^{\mspace{1mu}\text{cl}}(\vect{x}, \vect{p}; \eta, \epsilon \, |\, \tau) & = D_n^{\mspace{1mu}\text{cl}}\bigl(-\vect{x}/\tau , -\tau\,\vect{p} ; -\eta/\tau , -\epsilon/\tau \,\big|\, {-}1/\tau \bigr)  \nonumber \\
		&= 	
		\!\sum_{|I| = n} \!\! A_I\bigl(-\vect{x}/\tau ; -\eta/\tau \,\big|\, {-}1/\tau \bigr) \, \gamma_I  \nonumber \\
		& = \!\sum_{|I| = n} \!\!
		A_I(\vect{x} ;\eta \,|\, \tau) \, \exp \Biggl(\frac{ \I \mspace{2mu}\pi}{\tau} \biggl(n\,(N-n) \, \eta^2 + 2 \, \eta \!\!\sum_{j \in I \niton k}\!\! (x_j - x_{k}) \biggr) \Biggr) \, \gamma_I \\
		&= \E^{\I \mspace{2mu} \pi \, n \, (N-n) \, \eta^2/\tau} \!\sum_{|I| = n}\!\! A_I(\vect{x};\eta \,|\, \tau) \, \exp \Biggl( \frac{2 \pi \mspace{2mu} \I \, \eta}{\tau} \biggl( N \sum_{ j \in I} x_{j} \: - n \, |\vect{x}| \biggr) + \epsilon \sum_{j \in I}  p_{j} \Biggr) \, .  \nonumber
	\end{align} 
	The exponential factor in the sum can be made independent of $I$ by 
	shifting the momenta $\vect{p}$ as in \eqref{eq:pj_shift}.
	Besides the symplectic structure, this shift preserves the value $\sum_j p_j$ of the total momentum and thus of 
	$D_N^{\mspace{1mu}\text{cl}} = \gamma_1 \cdots \gamma_N$.
	By incorporating the momentum shift \eqref{eq:pj_shift} we are led to define the modified action \eqref{eq:S_T_extended_with_shift} for which \eqref{eq:S.Dn^cl} holds as claimed with $c_n(\eta)\coloneqq \E^{\I \mspace{2mu} \pi \, n \, (N-n) \, \eta^2/\tau}$.
\end{proof}

	We have thus shown how the extended modular action \eqref{eq:S_T_extended} on $(\vect{x},\vect{p} ; \eta , \epsilon \,|\, \tau)$ relates different members of the family of Ruijsenaars systems parametrised by $(\eta,\epsilon \,|\, \tau) \in \C^2 \times \mathbb{H}$. More precisely, we are free to rescale the coordinates ($\vect{x},\vect{p}$ classically; $\vect{x}$ quantum-mechanically) and other parameters in such a way that the resulting system is defined on the same lattice. Hence physically \emph{in}equivalent Ruijsenaars systems are parametrised by the quotient $\bigl( \C^2 \times \mathbb{H} \bigr) / \mspace{2mu} \mathrm{PSL}(2,\Z)$. Put differently, for any fixed $\tau \in \mathbb{H}\mspace{1mu}/\mspace{1mu}\mathrm{PSL}(2,\mathbb{Z})$ in the fundamental domain, the modular action generates a family of physically \emph{equivalent} Ruijsenaars systems. At the classical level, this action allows one to construct a modular family of (discrete) equilibrium configurations. This is our next topic.

\subsection{Classical equilibrium configurations} \label{sec:equilibria}

Due to the Poisson commutativity~\eqref{eq:class integrab}, each of the integrals of motion~\eqref{eq:class_RS_fns} generates a time flow $\partial/\partial t_n = \{ \, \cdot \, , D_n^{\mspace{1mu}\text{cl}} \mspace{1mu}\}$ under which all other $D_m^{\mspace{1mu}\text{cl}}$ are preserved. The corresponding velocities
are
\begin{subequations} \label{eq:freezing}
	\begin{align} \label{eq:freezing_xj}
		\frac{\partial x_i}{\partial t_n} & = \{x_i,D_n^{\mspace{1mu}\text{cl}}\mspace{1mu}\} 
		= + \frac{\partial D_n^{\mspace{1mu}\text{cl}}}{\partial p_i} 
		= \epsilon \! \sum_{I \mspace{2mu}:\, I \ni i} \!\! A_I(\vect{x}) \, \gamma_I
		\intertext{
			while the associated jerks (jolts) are}
		\label{eq:freezing_pj}
		\frac{\partial p_j}{\partial t_n} & = \{p_j,D_n^{\mspace{1mu}\text{cl}} \mspace{1mu}\} 
		= -\frac{\partial D_n^{\mspace{1mu}\text{cl}}}{\partial x_j} 
		= -\sum_{I} \partial_{x_j} A_I(\vect{x}) \, \gamma_I \, ,
	\end{align}
\end{subequations}
where the sums run over all $n$-element subsets $I \subset \{1,\dots,N\}$, for \eqref{eq:freezing_xj} subject to $i \in I$. 
We are interested in classical equilibria:

\begin{def*}
	A \emph{classical equilibrium configuration} is a point $(\vect{x}^\star, \vect{p}^\star) \in M_\mathbb{C}$ at which the functions \eqref{eq:freezing} are fixed to values
	\begin{equation} \label{eq:freezing_stationary}
		\frac{\partial x_i^\star}{\partial t_n} = v_n^\star\, , \quad \frac{\partial p_j^\star}{\partial t_n} = 0\, , \qquad 1 \leqslant n \leqslant N \, ,
	\end{equation}
	for constants $v_n^\star \in \mathbb{C}$ depending on the point $(\vect{x}^\star, \vect{p}^\star)$ and the
	parameters $\eta,\epsilon,\tau$, but \emph{not} on 
	$1\leqslant i \leqslant N$.
	We write $M_\mathbb{C}^\star \subset M_\mathbb{C}$ for the set of all such classical equilibrium configurations.
\end{def*}
In particular, by \eqref{eq:freezing_xj}, the partial sums 
\begin{equation} \label{eq:v_n*}
	v_n^\star = \frac{\partial x_i^\star}{\partial t_n}
	= \epsilon \! \sum_{I \mspace{2mu}:\, I \ni i} \!\! A_I(\vect{x}^\star) \, \gamma_I^\star \, , \qquad 
	\gamma_I^\star \coloneqq \prod_{i\in I} \! \gamma_i^\star \, , \quad \gamma_i^\star \coloneqq \E^{\epsilon\,p_i^\star} \, ,
\end{equation} 
must be independent of $i$. For the time flow with $n=1$ this requires all summands to coincide, so that
\begin{equation} \label{eq:v_1*}
	A_i(\vect{x}^\star) \, \gamma_i^\star = v_1^\star/\epsilon \, , \quad i \in I \, .
\end{equation}
\begin{rmk*}
	Classical equilibrium configurations are `frozen' in the sense that they remain stationary in the linearly co-moving frame with (constant) velocity $v_n^\star$. 
\end{rmk*}
\begin{thm} \label{thm:equilibria_modular}
	Given parameters $(\eta,\epsilon\,|\,\tau) \in \mathbb{C}^2 \times \mathbb{H}$, suppose we have an equilibrium configuration $(\vect{x}^\star, \vect{p}^\star)$ for $ D_n^{\mspace{1mu}\mathrm{cl}}(\vect{x}, \vect{p}; \eta, \epsilon \, |\, \tau)$ for which the associated constant velocities obey the 
	symmetry properties
	\begin{equation} \label{eq:vn*_properties}
		v_{N-n}^\star(\eta) = v_n^\star(-\eta) = v_n^\star(\eta) \, , \qquad 1\leqslant n\leqslant N \, .
	\end{equation}
	Then, for any $\elt\in \mathrm{SL}(2,\mathbb{Z})$ acting as in Theorem~\ref{thm:RS_modular}, the dynamical system defined by the functions $\elt\cdot D_n^{\mspace{1mu}\mathrm{cl}} (\vect{x}, \vect{p} ; \eta, \epsilon \, |\, \tau)$ 
	has an equilibrium configuration at $\elt \cdot (\vect{x}^\star, \vect{p}^\star)$ that again obeys \eqref{eq:vn*_properties}. 
\end{thm}

\begin{proof}[Proof.]
	To show this, it suffices to do so for the two generators. Since the action of $T$ changes neither the symplectic structure nor the hamiltonians, cf.~\eqref{eq:T_action_on_DRS}, it obviously preserves equilibrium configurations. So we concentrate on $S$. Let us denote by $\vect{x}' \coloneqq S\cdot \vect{x} = -\vect{x}/\tau$ and $\vect{p}' \coloneqq S \cdot \vect{p} = -\tau\,\vect{p} - 2\pi \mspace{2mu}\I \, \eta \, \epsilon^{-1} \sum_{j=1}^N \bigl( |\vect{x}| - N\mspace{1mu}x_{j} \bigr)\, \vect{e}_j$ the coordinates on $M_\C$ obtained by applying the action of $S$. Since $(\vect{x},\vect{p}) \longmapsto (\vect{x}',\vect{p}')$ is a canonical transformation, for any two functions $f,g \colon M_\C \longrightarrow \C$ the Poisson bracket in the transformed coordinates $\vect{x}',\vect{p}'$ can be expressed in terms of that in terms of the original coordinates as
	\begin{equation} \label{eq:fg_brackets}
		\pigl\{ f(\vect{x}',\vect{p}') , g(\vect{x}',\vect{p}') \pigr\}' = \, \pigr\{ f\bigl( \vect{x}'(\vect{x},\vect{p}) , \vect{p}'(\vect{x},\vect{p})\bigr) , g\bigl( \vect{x}'(\vect{x},\vect{p}) , \vect{p}'(\vect{x},\vect{p}) \bigr) \pigr\} \, , 
	\end{equation}
	and likewise for functions on $M_\mathbb{C} \times \C^2 \times \mathbb{H}$.
	For the transformed velocities~\eqref{eq:freezing_xj} we compute
	\begin{equation}
		\begin{aligned}
			\frac{\partial x_i'}{\partial t_n} = {}& \bigl\{x_i',D_n^{\mspace{1mu}\text{cl}}(\vect{x}', \vect{p}'; \eta', \epsilon' \, |\, \tau') \bigr\}' && \\
			= {}& \bigl\{x_i'(\vect{x}), D_n^{\mspace{1mu}\text{cl}}(\vect{x}'(\vect{x}), \vect{p}'(\vect{p}) ; \eta', \epsilon' \, |\, \tau')\bigr\} && \text{by \eqref{eq:fg_brackets}}  \\
			= {}& \frac{c_n(\eta)}{-1/\tau} \, \bigl\{x_i,  D_n^{\mspace{1mu}\text{cl}}(\vect{x}, \vect{p} ; \eta, \epsilon \, |\, \tau)\bigr\} && \text{by definition of $x'_i$ and \eqref{eq:S.Dn^cl}} \\
			= {}& \frac{c_n(\eta)}{-1/\tau} \, \frac{\partial D_n^{\mspace{1mu}\text{cl}}(\vect{x}, \vect{p};\eta, \epsilon \, |\, \tau)}{\partial p_i}  \, . && \\
		\end{aligned}
	\end{equation}
	Evaluating the coordinates at the equilibrium $(\vect{x}^\star\!\!,\, \vect{p}^\star)$, we obtain new velocities $v'_n(\eta) = \frac{c_n(\eta)}{-1/\tau} \, v_n(\eta)$. Similarly, for the transformed jerks~\eqref{eq:freezing_pj} we calculate
	\begin{equation}
		\begin{aligned}
			\frac{\partial p_i'}{\partial t_n} &= \bigl\{p_i',D_n^{\mspace{1mu}\text{cl}}(\vect{x}', \vect{p}';\eta', \epsilon' \, |\, \tau')\bigr\}' \\
			& = \bigl \{p_i'(\vect{x},\vect{p}) ,  D_n^{\mspace{1mu}\text{cl}}(\vect{x}'(\vect{x},\vect{p}), \vect{p}'(\vect{x},\vect{p}) ; \eta', \epsilon' \, |\, \tau') \bigr\}  \\
			& =  -c_n(\eta) \, \Bigl\{\tau \, p_i + \frac{2\pi \I \eta}{\epsilon} (|\vect{x}| - N x_i) ,D_n^{\mspace{1mu}\text{cl}}(\vect{x}, \vect{p};\eta, \epsilon \, |\, \tau) \Bigr\} \\
			& = -c_n(\eta) \Biggl( \frac{\partial D_n^{\mspace{1mu}\text{cl}}(\vect{x}, \vect{p};\eta, \epsilon \, |\, \tau)}{\partial x_i} + \frac{2\pi \I \eta}{\epsilon} \sum_{j=1}^N (1 - \delta_{ji} \, N) \times - \frac{\partial D_n^{\mspace{1mu}\text{cl}}(\vect{x}, \vect{p};\eta, \epsilon \, |\, \tau)}{\partial p_i} \Biggr) \, . 
		\end{aligned}
	\end{equation}
	At equilibrium this gives 
	\begin{equation}
		\frac{\partial p_i'}{\partial t_n}\bigg|_{(\vect{x}^\star\!\!,\, \vect{p}^\star)} = -c_n(\eta) \, \biggl(0 - \frac{2\pi \I \eta}{\epsilon} \, n(N-n) \, \bigl(v_n^\star(\eta) - v_{N-n}^\star(-\eta) \bigr) \biggr) = 0\, , 
	\end{equation}
	as desired. We conclude that the transformed hamiltonians $S\cdot D_n^\text{cl}(\vect{x},\vect{p} ; \eta, \epsilon \, | \, \tau)$ have an equilibrium configuration at $S \cdot (\vect{x}^\star, \vect{p}^\star)$. Moreover, the new velocities $c_n(\eta)/(-1/\tau) \, v_n(\eta)$ also obey the symmetry property \eqref{eq:vn*_properties}. Hence it follows that for any $B \in \mathrm{PSL}(2,\Z)$ the dynamical system $\elt\cdot D_n^{\mspace{1mu}\text{cl}} (\vect{x}, \vect{p},\eta, \epsilon \, |\, \tau)$ also has an equilibrium configuration. 
\end{proof}

\begin{cor} \label{cor:equilibria_modular}
	The subset of $M_\mathbb{C}^\star \subset M_\mathbb{C}$ consisting of classical equilibria obeying \eqref{eq:vn*_properties} is stable under the $\mathrm{SL}(2,\mathbb{Z})$-action from Theorem~\ref{thm:RS_modular}.
\end{cor}

\subsection{A modular family of classical equilibrium configurations}\label{sec:equilibria_modular}

We would like to find all classical equilibrium configurations of the elliptic Ruijsenaars--Schneider system.
Actually, we will settle for a little less: we will exploit 
Theorems~\ref{thm:RS_modular} and \ref{thm:equilibria_modular} to generate a whole family of classical equilibrium configurations starting, for given parameters $\eta,\epsilon,\tau$, from a simple well-known `seed'
$(\vect{x}^\star,\vect{p}^\star)$ for which \eqref{eq:vn*_properties} holds.%

Let $\omega \in \mathbb{H}$. Consider the standard classical equilibrium configuration $x_i^\star{}^{\, (\SLid)} = i/N$, $p_j^\star{}^{\,(\SLid)} = 0$, with particles that are fixed equidistantly along the real cycle of the torus. To guarantee a more uniform notation it is prudent to choose an analogous dependence of the parameters on $N$ as well. Thus we start from 
\begin{equation} \label{eq:freezing_type1}
	\begin{aligned} 
		x_i^\star{}^{\,(\SLid)} & = \frac{i}{N} \, , \quad && p_j^\star{}^{\,(\SLid)} = 0 \, , \\
		\eta^{(1)} & = \frac{\eta}{N}\, , && \ \: \epsilon^{(\SLid)} = \frac{\epsilon}{N} \, ,
	\end{aligned}
	\qquad\quad \tau^{(\SLid)} = \frac{\omega}{N} \, ,
\end{equation}
with the superscript indicating the neutral element $\SLid \in \text{PSL}(2,\Z)$. It is straightforward to check that the associated velocities satisfy the symmetry property \eqref{eq:vn*_properties}.
By applying the $S$-transformation we obtain the new solution
\begin{equation} \label{eq:freezing_type2}
	\begin{aligned}
		x_i^\star{}^{\,(S)} & = -\frac{i}{\omega} \, , \quad && p_j^\star{}^{\,(S)} = -(N+1-2j) \, \frac{\pi \mspace{2mu}\I \, \eta}{\epsilon} \, ,\\
		\eta^{(S)} & = -\frac{\eta}{\omega} \, , && \ \: \epsilon^{(S)} = -\frac{\, \epsilon}{\omega} \, ,
	\end{aligned} 
	\qquad\quad \tau^{(S)} = -\frac{N}{\omega} \, .
\end{equation}
In this equilibrium configuration the particles are equally spaced along the complex (e.g.\ imaginary) cycle of the torus, see Fig.~\ref{fg:S-trafo}.

\begin{figure}
	\centering
	\begin{tikzpicture}[xscale=0.75,yscale=.6]
		\draw[dotted, darkgray] (5,-.15) -- (5,5.5);
		\draw[dotted, darkgray] (-.15,5) -- (5.5,5);
		\draw[dotted, lightgray] (10,-.15) -- (10,5.5);
		\draw[dotted, lightgray] (5.5,0) -- (10.5,0);
		\draw[dotted, lightgray] (5.5,5) -- (10.5,5);
		\draw[dotted, white!50!lightgray] (15,-.15) -- (15,5.5);
		\draw[dotted, white!50!lightgray] (10.5,0) -- (15.5,0);
		\draw[dotted, white!50!lightgray] (10.5,5) -- (15.5,5);
		\draw[thick] (-.5,0) -- (5.2,0);
		\draw[thick] (0,-.5) -- (0,5.2);
		\node at (5,0) [yshift=-.25cm]{$1$};
		\node at (0,5) [xshift=-.22cm]{$\tau$};
		%
		\draw[ultra thin, white!50!lightgray] (0-.25,0-.25/3) -- (5,5/3) (-.25,5/3-.25/3) -- (5+.25,2*5/3+.25/3) (-.25,2*5/3-.25/3) -- (5+.25,5+.25/3);
		\draw[ultra thin, dashed, white!50!lightgray] (5,5/3) -- (15+.25,5+.25/3);
		\draw[<-,lightgray] (5+.5,5/3+.5/3) -- (5+1,5/3+1/3) node[shift={(.82cm,.04cm)}] {$\elt=T^3\mspace{2mu}S$}; 
		\foreach \x in {.5,1,...,5} \fill[Red] (\x,0) circle (2pt);
		\draw[<-,Red] (5.5,0) -- (6,0) node[xshift=.55cm] {$\elt=\SLid$};
		\foreach \x in {.5,1,...,5} \fill[Green] (0,\x) circle (2pt);
		\draw[<-,Green] (0,5.5) -- (0,6) node[yshift=.17cm] {$\elt=S$}; 
		\draw[ultra thin, lightgray] (0-.25,0-.25) -- (5+.25,5+.25);
		\foreach \x in {.5,1,...,5} \fill[Cyan] (\x,\x) circle (2pt);
		\draw[<-,Cyan] (5.5,5.5) -- (5.85,5.85) node[shift={(.76cm,.15cm)}] {$\elt=T\,S$}; 
		\draw[ultra thin, lightgray] (0,0) -- (5,2.5) (0,2.5) -- (5+.25,5+.125);
		\draw[ultra thin, dashed, lightgray] (-1,2.5-.5) -- (0,2.5) (5,2.5) -- (10+.25,5+.125);
		\foreach \x in {.5,1,...,2.5} \fill[Dandelion] (2*\x,\x) circle (2pt);
		\foreach \x in {3,3.5,...,5} \fill[white!80!Dandelion] (2*\x,\x) circle (2pt);
		\foreach \x in {3,3.5,...,5} \fill[Dandelion] (2*\x-5,\x) circle (2pt);
		\draw[<-,Dandelion] (5+.5,2.5+.25) -- (5+.9,2.5+.45) node[shift={(.85cm,.1cm)}] {$\elt=T^2\mspace{2mu}S$}; 
		\draw[ultra thin, white!60!lightgray] (0-.125,0-.25) -- (2.5,5) (2.5-.125,-.25) -- (5+.125,5+.25);
		\draw[ultra thin, dashed, white!60!lightgray] (2.5,5) -- (3.25+.125,6.5+.25);
		\draw[<-, lightgray] (2.5+.25,5+.5) -- (2.5+.475,5+.95) node[shift={(.15cm,.22cm)}] {$\elt=S\mspace{2mu}T^2\mspace{2mu}S$}; 
		\draw[fill=white] (0,0) circle (2pt);
		\draw[Dandelion,fill=Green] (0,2.5) circle (2pt);
		\fill[Cyan] (5,5) circle (2pt);
		\fill[Dandelion] (5,5) -- +(37.5:2pt) arc (37.5:37.5-180:2pt) -- cycle; 
		\node at (0,0) [shift={(-.17cm,-.22cm)}]{$0$};
	\end{tikzpicture}
	\caption{Examples of equilibrium positions $x_j^\star{}^{\mspace{2mu}(\elt)}$ for simple $\elt \in \mathrm{PSL}(2,\mathbb{Z})$, where we take $\tau \in \I \, \mathbb{R}_{>0}$. For freezing, the choices $\elt = \SLid$ (\textcolor{Red}{$x_j^\star{}^{\mspace{2mu}(\SLid)} = j/N$} real) and especially $\elt = S$ (\textcolor{Green}{$x_j^\star{}^{\mspace{2mu}(S)} = \tau \, j/N$} imaginary) are particularly important. In general, equilibrium positions lie equispaced on any line $[0,t]$ (for $t\in\Lambda_\tau \setminus \{0\}$) that does not intersect any other point in $\Lambda_\tau$}
	\label{fg:S-trafo}
\end{figure}

Applying the $T$-transformation to any solution only
shifts the lattice parameter, which, as discussed above, does not change the dynamics. Nevertheless, any subsequent application of the $S$-transformation propagates the shifted lattice parameter to all other parameters. In particular, this tilts the line in the complex plane on which the equilibrium positions $x_i^\star$ lie. The associated spin-chain hamiltonians that we will construct below differ accordingly. 

More generally, we obtain a new solution from the `seed' configuration \eqref{eq:freezing_type1} for most products of $S,T$ and their inverses;
the only duplicate solutions arise from elements related as $\elt = \elt' \, T^n$ for some $n$,\,%
\footnote{\ Such elements lead to different values of $\tau \in \mathbb{H}$ but the same point $(\vect{x}^\star,\vect{p}^\star)\in M_\mathbb{C}^\star$.} or due to the relations of $\mathrm{PSL}(2,\Z)$. Hence the orbit under the modular group produces a whole family of equilibrium configurations, 
\begin{equation} \label{eq:modular_family_equilib}
	\text{PSL}(2,\mathbb{Z}) \cdot \bigl( \vect{x}^\star{}^{\,(1)}, \vect{p}^\star{}^{\,(1)}  \, ; \eta^{(1)}, \epsilon^{(1)} \,\big|\, \tau^{(1)} 
	\bigr) 
	\,\subset\, M_\mathbb{C}^\star \times \mathbb{C}^2 \times \mathbb{H} \,\subset\, M_{\mathbb{C}} \times \mathbb{C}^2 \times \mathbb{H} \, .
\end{equation}

\begin{rmk*}
	\textbf{a.} Note that the equilibrium \emph{positions} $\vect{x}^\star$ are independent of the deformation parameter~$\eta$, i.e.\ the same as in the `non-relativistic' (Calogero--Sutherland--Moser) limit ($\epsilon \propto \eta/g$, $\eta\to 0$). In contrast, the associated momenta $\vect{p}^\star$ may depend on $\eta$ due to the canonical transformation \eqref{eq:pj_shift} needed to preserve the equilibrium condition. Notably, since the non-relativistic limit requires setting $\epsilon \propto \eta/g$, the shift seems to be required in the elliptic Calogero--Sutherland--Moser case as well.
	
	\textbf{b.} While it seems to be common lore at least for the elliptic Calogero--Sutherland system \cite{doreyEllipticSuperpotentialSoftly1999,bourgetDualityModularityElliptic2015,bourgetN1gauge2016}, we do not know how to prove that no further (discrete) classical equilibria exist for the elliptic Ruijsenaars--Schneider system. 
\end{rmk*}

\section{Elliptic spin-Ruijsenaars system} \label{sec:ell_sRuij}

In this section we return to the quantum level, and introduce the matrix-valued generalisations of the Ruijsenaars operators~\eqref{eq:quantum_RS_ops}--\eqref{eq:D_-n} that will play a central role in what follows. 

Fix $r\in \mathbb{Z}_{\geqslant 1}$. We
generalise $\mathrm{Fun}(\vect{x})$ to the space  $\mathrm{Fun}(\vect{x}) \otimes V^{\otimes N}$ of vector-valued functions, where each particle carries a `spin' that lives in a complex vector space $V \cong \mathbb{C}^r$.
The case $r = 1$ gives back the scalar (spinless) setting of \textsection\ref{sec:ell_Ruijs}.
The \emph{integrable} matrix-valued difference operators that we will consider are built from \textit{R}-matrices~\cite{MZ_23a,klabbers2024deformed}. In view of the coefficients~\eqref{eq:A_coeffs} it is natural to consider \emph{elliptic} \textit{R}-matrices, which come in two well-known variants: 
\begin{itemize}
	\item \textbf{Vertex-type:} Baxter--Belavin 
	\textit{R}-matrix \cite{baxter1972one,belavinDynamicalSymmetryIntegrable1981};
	\item \textbf{Face-type:} Felder's dynamical
	\textit{R}-matrix \cite{baxter1973eight,felder1994elliptic}. 
\end{itemize}
The resulting matrix-valued generalisations of \eqref{eq:quantum_RS_ops} are the spin-Ruijsenaars systems of Matushko and Zotov \cite{MZ_23a,matushko2024supersymmetric} and our dynamical spin-Ruijsenaars systems \cite{klabbers2024deformed}, respectively. For $r>1$ these two models are different, cf.\ \textsection5.2 in \cite{klabbers2025landscapes}, yet they look very similar. A uniform description goes as follows.

\subsection{Deformed spin permutations} \label{sec:P_w(x)}

We will define a family of operators on $V^{\otimes N}$ indexed by the symmetric group $S_N$ that \mbox{($q$-)}deforms the natural $S_N$-action on $V^{\otimes N}$. We start with the analogues of simple transpositions.

\begin{def*}[Cherednik \cite{cherednikQuantumKnizhnikZamolodchikovEquations1992}]
	A \emph{generalised $\check{R}$-matrix} is a family of linear operators $P_{i,i+1}(u) = P_{i,i+1}(u;\eta \,|\, \tau)$ on $V^{\otimes N}$ obeying the unitarity condition
	\begin{equation} \label{eq:unitarity}
		P_{i,i+1}(u)\, P_{i,i+1}(-u) = \mathrm{id} \,,
	\end{equation} 
	the (`braided') Yang--Baxter equation\,%
	\footnote{\ The braided setting simplifies keeping track of the shifts of the dynamical parameters in the face-type setting.}
	\begin{equation} \label{eq:braided_YBe}
		\begin{aligned} 
			P_{i,i+1}(u) \, P_{i+1\mspace{-1mu},\mspace{1mu}i+2}(u+v) \, P_{i,i+1}(v) 
			= P_{i+1\mspace{-1mu},\mspace{1mu}i+2}(v) \, P_{i,i+1}(u+v) \, P_{i+1\mspace{-1mu},\mspace{1mu}i+2}(u) \, ,
		\end{aligned}
	\end{equation}
	and the commutativity
	\begin{equation} \label{eq:commutativity} 				
		[P_{i,i+1}(u),P_{j\mspace{-1mu},\mspace{1mu}j+1}(v)]=0 \, , \qquad |i-j| > 1 \, .
	\end{equation}
\end{def*}
We are particularly interested in the following two examples. 

\begin{ex*}
	Let $\mathbb{1}$ denote the identity on $V$, and $P$ the flip on $V\otimes V$.
	\begin{itemize}
		\item \textbf{Vertex-type.} Let $R(u) = R(u;\eta\,|\,\tau)$ be the Baxter--Belavin \textit{R}-matrix, which is Baxter's eight-vertex \textit{R}-matrix if $r=2$, see \textsection\ref{app:R-mat_vx}. Set $\check{R}(u) = P \, R(u)$. Then 
		\begin{equation} \label{eq:P^v}
			P^\textsc{v}_{i,i+1}(u) = \mathbb{1}^{\otimes( i-1)} \otimes \check{R}(u;\eta \,|\,\tau) \otimes \mathbb{1}^{\otimes (N-i-1)} \, .
		\end{equation}
		is a generalised $\check{R}$-matrix.
		\item \textbf{Face-type.} Let $R(u,\vec{a}\mspace{2mu}) = R(u,\vec{a} ;\eta \,|\,\tau)$ be Felder's dynamical \textit{R}-matrix of type $\mathfrak{gl}_r$, involving `dynamical' parameters $\vec{a}=(a_1,\dots,a_r)$, see \textsection\ref{app:R-mat_face}. It obeys the \emph{dynamical} Yang--Baxter equation, in which the dynamical parameters $\vec{a}$ are shifted depending on the weight of the factors of $V$ to the left of the \textit{R}-matrix. Namely, write $\ketbra{\vec{\mu}\mspace{2mu}}{\vec{\mu}\mspace{2mu}}$ for the projection onto the weight-$\vec{\mu}$ subspace of $V^{\otimes (i-1)}$. Again putting $\check{R}(u,\vec{a}\mspace{2mu}) = P \, R(u,\vec{a}\mspace{2mu})$, we now take
		\begin{equation} \label{eq:P^f}
			P^{\mspace{1mu}\textsc{f}}_{i,i+1}(u) = \sum_{\vec{\mu}} \,
			\ketbra{\vec{\mu}\mspace{2mu}}{\vec{\mu}\mspace{2mu}} \otimes \check{R}\bigl(u,\vec{a}-\eta \, \vec{\mu}\mspace{2mu};\eta \,\big|\,\tau\bigr) \otimes \mathbb{1}^{\otimes (N-i-1)} \, .
		\end{equation}	
		Since $\sum_{\vec{\mu}} \, \ketbra{\vec{\mu}\mspace{2mu}}{\vec{\mu}\mspace{2mu}} = \mathbb{1}^{\otimes (i-1)}$ is (a resolution of) the identity, \eqref{eq:P^f} acts nontrivially only on the $i$th and $i+1$st factors of $V^{\otimes N}$. For a more concrete description of the meaning of \eqref{eq:P^f} for $r=2$ see e.g.\ \textsection2.1 and \textsection{}B in \cite{klabbers2024deformed}. The shifts of $\vec{a}$ ensure that \eqref{eq:braided_YBe} is equivalent to the dynamical Yang--Baxter equation. The operators \eqref{eq:P^f} 
		are also a generalised $\check{R}$-matrix.
	\end{itemize}
\end{ex*}
Since both of these examples moreover obey the condition
\begin{equation} \label{eq:initial}
	P_{i,i+1}(u)\big|_{\eta=0} = P_{i,i+1} \coloneqq \mathbb{1}^{\otimes (i-1)} \otimes P \otimes \mathbb{1}^{\otimes(N-i-1)} \, ,
\end{equation} 
we think of $P_{i,i+1}(u)$ as a ($q$-)deformed nearest-neighbour spin \textit{P}ermutation operator.

We will use the standard graphical notation: each copy of the vector space $V$ (containing a spin) is drawn as a vertical line, which carries an `inhomogeneity parameter'. 
The `deformed spin permutation' $P_{i,i+1}(u)$ is a crossing of the $i$ and $i+1$st lines, which each carry along their inhomogeneities,
\begin{equation} \label{eq:deformed_permutation_diagram}
	P_{i,i+1}(u) = 
	\tikz[baseline={([yshift=-.5*11pt*0.3]current bounding box.center)},xscale=.6,yscale=0.3,font=\footnotesize]{
		\draw[->] (0,0) -- (0,3);
		\foreach \x in {-1,...,1} \draw (.75+.2*\x,1.5) node{$\cdot\mathstrut$};	
		\draw[->] (1.5,0) -- (1.5,3);
		\draw[rounded corners=2pt,->] (3.5,0) node[below]{$\,x\smash{'}$} -- (3.5,1) -- (2.5,2) -- (2.5,3) node[above]{$x\smash{'}$};
		\draw[rounded corners=2pt,->] (2.5,0) node[below]{$x$} -- (2.5,1) -- (3.5,2) -- (3.5,3) node[above]{$x$};
		\draw[->] (4.5,0) -- (4.5,3);
		\foreach \x in {-1,...,1} \draw (5.25+.2*\x,1.5) node{$\cdot\mathstrut$};	
		\draw[->] (6,0) -- (6,3);
		\node at (-.5,1.5) {\smash{\textcolor{lightgray}{$\vec{a}$}}\vphantom{$a$}};
	} \ \ , \quad u = x - x' \, .
\end{equation}
The only difference between the vertex- and face-type examples for $P(u)$ is that, in the latter case, the left-most face is decorated with the dynamical parameters $\vec{a}$ (indicated in gray), determining the dynamical parameters on all other faces, see \textsection\ref{app:R-mat_face}. Then \eqref{eq:unitarity} with $u = x - x'$ becomes
\begin{gather} \label{eq:unitarity_diagram}
	\tikz[baseline={([yshift=-.5*11pt*0.4]current bounding box.center)},xscale=.5,yscale=0.25,font=\footnotesize]{
		\node at (-3,2.5) {\smash{\textcolor{lightgray}{$\vec{a}$}}\vphantom{$a$}};
		\draw[->] (-2.5,0) -- (-2.5,5);
		\foreach \x in {-1,...,1} \draw (-2.5+.75+.2*\x,2.5) node{$\cdot\mathstrut$};	
		\draw[->] (-1,0) -- (-1,5);
		\draw[rounded corners=2pt,->] (1,0) node[below]{$\,x\smash{'}$}-- (1,1) -- (0,2) -- (0,3) -- (1,4) -- (1,5) node[above]{$x\smash{'}$};
		\draw[rounded corners=2pt,->] (0,0) node[below]{$x$} -- (0,1) -- (1,2) -- (1,3) -- (0,4) -- (0,5) node[above]{$x$};
		\draw[->] (2,0) -- (2,5);
		\foreach \x in {-1,...,1} \draw (2+.75+.2*\x,2.5) node{$\cdot\mathstrut$};	
		\draw[->] (3.5,0) -- (3.5,5);
	} \ \ = \,
	\tikz[baseline={([yshift=-.5*11pt*0.4]current bounding box.center)},xscale=.5,yscale=0.25,font=\footnotesize]{
		\node at (-3,1.5) {\smash{\textcolor{lightgray}{$\vec{a}$}}\vphantom{$a$}};
		\draw[->] (-2.5,0) -- (-2.5,3);
		\foreach \x in {-1,...,1} \draw (-2.5+.75+.2*\x,1.5) node{$\cdot\mathstrut$};	
		\draw[->] (-1,0) -- (-1,3);
		\draw[rounded corners=2pt,->] (1,0) node[below]{$\,x\smash{'}$}
		-- (1,3) node[above]{$\,x\smash{'}$};
		\draw[rounded corners=2pt,->] (0,0) node[below]{$x$} -- (0,3) node[above]{$x$};
		\draw[->] (2,0) -- (2,3);
		\foreach \x in {-1,...,1} \draw (2+.75+.2*\x,1.5) node{$\cdot\mathstrut$};	
		\draw[->] (3.5,0) -- (3.5,3);
	} \ \ , 
\end{gather}
while \eqref{eq:braided_YBe} with $v = x' - x''$ is
\begin{equation}
	\begin{aligned}
		\tikz[baseline={([yshift=-.5*11pt*0.4]current bounding box.center)},xscale=.5,yscale=0.25,font=\footnotesize]{
			\node at (-3,2.5) {\smash{\textcolor{lightgray}{$\vec{a}$}}\vphantom{$a$}};
			\draw[->] (-2.5,0) -- (-2.5,5);
			\foreach \x in {-1,...,1} \draw (-2.5+.75+.2*\x,2.5) node{$\cdot\mathstrut$};	
			\draw[->] (-1,0) -- (-1,5);
			\draw[rounded corners=2pt,->] (0,0) node[below]{$x$} -- (0,1) -- (2,3) -- (2,5) node[above]{$x$};
			\draw[rounded corners=2pt,->] (1,0) node[below]{$x\smash{'}\,$} 
			-- (1,1) -- (0,2) -- (0,3) -- (1,4) -- (1,5) node[above]{$\;x\smash{'}$};
			\draw[rounded corners=2pt,->] (2,0) node[below]{$\,x\smash{''}$} -- (2,2) -- (0,4) -- (0,5) node[above]{$x\smash{''}\,$};
			\draw[->] (3,0) -- (3,5);
			\foreach \x in {-1,...,1} \draw (3+.75+.2*\x,2.5) node{$\cdot\mathstrut$};	
			\draw[->] (4.5,0) -- (4.5,5);
		} \ \ = \,
		\tikz[baseline={([yshift=-.5*11pt*0.4]current bounding box.center)},xscale=-.5,yscale=0.25,font=\footnotesize]{
			\node at (5,2.5) {\smash{\textcolor{lightgray}{$\vec{a}$}}\vphantom{$a$}};
			\draw[->] (-2.5,0) -- (-2.5,5);
			\foreach \x in {-1,...,1} \draw (-2.5+.75+.2*\x,2.5) node{$\cdot\mathstrut$};	
			\draw[->] (-1,0) -- (-1,5);
			\draw[rounded corners=2pt,->] (0,0) node[below]{$\,x\smash{''}$}
			-- (0,1) -- (2,3) -- (2,5) node[above]{$x\smash{''}\,$};
			\draw[rounded corners=2pt,->] (1,0) node[below]{$x\smash{'}\,$} -- (1,1) -- (0,2) -- (0,3) -- (1,4) -- (1,5) node[above]{$\,x\smash{'}$};
			\draw[rounded corners=2pt,->] (2,0) node[below]{$x$} -- (2,2) -- (0,4) -- (0,5) node[above]{$x$};
			\draw[->] (3,0) -- (3,5);
			\foreach \x in {-1,...,1} \draw (3+.75+.2*\x,2.5) node{$\cdot\mathstrut$};	
			\draw[->] (4.5,0) -- (4.5,5);
		} \ \ .
	\end{aligned}
\end{equation}
Note that, while the inhomogeneities are carried around by the lines, (the subscripts labelling) the vector spaces do not move, since we work with $\check{R}(x) = P \, R(x)$.
In the following, each diagram starts out (at the bottom) with inhomogeneity parameters given by the particle coordinates $x_1,\dots,x_N$.

\begin{def*}[Cherednik \cite{cherednikQuantumKnizhnikZamolodchikovEquations1992}; see also \textsection\ref{app:deformed_spin_perm gen}]
	We define operators $P_w(\vect{x})$ for any $w\in S_N$ 
	starting from
	\begin{equation} \label{eq:id}
		P_e(\vect{x}) = \mathrm{id} = 
		\tikz[baseline={([yshift=-.5*11pt*0.4]current bounding box.center)},xscale=.5,yscale=0.25,font=\footnotesize]{
			\node at (-.5,1.5) {\smash{\textcolor{lightgray}{$\vec{a}$}}\vphantom{$a$}};
			\draw[->] (0,0) node[below]{$x_1$}
			-- (0,3) node[above]{$x_1$};
			\draw[->] (1,0) node[below]{$x_2$} -- (1,3) node[above]{$x_2$};
			\draw[->] (2,0) -- (2,3);
			\foreach \x in {-1,...,1} \draw (2+.75+.2*\x,1.5) node{$\cdot\mathstrut$};	
			\draw[->] (3.5,0) node[below]{$x_N$} -- (3.5,3) node[above]{$x_N$};
		}
	\end{equation}
	and using recursion via the cocycle condition
	\begin{equation} \label{eq:cocycle}
		P_{w\,(i,i+1)}(\vect{x}) = P_{w}(x_1,\dots,x_{i+1},x_i,\dots,x_N) \, P_{i,i+1}(x_i - x_{i+1}) \, ,
	\end{equation}
\end{def*} 
The result is well defined thanks to \eqref{eq:unitarity}--\eqref{eq:commutativity}, see \textsection\ref{app:deformed_spin_perm gen}. Note that most $P_w(\vect{x})$ do not actually depend on all $x_1,\dots,x_N$, as the examples \eqref{eq:id} and $P_{(i,i+1)}(\vect{x}) = P_{i,i+1}(x_i - x_{i+1})$ illustrate. The swap $x_i \leftrightarrow x_{i+1}$ in \eqref{eq:cocycle} takes into account how the parameters move around in the graphical notation. A general formula for $P_w(\vect{x})$ can be found in \textsection\ref{app:deformed_spin_perm gen}.

We will be interested in a subset of permutations that can be concisely defined as follows.
\begin{def*}[see also \textsection\ref{app:grassmann}]
	For $1\leqslant n\leqslant N$, given an $n$-element subset $I=\{i_1 < \dots < i_n\} \subseteq \{1,\dots,N\}$, consider the (Grassmannian) permutation $w_I \in S_N$ that sends $k \longmapsto i_k$ for all $1\leqslant k \leqslant n$ while permuting neither any of $\{1,\dots,n\}$ amongst each other, nor any of $\{n+1,\dots,N\}$. From it, we define 
	\begin{equation} \label{eq:P_I}
		P_I(\vect{x}) \coloneqq P_{w_I^{-1}}(\vect{x}) \, , \qquad P_{-I}(\vect{x}) \coloneqq P_{\{1,\dots,N\} \setminus I}(\vect{x}) \, .
	\end{equation}
\end{def*}

As $n<N/2$ increases the $P_I(\vect{x})$ become more complicated, see \eqref{eq:P_I decomp} below, until $n=N/2$ where they start to simplify again and it is more convenient to switch to the $P_{-I}(\vect{x})$.

\begin{ex*}
	As discussed in more detail in \textsection\ref{app:cycles}, at $n=1$ the cycle $w_{\{i\}} = (i~i-1\,\dots\,1)$ gives 
	\begin{gather} \label{eq:P_1...i}
		P_{\{i\}}(\vect{x}) = P_{(1~\dots~i-1~i)}(\vect{x}) = P_{12}(x_1 - x_{i}) \cdots P_{i-1,i}(x_{i-1} - x_i) =
		\tikz[baseline={([yshift=-.5*11pt*0.3]current bounding box.center)},xscale=-.5,yscale=0.25,font=\footnotesize]{
			\draw[->] (-1,0) node[below]{$x_N$} -- (-1,5) node[above]{$x_N$};
			\foreach \x in {-1,...,1} \draw (-.25+.2*\x,2.5) node{$\cdot\mathstrut$};	
			\draw[->] (.5,0) -- (.5,5);
			\draw[rounded corners=2pt,->] (1.5,0) node[below]{$x_i$} -- (1.5,1) -- (4.5,4) -- (4.5,5) node[above]{$x_i$};
			\draw[rounded corners=2pt,->] (2.5,0) node[below]{$x_{i-1}\;$} -- (2.5,1) -- (1.5,2) -- (1.5,5) node[above]{$\,x_{i-1}$};
			\draw[rounded corners=2pt,->] (3.5,0) node[below]{$\!\!\!\!\vphantom{x_i}\dots$} -- (3.5,2) -- (2.5,3) -- (2.5,5) node[above]{$\!\dots$};
			\draw[rounded corners=2pt,->] (4.5,0) node[below]{$x_1$} -- (4.5,3) -- (3.5,4) -- (3.5,5) node[above]{$x_1$};
			\node at (5,2.5) {\smash{\textcolor{lightgray}{$\vec{a}$}}\vphantom{$a$}};
		} , \\
		\label{eq:P_N...i}
		P_{-\{i\}}(\vect{x}) = P_{(N~N-1~\dots~i)}(\vect{x}) = P_{N-1,N}(x_i - x_{N}) \cdots P_{i,i+1}(x_i - x_{i+1}) = 
		\tikz[baseline={([yshift=-.5*11pt*0.3]current bounding box.center)},xscale=.5,yscale=0.25,font=\footnotesize]{
			\draw[->] (-1,0) node[below]{$x_1$} -- (-1,5) node[above]{$x_1$};
			\foreach \x in {-1,...,1} \draw (-.25+.2*\x,2.5) node{$\cdot\mathstrut$};	
			\draw[->] (.5,0) -- (.5,5);
			\draw[rounded corners=2pt,->] (1.5,0) node[below]{$x_i$} -- (1.5,1) -- (4.5,4) -- (4.5,5) node[above]{$x_i$};
			\draw[rounded corners=2pt,->] (2.5,0) node[below]{$\;x_{i+1}$} -- (2.5,1) -- (1.5,2) -- (1.5,5) node[above]{$x_{i+1}\,$};
			\draw[rounded corners=2pt,->] (3.5,0) node[below]{$\vphantom{x_i}\dots\!\!\!$} -- (3.5,2) -- (2.5,3) -- (2.5,5) node[above]{$\dots\!$};
			\draw[rounded corners=2pt,->] (4.5,0) node[below]{$x_N$} -- (4.5,3) -- (3.5,4) -- (3.5,5) node[above]{$x_N$};
			\node at (-1.5,2.5) {\smash{\textcolor{lightgray}{$\vec{a}$}}\vphantom{$a$}};
		} .
	\end{gather}
\end{ex*}

\subsection{Matrix-valued Ruijsenaars operators} \label{sec:Dtildes}
In terms of the deformed spin permutations, the 
matrix-valued generalisations of the scalar Ruijsenaars operators \eqref{eq:quantum_RS_ops}--\eqref{eq:D_-n} that we will consider are as follows.
\begin{def*}
	Given a generalised $\check{R}$-matrix, define \emph{spin-Ruijsenaars operators}
	\begin{equation} \label{eq:Ruij_spin_pm n}
		\widetilde{D}_{\pm n}(\vect{x}; \hbar,\eta,\epsilon\textcolor{lightgray}{,\vec{a}} \,|\, \tau) 
		\coloneqq \sum_{|I| = n} \! A_{\pm I}(\vect{x}) \, P_{\pm I}(\vect{x})^{-1} \, \Gamma_{\pm I} \, P_{\pm I}(\vect{x}) \, , \qquad 1 \leqslant n \leqslant N \, ,  
	\end{equation}
	where $A_{-I}(\vect{x}) = A_{I}(-\vect{x}) = A_I(\vect{x})|_{\eta \mspace{2mu}\mapsto\mspace{1mu} -\eta}$ and $\Gamma_{-I} = \Gamma_I^{-1} = \Gamma_{I}|_{\epsilon \mspace{2mu}\mapsto\mspace{1mu} -\epsilon}$ as in the scalar case.
\end{def*}

These operators depend on the parameters from the scalar case, along with any further parameters from the $R$-matrix (e.g.\ $\vec{a}$ in the face-type case); again, we will often suppress this from our notation.

\begin{ex*}
	Due to \eqref{eq:P_1...i}, the 
	first spin-Ruijsenaars operator 
	takes the form
	\begin{equation} \label{eq:Dtilde_1}
		\begin{aligned}
			\widetilde{D}_{1} & = \sum_{i=1}^N A_{i}(\vect{x}) \, P_{(1~\dots~i)}(\vect{x})^{-1} \, \Gamma_i \, P_{(1~\dots~i)}(\vect{x}) \\[-1ex]
			& = \sum_{i=1}^N A_{i}(\vect{x}) \times\! 		
			\tikz[baseline={([yshift=-.5*11pt*.2-1pt]current bounding box.center)},xscale=.5,yscale=.25,font=\footnotesize]{
				\draw[->] (10.5,0) node[below]{$x_N$} -- (10.5,10) node[above]{$x_N$};
				\draw[->] (9,0) -- (9,10);
				\foreach \x in {-1,...,1} \draw (9.25+.2*\x,-1) node{$\cdot\mathstrut$};
				\foreach \x in {-1,...,1} \draw (9.25+.2*\x,11) node{$\cdot\mathstrut$};
				\draw[very thick, rounded corners=2pt] (8,0) 	
				node[below]{$x^{\smash{\vect{-}}}_i$}
				-- (8,1) -- (5,4) -- (5,5) node{$\mathllap{\epsilon\,}\tikz[baseline={([yshift=-.5*11pt*.25]current bounding box.center)},scale=.35]{\fill[black] (0,0) circle (.2)}$};
				\draw[rounded corners=2pt,->] (5,5) -- (5,6) -- (8,9) -- (8,10) node[above]{$\smash{x_i}\vphantom{x_i}$};
				\draw[rounded corners=2pt,->] (7,0) -- (7,1) -- (8,2) -- (8,5) -- (8,8) -- (7,9) -- (7,10);
				\draw[rounded corners=2pt,->] (6,0) -- (6,2) -- (7,3) -- (7,5) -- (7,7) -- (6,8) -- (6,10);
				\foreach \x in {-1,...,1} \draw (6.5+.2*\x,-1) node{$\cdot\mathstrut$};
				\foreach \x in {-1,...,1} \draw (6.5+.2*\x,11) node{$\cdot\mathstrut$};
				\draw[rounded corners=2pt,->] (5,0) node[below]{$x_1$} -- (5,3) -- (6,4) -- (6,5) -- (6,6) -- (5,7) -- (5,10) node[above]{$x_1$};
				\foreach \x in {-1,...,1} \draw (9.75+.2*\x,5) node{$\cdot\mathstrut$};		
				\node at (4.4,2.5) {\smash{\textcolor{lightgray}{$\vec{a}$}}\vphantom{$a$}};
			} 
			\qquad\quad
			\tikz[baseline={([yshift=-.5*11pt*.2-2pt]current bounding box.center)},xscale=.5,yscale=.25,font=\footnotesize]{
				\fill[black] (0,5) node {$\mathllap{\epsilon\,}\tikz[baseline={([yshift=-.5*11pt*.35]current bounding box.center)},scale=.35]{\fill[black] (0,0) circle (.2)}$}; 
				\draw[very thick] (0,3.5) node[below] 
				{$x^{\smash{\vect{-}}}_i$}
				-- (0,5.13);
				\draw[->] (0,5.13) -- (0,6.75) node[above] {$x_i$};
			} \!\! = \Gamma_i \, , \qquad x^{\smash{\vect{-}}}_i \equiv x_i -\I\,\hbar\,\epsilon \\[-1ex]
			& = \sum_{i=1}^N A_{i}(\vect{x}) \, P_{(1~\dots~i)}(\vect{x})^{-1} P_{(1~\dots~i)}(x_1,\dots,x_i^-, \dots, x_N) \; \Gamma_i \, ,
		\end{aligned}
	\end{equation}
	where in the last line we pushed the difference operator to the right. 
	By \eqref{eq:P_N...i}, its `mirror image' is
		\begin{equation} \label{eq:tildeD_-1}
			\begin{aligned} 
				\widetilde{D}_{-1} & = \sum_{i=1}^N A_{-i}(\vect{x}) \, P_{(N~N-1~\dots~i)}(\vect{x})^{-1} \, \Gamma_i^{-1} \, P_{(N~N-1~\dots~i)}(\vect{x}) \\[-1ex]
				& = \sum_{i=1}^N A_{-i}(\vect{x}) \times \! \tikz[baseline={([yshift=-.5*11pt*.2-1pt]current bounding box.center)},xscale=-.5,yscale=.25,font=\footnotesize]{
					\draw[->] (10.5,0) node[below]{$x_1$} -- (10.5,10) node[above]{$x_1$};
					\draw[->] (9,0) -- (9,10);
					\foreach \x in {-1,...,1} \draw (9.25+.2*\x,-1) node{$\cdot\mathstrut$};
					\foreach \x in {-1,...,1} \draw (9.25+.2*\x,11) node{$\cdot\mathstrut$};
					\draw[very thick, rounded corners=2pt] (8,0) 	
					node[below]{$x^{\smash{\vect{+}}}_i$}
					-- (8,1) -- (5,4) -- (5,5) node{$\tikz[baseline={([yshift=-.5*11pt*.25]current bounding box.center)},scale=.35]{\fill[black] (0,0) circle (.2)}\mathrlap{\;-\epsilon}$};
					\draw[rounded corners=2pt,->] (5,5) -- (5,6) -- (8,9) -- (8,10) node[above]{$\smash{x_i}\vphantom{x_i}$};
					\draw[rounded corners=2pt,->] (7,0) -- (7,1) -- (8,2) -- (8,5) -- (8,8) -- (7,9) -- (7,10);
					\draw[rounded corners=2pt,->] (6,0) -- (6,2) -- (7,3) -- (7,5) -- (7,7) -- (6,8) -- (6,10);
					\foreach \x in {-1,...,1} \draw (6.5+.2*\x,-1) node{$\cdot\mathstrut$};
					\foreach \x in {-1,...,1} \draw (6.5+.2*\x,11) node{$\cdot\mathstrut$};
					\draw[rounded corners=2pt,->] (5,0) node[below]{$x_N$} -- (5,3) -- (6,4) -- (6,5) -- (6,6) -- (5,7) -- (5,10) node[above]{$x_N$};
					\foreach \x in {-1,...,1} \draw (9.75+.2*\x,5) node{$\cdot\mathstrut$};		
					\node at (11,3.5) {\smash{\textcolor{lightgray}{$\vec{a}$}}\vphantom{$a$}};
				}
				\qquad\quad
				\tikz[baseline={([yshift=-.5*11pt*.2-2pt]current bounding box.center)},xscale=.5,yscale=.25,font=\footnotesize]{
					\fill[black] (0,5) node {$\tikz[baseline={([yshift=-.5*11pt*.35]current bounding box.center)},scale=.35]{\fill[black] (0,0) circle (.2)}\mathrlap{\;-\epsilon}$}; 
					\draw[very thick] (0,3.5) node[below] 
					{$x^{\smash{\vect{+}}}_i$}
					-- (0,5.13);
					\draw[->] (0,5.13) -- (0,6.75) node[above] {$x_i$};
				} \ \ = \Gamma_i^{-1} \, , \qquad x^{\smash{\vect{+}}}_i \equiv x_i +\I\,\hbar\,\epsilon \\[-1ex]
				& = \sum_{i=1}^N A_{-i}(\vect{x}) \, P_{(N~N-1~\dots~i)}(\vect{x})^{-1}  P_{(N~N-1~\dots~i)}(x_1,\dots,x_i^+,\dots,x_N) \; \Gamma_i^{-1} \, . 
			\end{aligned}
		\end{equation}
		Higher $\widetilde{D}_n$ can be depicted and explicitly given similarly, see e.g.\ \eqref{eq:second_difference_op} below for $n=2$. A closed form expression (reduced decomposition) for the general deformed spin permutations $P_I$ will be given in \eqref{eq:P_I decomp}.
\end{ex*}

Set $\widetilde{D}_0 \coloneqq \mathrm{id}$. The following result is a variant of Lemma~\ref{lem:D_-n} in the matrix-valued setting.
	\begin{lem}
		The spin-Ruijsenaars operators $\widetilde{D}_{-n}$ are related to the $\widetilde{D}_m$ by 
		\begin{equation}
			\widetilde{D}_{-n} = \widetilde{D}_N^{-1}\, \widetilde{D}_{N-n}  \, , \qquad 1\leqslant n \leqslant N  \, .
		\end{equation}
	\end{lem}
	\begin{proof}
		The $N$th spin-Ruijsenaars operator $\widetilde{D}_N = D_N \, \mathrm{id}$ acts as the identity on spins, and is again invertible. Now follow the proof of Lemma~\ref{lem:D_-n}. For the deformed spin permutations, observe that $D_N$ commutes with all $P_{\pm I}(\vect{x})$, since the latter depend on coordinate differences only by construction~\eqref{eq:cocycle}, and recall that $P_{\{1,\dots,N\}\setminus I}(\vect{x}) = P_{-I}(\vect{x})$ by definition~\eqref{eq:P_I}.
	\end{proof}

For certain choices of \textit{R}-matrices, the spin-Ruijsenaars
system is quantum integrable in the sense that the operators \eqref{eq:Ruij_spin_pm n} commute pairwise like in \eqref{eq:commutativity_scalar_quantum},
\begin{equation} \label{eq:Dtilde commute}
	\bigl[ \widetilde{D}_n, \widetilde{D}_m \bigr] = 0 \, , \qquad  -N \leqslant n ,m\leqslant N \, .
\end{equation}
In the case of elliptic coefficients $A_I$ given by \eqref{eq:A_coeffs}, the \textit{R}-matrices need to be elliptic too. 

	\begin{thm*}[Matushko--Zotov~\cite{MZ_23a}]
		The vertex-type elliptic spin-Ruijsenaars operators \eqref{eq:Ruij_spin_pm n}, with Baxter--Belavin generalised $\check{R}$-matrix \eqref{eq:P^v}, obey \eqref{eq:Dtilde commute}. 
	\end{thm*}
	In \cite{klabbers2024deformed} we announced the analogous result
	\begin{thm*}
		The face-type elliptic spin-Ruijsenaars operators \eqref{eq:Ruij_spin_pm n}, with Felder's dynamical generalised \textit{R}-matrix \eqref{eq:P^f}, obey \eqref{eq:Dtilde commute}. 
	\end{thm*}
	Our proof goes well beyond this paper, and will appear elsewhere.

\section{Freezing} \label{sec:freezing}

\noindent A suitable expansion of our quantum many-body system with spins yields a family of commuting operators that act (nontrivially) on spins only via a method called `freezing' \cite{Pol_93}. This is a two-step process, 
\begin{equation} \label{eq:2-step-process}
	\begin{array}{c}
		\text{quantum} \\ \text{many-body system} \\ \text{with spins} \\ 
	\end{array}
	\longmapsto
	\begin{array}{c}
		\text{hybrid} \\ \text{many-body system} 
	\end{array}
	\longmapsto
	\begin{array}{c}
		\text{long-range} \\ \text{(quantum) spin chain}\, .
	\end{array}
\end{equation}	
Loosely speaking, the physical intuition behind freezing is as follows. Starting from a (quantum) spin-Ruijsenaars system, one takes a limit in which the potential energy dominates the kinetic energy. It can be viewed as a strong-coupling or partial (semi)classical limit. 
In this limit, the system decouples into a `hybrid system', consisting of a classical Ruijsenaars--Schneider system that governs the dynamics (times the identity on spins) plus a quantum part with spin interactions (but no difference/differential operators). The dynamics completely disappear at any classical equilibrium configuration, yielding a system of quantum-mechanical spins at fixed positions: this is the spin chain. While it is possible and interesting to interpret the intermediate hybrid system as a (partially quantum) many-body system in its own right \cite{liashyk2024classical}, this is \emph{not} essential for freezing \cite{chalykh2024integrability}: the (quantum) integrability of the initial quantum many-body system with spins contains everything needed to prove the (quantum) integrability of the resulting long-range spin chain. We will return to the hybrid systems in \textsection\ref{sec:hybrid_systems}.

The aim of this section is to describe how the freezing process can be made precise. 
To state our main result we need some notation.
	\begin{def*}
		For any $\elt \in \text{PSL}(2,\mathbb{Z})$ we define the \emph{evaluation map}
		\begin{equation} \label{eq:evaluation}
			\begin{aligned}
				\mathrm{ev}_{\mspace{-2mu}\elt} \colon (\vect{x}, \vect{p} ; \eta, \epsilon \textcolor{lightgray}{,\vec{a}} \,|\, \tau) \longmapsto {} & \pigl(\vect{x}^\star{}^{(\elt)}, \vect{p}^\star{}^{(\elt)} ; \eta^\star{}^{(\elt)}, \epsilon^\star{}^{(\elt)} \textcolor{lightgray}{,\vec{a}^{\,\star}{}^{(\elt)}} \,\pig|\, \tau^{(\elt)} \pigr)  \\
				& = \elt \cdot \pigl(\vect{x}^\star{}^{(1)}, \vect{p}^\star{}^{(1)} ; \eta^\star{}^{(1)}, \epsilon^\star{}^{(1)} \textcolor{lightgray}{,\vec{a}^{\,\star}{}^{(1)}} \,\pig|\, \tau^{(1)} \pigr) , \quad \elt \in \text{PSL}(2,\Z) \, ,
			\end{aligned}
		\end{equation}
		that restricts the coordinates
		to the classical equilibrium configuration generated by $\elt \in \text{PSL}(2,\Z)$ and specialises the parameters accordingly as per \textsection\ref{sec:equilibria_modular}. 
	\end{def*}
	From this,
	\begin{def*}
		For any fixed $\elt \in \text{PSL}(2,\mathbb{Z})$ we define the \emph{spin-chain hamiltonians} as
		\begin{equation} \label{eq:spin chain hamiltonians}
			H_{n,\elt} \coloneqq \mathrm{ev}_{\mspace{-2mu}\elt} \biggl(
			\widetilde{D}_n^{(1)} 
			\Big|_{\substack{ \Gamma_i \longmapsto \gamma_i \\ \hbar \longmapsto 0}}  \biggr)  \; \in \mathrm{End}(V^{\otimes N}) \, , \qquad -N \leqslant n \leqslant N \, .
		\end{equation}
	\end{def*}
	It will shortly become clear that the replacement really is a partial classical limit. The main result of this section is
	\begin{thm} \label{thm:HnB commute}
		Consider a family of 
		spin-Ruijsenaars operators as in \textsection\ref{sec:Dtildes} that commute pairwise. For any choice of classical equilibrium labeled by $\elt \in \mathrm{PSL}(2,\mathbb{Z})$ the spin-chain hamiltonians obtained by freezing obey
		\begin{equation} \label{eq:spin chain integrable}
			\bigl[H_{n,\elt},H_{m,\elt}\bigr] = 0 \, , \qquad -N \leqslant n, m \leqslant N \, . 
		\end{equation}
	\end{thm}
	This result can also be proven by a direct application of Theorem~5.5 of \cite{chalykh2024integrability}, cf.\ the start of \textsection\ref{sec:freezing_hybrid}. Our proof in \textsection\ref{sec:proof} is closer to e.g.~\cite{TH_95,lamers2022spin,MZ_23a}, and shows precisely how the different orders conspire to produce commuting spin-chain hamiltonians.
We follow \cite{Ugl_95u,lamers2022spin}, supplemented by further insights from \cite{MZ_23a} for the elliptic case, and 
\cite{chalykh2024integrability} for the setting of deformation quantisation.

\subsection{Partial classical limit} \label{sec:partial class lim}
For the purpose of freezing, the formalism of deformation quantisation from \textsection\ref{sec:deformation_quantisation_scalar} is adapted to the spin case as follows.

For $r\geqslant 2$ we upgrade the space $\alg_\hbar$ from \eqref{eq:A_hbar} to its matrix-valued variant\,%
\footnote{\ \label{fn:grp_alg} One may replace $\text{Mat}(r^N,\mathbb{C}) \cong \mathrm{End}(V^{\otimes N})$ for $V\cong \mathbb{C}^r$ by any representation $U$ of $S_N$, or, if one prefers to work abstractly rather than in a representation, the group algebra $\mathbb{C}S_N$, cf.\ e.g.\ \cite{chalykh2024integrability}. In this language, the bracket \eqref{eq:PB_spin} is simply $\{a,b\otimes w\} = \{a,b\} \otimes w$, for $a,b\in \mathrm{Fun}(\vect{x})[\gamma_1^{\pm1},\dots, \gamma_N^{\pm1}]$ and $w \in \mathbb{C}S_N$.}
\begin{equation} \label{eq:Ahbar_spin}
	\widetilde{\alg}_\hbar \coloneqq \mathrm{Fun}(\vect{x})\otimes \text{Mat}(r^N, \C)\otimes \C[\Gamma_1^{\pm1},\dots, \Gamma_N^{\pm1}] \otimes \C[\mspace{-2mu}[\hbar]\mspace{-2mu}] \, ,
\end{equation}
which again is noncommutative and thus has a nontrivial Poisson bracket given by the commutator. 

For our purposes, the classical limit~\eqref{eq:cl_def} from the spinless case has to be replaced by
\begin{def*}
	Considering the \emph{non}-commutative algebra
	\begin{equation}
		\widetilde{\alg}_0 \coloneqq 	\mathrm{Fun}(\vect{x})\otimes \text{Mat}(r^N, \C)[\gamma_1^{\pm1},\dots, \gamma_N^{\pm1}]\, , 
	\end{equation}
	we define the \emph{partial classical limit} as the surjective algebra homomorphism
	\begin{equation} \label{eq:cl_mapping_spin}
		\tilde{c}_0 \coloneqq c_0 \otimes \mathrm{id} \colon \widetilde{\alg}_\hbar \longtwoheadrightarrow \widetilde{\alg}_0 \, , \qquad 
		f(\vect{x}) \longmapsto f(\vect{x}) \, , \quad 
		\quad \Gamma_i \longmapsto \gamma_i \, , \quad \hbar \longmapsto 0\, ,
	\end{equation}
	which acts as the identity on the factor $\text{Mat}(r^N, \C)$.
\end{def*}
This is the precise meaning of the replacement in the definition~\eqref{eq:spin chain hamiltonians} of the spin-chain hamiltonians, i.e.\ $H_{n,\elt} = \mathrm{ev}_\elt\bigl(\tilde{c}_0(\widetilde{D}_n^{(1)})\bigr)$.
	We stress that 
\eqref{eq:cl_mapping_spin} is only a \emph{partial} classical limit: the dependence on $\vect{x}$ and $\vect{p}$ becomes classical, but the spin part
remains fully quantum mechanical. Like 
for the spinless case $r=1$, its kernel is $\hbar\, \widetilde{\alg}_\hbar$, but this time the corresponding map $\widetilde{\alg}_\hbar/\hbar \mspace{2mu}\widetilde{\alg}_\hbar \overset{\!\sim}{\longrightarrow} \widetilde{\alg}_0$ is an isomorphism of \emph{non}commutative algebras. 
As in the spinless case, choose an extension of \eqref{eq:cl_mapping_spin} to a $\mathbb{C}[\mspace{-2mu}[\hbar]\mspace{-2mu}]$-linear map
\begin{equation} \label{eq:ctilde_hbar}
	\tilde{c}_\hbar \colon \widetilde{\alg}_\hbar \overset{\sim}{\longrightarrow} \widetilde{\alg}_0[\mspace{-2mu}[\hbar]\mspace{-2mu}] \, ,
\end{equation}
which generalises \eqref{eq:cl_mapping_isom_hbar} to
arbitrary $r\geqslant 1$. Due to the noncommutativity for $r\geqslant 2$, the definition \eqref{eq:Pbracket_scalar} of the Poisson bracket in the scalar case does not straightforwardly generalise to $\widetilde{\alg}_0$; cf.~\cite{mikhailovCommutativePoissonAlgebras2024,liashyk2024classical} and \textsection\ref{sec:hybrid_systems}. For the purpose of freezing, though, it will be enough to consider the centre $Z(\widetilde{\alg}_0)$, which can be identified with the commutative subalgebra $\alg_0 \, \mathrm{id} \subset \widetilde{\alg}_0$.
It acquires a Poisson bracket as in \eqref{eq:Pbracket_scalar}: for any $a,b \in Z(\widetilde{\alg}_0)$ we set  
\begin{equation} \label{eq:PB_spin}
	\{ a,b \} \coloneqq -\I \, \hbar^{-1} \, \tilde{c}_\hbar \pigl( \bigl[ \tilde{c}_\hbar^{\mspace{2mu}-1}(a),\tilde{c}_\hbar^{\mspace{2mu}-1}(b) \bigr] \pigr) \ \, \mathrm{mod} \, \hbar \, ,
\end{equation}
retrieving on $Z(\widetilde{\alg}_0)$ the Poisson structure \eqref{eq:Pbracket_scalar} (times the identity matrix).

If $a \in Z(\widetilde{\alg}_0)$ but $\tilde{b} \in \widetilde{\alg}_0$, then $a$ does not see the matrix structure of $\tilde{b}$, so the commutator $\tilde{c}_\hbar\bigl( [ \tilde{c}_\hbar^{\mspace{2mu}-1}(a),\tilde{c}_\hbar^{\mspace{2mu}-1}(\tilde{b}) ]\bigr)$ remains of order $\hbar$. Hence the formula~\eqref{eq:PB_spin} can be used to define an action of $Z(\widetilde{\alg}_0)$ on $\widetilde{\alg}_0$ 
given by $a \cdot \tilde{b} \coloneqq \{a,\tilde{b}\}$ for $a\in \alg_0 \cong Z(\widetilde{\alg}_0)$ and $\tilde{b} \in\widetilde{\alg}_0$. 
This turns $\widetilde{\alg}_0$ into a Poisson module over its centre $ Z(\widetilde{\alg}_0)$. Recall that, see e.g. \cite{mikhailovCommutativePoissonAlgebras2024},
	\begin{def*}
		A \emph{Poisson module} $\mathcal{B}$ over a 
		(possibly noncommutative)
		Poisson algebra $(\mathcal{A},\{\,\cdot\,,\,\cdot\,\})$ is a vector space $\mathcal{B}$ that is both a 
		bimodule for the associative algebra $\mathcal{A}$ and a Lie-algebra module for the Poisson bracket, with the compatibility conditions $\{ a , a' \, b\} = \{a,a'\} \, b + a' \, \{a,b\}$, $\{ a , b \, a'\} = \{a,b\} \, a' + b \, \{a,a'\}$ and $\{a \, a' , b \} = a \, \{ a',b\} + \{a,b\} \, a'$, i.e.\
		the Lie-algebra action by the Poisson bracket is a derivation for the 
		the $\mathcal{A}$-bimodule structure on $\mathcal{B}$ and for the product on $\mathcal{A}$. If $(\mathcal{A},\{\cdot,\cdot\})$ is commutative, the first two compatibility conditions coincide.
	\end{def*}
	In our case, $\widetilde{\alg}_0 \cong \widetilde{\alg}_\hbar/\hbar \mspace{2mu}\widetilde{\alg}_\hbar$ is itself a (noncommutative) algebra, with product inherited from $\widetilde{\alg}_\hbar$, coinciding with the (noncommutative) restriction $m_0$ of the star-product on $\widetilde{\alg}_0[\mspace{-2mu}[\hbar]\mspace{-2mu}]$. The Poisson action of $Z(\widetilde{\mathcal{A}}_0)$ furthermore acts by derivations for this product, i.e.\ $\{a,\tilde{b}\,\tilde{b}'\} = \{a, \tilde{b}\} \, \tilde{b}' + \tilde{b} \, \{a, \tilde{b}'\}$.

\subsection{Proof of Theorem~\ref{thm:HnB commute}} \label{sec:proof}

To view the spin-Ruijsenaars operators $\widetilde{D}_{\pm n}$ from \eqref{eq:Ruij_spin_pm n} as elements of $\widetilde{\alg}_\hbar$ we (formally) expand in the \emph{explicitly} appearing $\hbar$. 
For example, expanding \eqref{eq:Dtilde_1} gives
\begin{equation} \label{eq:Dtilde_1_expansion}
	\begin{aligned}
		\widetilde{D}_{1} & = \sum_{i=1}^N A_{i}(\vect{x}) \, P_{(1~\dots~i)}(\vect{x})^{-1} P_{(1~\dots~i)}(x_1,\dots,x_{i-1},x_i - \I \, \hbar \, \epsilon, x_{i+1}, \dots, x_N) \; \Gamma_i \\
		& = \sum_{k\geqslant 0} \widetilde{D}_{1}^{(k)} \, \hbar^{\mspace{1mu}k} \, , \qquad 
		\widetilde{D}_{1}^{(k)}\coloneqq \sum_{i} \frac{(-\I \, \epsilon)^k}{k\mspace{1mu}!} \, A_i(\vect{x}) \, P_{(1~\dots~i)}(\vect{x})^{-1} \, \partial_i^{\mspace{2mu}k} P_{(1~\dots~i)}(\vect{x}) \, \Gamma_i^{\pm 1} \, .
	\end{aligned}
\end{equation}
We can similarly expand any spin-Ruijsenaars operators~\eqref{eq:Ruij_spin_pm n} as
\begin{equation} \label{eq:hbar_expansion}
	\widetilde{D}_{\pm n} = \sum_{I} 
	 \widetilde{A}_{\pm I}(\vect{x}) \, \Gamma_{\pm I} = \sum_{k \geqslant 0} \widetilde{D}_{\pm n}^{(k)} \; \hbar^{\mspace{1mu}k} \, , \quad 
	\widetilde{D}_{\pm n}^{(k)}\coloneqq \sum_{I} \widetilde{A}_{\pm I}^{(k)}(\vect{x}) \, 
	 \Gamma_{\pm I} \, ,
\end{equation}
where we use tildes to indicate matrix valuedness.

For the present discussion, the explicit form of the coefficients is not relevant, and only distracts from the argument. We return to the explicit form in \textsection\ref{sec:explicit_hams}.
For the present discussion we will only use the following property.
	\begin{def*}
		We call a matrix-valued difference operator of the form
		$\widetilde{D} = \sum_{I} \widetilde{A}_I(\vect{x}) \, 
		\Gamma_{I}$ \emph{semiclassically spin separated} if the matrix-valued coefficients $\widetilde{A}_I(\vect{x}) \in \mathrm{Fun}(\vect{x})\otimes \text{Mat}(r^N, \C)$ are such that 
		\begin{equation} \label{eq:Atilde=A id+Ohbar}
			\widetilde{A}_{I}(\vect{x}) = A_{I}(\vect{x}) \, \mathrm{id} + O(\hbar) \, ,  \quad \text{i.e.} \quad \widetilde{A}_{I}^{(0)}(\vect{x}) = A_{I}(\vect{x}) \, ,
		\end{equation}
		with $A_I$ the coefficients~\eqref{eq:A_coeffs} from the scalar Ruijsenaars operators.
	\end{def*}

Such an operator exhibits the `(charge-)spin separation'
\begin{equation} \label{eq:Dtilde=Dcl id+Ohbar}
	\widetilde{D} = D \, \mathrm{id} + \widetilde{D}^{(1)} \, \hbar + O(\hbar^2) \quad \text{ in } \widetilde{\alg}_\hbar \, , 
\end{equation}
with scalar-valued (spinless) difference operators $D = \sum_{I} A_I(\vect{x}) \, \Gamma_{I} \; \in \mathrm{Fun}(\vect{x})[\Gamma_1^{\pm1},\dots, \Gamma_N^{\pm1}]$.

	\begin{ex*}
		The spin-Ruijsenaars operators from \textsection\ref{sec:Dtildes} are semiclassically spin separated.
		Indeed, the expansion of \eqref{eq:Ruij_spin_pm n} as a formal power series in $\hbar$ starts as
		\begin{equation} \label{eq:Dtilde_n_expansion}
			\begin{aligned}
				\widetilde{D}_{n} & = \sum_{I} A_{I}(\vect{x}) \, P_{I}(\vect{x})^{-1} P_{I}\biggl(\vect{x} - \I\, \epsilon \, \hbar {\textstyle \sum\limits_{i \in I}} \vect{e}_i \biggr) \,\Gamma_{I} \\
				& = D_{\pm n} \, \mathrm{id} \, - \, \I \, \epsilon \sum_{I} A_{I}(\vect{x}) \, P_{I}(\vect{x})^{-1} \sum_{i \in I} \partial_{x_i} P_{I}(\vect{x}) \; \Gamma_{I} \; \hbar + O(\hbar^2) \, ,
			\end{aligned}
		\end{equation}
		with $D_{\pm n}$ the scalar Ruijsenaars operators \eqref{eq:quantum_RS_ops} and $\widetilde{A}_I^{(1)}(\vect{x}) = - \I \, \epsilon \, A_{I}(\vect{x}) \, P_{I}(\vect{x})^{-1} \sum_{i \in I} \partial_{x_i} P_{I}(\vect{x})$.
	\end{ex*}

Note that the term at order $\hbar^0$ is \emph{not} the classical limit, since $D_{\pm n} \in \alg_\hbar$ is still fully quantum mechanical. Moreover, the term at order $\hbar^1$ is not a semiclassical limit in the usual sense either. Indeed, while $\tilde{c}_0\bigl(\widetilde{D}_{\pm n}^{(0)}\bigr) = D_{\pm n}^{\mspace{1mu}\text{cl}}\, \mathrm{id}$ could be viewed as a classical quantity, $\widetilde{D}_{\pm n}^{(1)}$ with $n\neq N$ will contain deformed spin permutations (quantum $R$-matrices) in addition to the difference operators~$\Gamma_i$; even if we apply the partial classical limit $\tilde{c}_0$ the result remains fully quantum mechanical in terms of spins. We will compute these $\tilde{c}_0\bigl(\widetilde{D}_n^{(1)}\bigr) \in \widetilde{\alg}_0$ in \textsection\ref{sec:explicit_hams}. 

We observe that
\begin{equation} \label{eq:Dtilde_N = DN}
	\widetilde{D}_N = D_N \, \mathrm{id} \qquad \text{i.e.} \qquad \widetilde{D}_N^{(k)} = 0 \, , \quad k \geqslant 1 \, ,
\end{equation}
and therefore
\begin{equation} \label{eq:D_N-n^k}
	\widetilde{D}_{N-n}^{(k)} = \bigl( \widetilde{D}_N \, \widetilde{D}_{-n} \bigr)^{\mspace{-2mu}(k)} = \widetilde{D}_N^{(0)} \, \widetilde{D}_{-n}^{(k)} \, , \quad k \geqslant 0 \, .
\end{equation}

	\begin{prp} \label{prp:partial class lim}
		Consider any two operators $\widetilde{D}_n$ and $\widetilde{D}_m$ of the form \eqref{eq:hbar_expansion} and suppose that \eqref{eq:Atilde=A id+Ohbar} holds. Then the partial classical limits obey
		\begin{equation} \label{eq:after_c0}
			0 = \I \, \pigl\{D^{\mspace{1mu}\mathrm{cl}}_n, \tilde{c}_0\bigl(\widetilde{D}_{m}^{(1)}\bigr) \pigr\} + \I \,  \pigl\{ \tilde{c}_0\bigl(\widetilde{D}_n^{(1)}\bigr), D^{\mspace{1mu}\mathrm{cl}}_m \pigr\}+ \pigl[ \tilde{c}_0\bigl(\widetilde{D}_n^{(1)}\bigr), \tilde{c}_0\bigl(\widetilde{D}_{m}^{(1)}\bigr) \pigr] \, ,
		\end{equation}
		where the Poisson brackets are those in \eqref{eq:PB_spin}.	
	\end{prp}

\begin{proof}[Proof.]
	Write $\widetilde{D}_n = \sum_I \widetilde{A}_I(\vect{x}) \, \Gamma_I$ and $\widetilde{D}_m = \sum_J \widetilde{A}_J(\vect{x}) \, \Gamma_J$, where, as before, sums over $n$- and $m$- element subsets $I$ and $J$, respectively, are understood.  From the $\hbar$-expansion of their commutator we obtain the following refinement of \eqref{eq:comm_Dn_expansion} in $\widetilde{\alg}_\hbar$:
	\begin{equation} \label{eq:comm_widetildeDn_expansion}
		\begin{aligned}
			0 = [\widetilde{D}_n,\widetilde{D}_m\mspace{1mu}] = {}& \sum_{I \mspace{-2mu},\mspace{2mu} J} \Biggl( \widetilde{A}_I(\vect{x}) \, \widetilde{A}_J\biggl(\vect{x} - \I \, \epsilon \,\hbar \, {\textstyle \sum\limits_{i \in I}} \vect{e}_i \biggr) - \widetilde{A}_J(\vect{x}) \, \widetilde{A}_I\biggl(\vect{x} - \I \, \epsilon\, \hbar \, {\textstyle \sum\limits_{j \in J}} \vect{e}_j \biggr) \Biggr) \, \Gamma_I \, \Gamma_J \\
			= {}& \sum_{I \mspace{-2mu},\mspace{2mu} J} \, \pigl[\widetilde{A}_I , \widetilde{A}_J \pigr] \, \Gamma_I \, \Gamma_J \ - \I \, \epsilon \, \hbar \sum_{I \mspace{-2mu},\mspace{2mu}J} \Biggl( \widetilde{A}_I \sum_{i \in I} \partial_i \mspace{2mu} \widetilde{A}_J - \widetilde{A}_J \sum_{j\in J} \partial_j \mspace{1mu} \widetilde{A}_I \Biggr) \, \Gamma_I \, \Gamma_J \\
			& - \frac{\epsilon^2 \, \hbar^2}{2} \sum_{I \mspace{-2mu},\mspace{2mu} J} \Biggl(\widetilde{A}_I \!\!\sum_{i ,\mspace{2mu} i'\in I}\!\! \partial_i \, \partial_{i'} \widetilde{A}_J - \widetilde{A}_J \!\!\sum_{j ,\, j'\in J}\!\! \partial_j \, \partial_{j'} \widetilde{A}_I \Biggr) \, \Gamma_I \, \Gamma_J \ + O(\hbar^3) \, .
		\end{aligned}
	\end{equation}
	Under the assumption~\eqref{eq:Atilde=A id+Ohbar} we have
	$\tilde{c}_0(\widetilde{D}_n) = D_n \, \mathrm{id} \in Z(\widetilde{\alg}_0)$. Moreover, the terms of order $\hbar^0$ in \eqref{eq:comm_widetildeDn_expansion} vanish, and the
	part of \eqref{eq:comm_widetildeDn_expansion} linear in $\hbar$ becomes precisely \eqref{eq:Poisson_DnDm} (times $\mathrm{id}$), 
	which implies that
	\begin{equation}
		\tilde{c}_0\pigl( -\I \, \hbar^{-1} \bigl[\widetilde{D}_n , \widetilde{D}_m \mspace{1mu} \bigr] \pigr) = \pigl\{ \tilde{c}_0\bigl(D_n\bigr), \tilde{c}_0\bigl(D_m\bigr) \!\pigr\} \, \mathrm{id} =0 \, .
	\end{equation} 
	We stress that the right-hand side features the Poisson bracket of the \emph{scalar} Ruijsenaars hamiltonians, times the identity on spins. In fact, by virtue of \eqref{eq:comm_Dn_expansion} the part of \eqref{eq:comm_widetildeDn_expansion} linear in $\hbar$ vanishes even \emph{before} applying $\tilde{c}_0$. Hence the first non-trivial terms occur at $\hbar^2$. They are given by
	\begin{equation} \label{eq:hsquared_dndm}
		\begin{aligned}
			0 = \hbar^{-2} \, \pigl[\widetilde{D}_n,\widetilde{D}_m \mspace{1mu}\pigr] = {} & 
			\sum_{I \mspace{-2mu},\mspace{2mu} J} \Bigl( \pigl[\widetilde{A}_I^{(2)} , \widetilde{A}_J^{(0)} 
			\pigr] + \pigl[\widetilde{A}_I^{(1)} , \widetilde{A}_J^{(1)} 
			\pigr] +\pigl[\widetilde{A}_I^{(0)} , \widetilde{A}_J^{(2)} 
			\pigr] \Bigr)\, \Gamma_I \, \Gamma_J \\
			& - \I \, \epsilon \, \sum_{I \mspace{-2mu},\mspace{2mu}J} \Biggl( \widetilde{A}_I^{(0)} \sum_{i \in I} \partial_i \mspace{1mu} \widetilde{A}_J^{(1)}  
			- \widetilde{A}_J^{(1)} \sum_{j\in J} \partial_j \mspace{1mu} \widetilde{A}_I^{(0)} \Biggr) \, \Gamma_I \, \Gamma_J \\
			& - \I \, \epsilon \, \sum_{I \mspace{-2mu},\mspace{2mu}J} \Biggl( \widetilde{A}_I^{(1)} \sum_{i\in I} \partial_i \mspace{1mu} \widetilde{A}_J^{(0)} - \widetilde{A}_J^{(0)} \sum_{j\in J} \partial_j \mspace{1mu} \widetilde{A}_I^{(1)} \Biggr) \, \Gamma_I \, \Gamma_J \\
			& - \frac{1}{2} \, \epsilon^2 \sum_{I \mspace{-2mu},\mspace{2mu} J} \Biggl(\widetilde{A}_I^{(0)} \!\!\sum_{i ,\mspace{2mu} i'\in I}\!\! \partial_i \, \partial_{i'} \widetilde{A}^{(0)}_J - \widetilde{A}^{(0)}_J \!\!\sum_{j ,\, j'\in J}\!\! \partial_j \, \partial_{j'} \widetilde{A}^{(0)}_I \Biggr) \, \Gamma_I \, \Gamma_J \\
			& + O(\hbar) \, . 
		\end{aligned}
	\end{equation}
	The first and third matrix commutator in the first line of the right-hand side vanish due to \eqref{eq:Atilde=A id+Ohbar}. The terms linear in $\epsilon$ on the next two lines come from the first-order bracket built from $m_1$, and will give rise to the Poisson bracket~\eqref{eq:PB_spin} upon applying $\tilde{c}_0$. The coefficient of $\epsilon^2$ on the fourth line, which is the contribution from the second-order bracket\,%
	\footnote{\ The overall sign is $(-\I)^2$, a convention consistent with our choice to expand the star commutator in powers of $-\I \,\hbar$, as in \eqref{eq:Pbracket_scalar}.}
	\begin{equation} \label{eq:m_2}
		\pigl\{ \tilde{c}_0\bigl( \widetilde{D}_n\bigr), \tilde{c}_0\bigl( \widetilde{D}_m \bigr) \pigr\}_2 \coloneqq
		- \Bigl( m_2\pigl(\tilde{c}_0 \bigl(\widetilde{D}_n\bigr), \tilde{c}_0 \bigl(\widetilde{D}_m\bigr)\pigr) - m_2\pigl(\tilde{c}_0 \bigl(\widetilde{D}_m\bigr), \tilde{c}_0 \bigl(\widetilde{D}_n\bigr) \pigr) \Bigr)\, ,
	\end{equation}
	vanishes thanks to the $\epsilon^2$-term of the identity \eqref{eq:comm_Dn_expansion} from the spinless case.\footnote{\ In the language of \textsection\ref{sec:hybrid_systems}, this is because $\widetilde{\alg}_0$ is `flat'.} 
	Therefore, comparing with \eqref{eq:Poisson_DnDm} we conclude that, after taking the classical limit,
	\begin{equation} 
		\begin{aligned}
			0 & = \tilde{c}_0 \pigl( \hbar^{-2} \, \bigl[\widetilde{D}_n,\widetilde{D}_{m}\bigr] \pigr) \\
			& = \I \, \pigl\{\tilde{c}_0( \widetilde{D}_n^{(0)}), \tilde{c}_0\bigl(\widetilde{D}_{m}^{(1)}\bigr) \pigr\} + \I \, \pigl\{ \tilde{c}_0\bigl(\widetilde{D}_n^{(1)}\bigr), \tilde{c}_0\bigl(\widetilde{D}_{m}^{(0)}\bigr) \pigr\}  + \pigl[ \tilde{c}_0\bigl(\widetilde{D}_n^{(1)}\bigr), c_0\bigl(\widetilde{D}_{m}^{(1)}\bigr) \pigr]  \\
			& = \I \, \pigl\{D^{\mspace{1mu}\text{cl}}_n, \tilde{c}_0\bigl(\widetilde{D}_{m}^{(1)}\bigr) \pigr\} + \I \,  \pigl\{ \tilde{c}_0\bigl(\widetilde{D}_n^{(1)}\bigr), D^{\mspace{1mu}\text{cl}}_m \pigr\}+ \pigl[ \tilde{c}_0\bigl(\widetilde{D}_n^{(1)}\bigr), \tilde{c}_0\bigl(\widetilde{D}_{m}^{(1)}\bigr) \pigr] \, ,
		\end{aligned}
	\end{equation}
	which is \eqref{eq:after_c0}.
\end{proof}

In particular, 
we obtain the non-trivial identity
that will come in handy momentarily:
	\begin{cor}
		The following identity holds:
		\begin{equation} \label{eq:DN_onDn}
			\sum_{i=1}^N \frac{\partial\, \tilde{c}_0\bigl(\widetilde{D}_n^{(1)}\bigr)}{\partial x_j} = 0 
			 \, , \qquad -N \leqslant n \leqslant N \, . 
		\end{equation}
	\end{cor}
	\begin{proof}[Proof.]
		Take $m=N$ in \eqref{eq:after_c0} and use \eqref{eq:Dtilde_N = DN}
		to get
		\begin{equation*}
			0 = \pigl\{ \tilde{c}_0(\widetilde{D}_n^{(1)}), \tilde{c}_0(D_N) \pigr\} = \tilde{c}_0(D_N) \sum_{i=1}^N \frac{\partial\, \tilde{c}_0\bigl(\widetilde{D}_n^{(1)}\bigr)}{\partial x_j} \, .
		\end{equation*} 
		But $\tilde{c}_0(D_N) = D_N^\mathrm{cl} = \gamma_1 \cdots \gamma_N$ is nonzero, as the multiplicative classical momenta $\gamma_j$ are valued in $\mathbb{C}^\times$.
	\end{proof}

With this partial (semi)classical expansion
in place, we now are ready to perform the second part of freezing, by evaluating \eqref{eq:after_c0} at one of the classical equilibria we obtained in \textsection\ref{sec:equilibria_modular}. We will want to include the case with spins. To this end, note that the dynamical parameter~$\vec{a}$ is readily included in our treatment of the equilibria since it occurs in the same way as the $x_i$ and $\eta$. Thus, $\vec{a}^{\,\star}{}^{(1)} = \vec{a}/N$ for \eqref{eq:freezing_type1} and $\vec{a}^{\,\star}{}^{(S)} = -\vec{a}/\omega$ for \eqref{eq:freezing_type2}.

Recall the evaluation map~\eqref{eq:evaluation} at a classical equilibrium labelled by $\elt \in \text{PSL}(2,\Z)$.
These maps may act on operators $\widetilde{O}$ by $\mathrm{ev}_{\mspace{-2mu}\elt} \circ \widetilde{O}$. We will write $\widetilde{O}_1 \overset{\mathrm{ev}}{=} \widetilde{O}_2$ as a shorthand for $\mathrm{ev}_{\mspace{-2mu}\elt} \circ \widetilde{O}_1 = \mathrm{ev}_{\mspace{-2mu}\elt} \circ\widetilde{O}_2$.

	\begin{lem}
		Upon evaluation, the Poisson brackets in \eqref{eq:after_c0} vanish: 
		\begin{equation} \label{eq:PB_at_equilibrium}
			\pigl\{D^{\mspace{1mu}\mathrm{cl}}_n, \tilde{c}_0\bigl(\widetilde{D}_{m}^{(1)}\bigr) \pigr\} = 0 \, .
		\end{equation}
	\end{lem}
	\begin{proof}[Proof.]
		By first applying the equilibrium conditions \eqref{eq:freezing_stationary} with the ($j$-independent) velocities \eqref{eq:freezing_xj} we compute
		\begin{equation*}
			\begin{aligned}
				\pigl\{D^{\mspace{1mu}\text{cl}}_n, \tilde{c}_0\bigl(\widetilde{D}_{m}^{(1)}\bigr) \pigr\} &
				= \sum_{j =1}^N \Biggl( \{ D_n^{\text{cl}}, p_j\}  \,\frac{\partial\, \tilde{c}_0 \bigl(\widetilde{D}_{m}^{(1)}\bigr)}{\partial p_j} \,
				+ \, \{D_n^\text{cl}, x_j\} \, \frac{\partial\, \tilde{c}_0 \bigl(\widetilde{D}_{m}^{(1)}\bigr)}{\partial x_j}  \Biggr)  \\
				& \overset{\mathrm{ev}}{=} 0 - v_n^\star \sum_{j=1}^N  \, \frac{\partial\, \tilde{c}_0 \bigl(\widetilde{D}_{m}^{(1)}\bigr)}{\partial x_j} \, .
			\end{aligned} 
		\end{equation*}
		Now use the identity \eqref{eq:DN_onDn}.\,%
		\footnote{\label{fn:difference}\ In fact, this this result does not rely on \eqref{eq:DN_onDn}: in terms of the relative positions $y_i \coloneqq x_i-x_{i+1}$ the equilibrium condition \eqref{eq:freezing_stationary} has vanishing velocities $\partial y_i/\partial t_n = \{y_i, D^{\mspace{1mu}\text{cl}}_n\} = 0$. As all our operators only depend on such differences, it follows directly that $\bigl\{D^{\mspace{1mu}\text{cl}}_n, \tilde{c}_0(\widetilde{D}_{m}^{(1)}) \bigr\} = 0$. We thank O.~Chalykh for pointing this out.}
	\end{proof}

\begin{proof}[Proof of Theorem~\ref{thm:HnB commute}.]
	Rearranging \eqref{eq:after_c0} and applying the evaluation map then yields 
	\begin{equation}
		\pigl[ \tilde{c}_0 \bigl(\widetilde{D}_n^{(1)} \bigr), \tilde{c}_0 \bigl(\widetilde{D}_{m}^{(1)}\bigr) \pigr] = -\I \, \pigl\{D^{\mspace{1mu}\text{cl}}_n, \tilde{c}_0 \bigl(\widetilde{D}_{m}^{(1)}\bigr) \pigr\} - \I \, \pigl\{\tilde{c}_0 \bigl(\widetilde{D}_{n}^{(1)}\bigr), D^{\mspace{1mu}\text{cl}}_m \pigr\} \overset{\mathrm{ev}}{=} 0 \, .
	\end{equation}
	After taking the classical limit there are no more difference operators, yielding operators that act trivially \emph{on functions}, and can be restricted to act on $V^{\otimes N}$ only. This justifies the definition of the spin-chain hamiltonians~\eqref{eq:spin chain hamiltonians} and concludes the proof.
\end{proof}

	\begin{cor}
		Theorem~\ref{thm:HnB commute} holds for any family semiclassically spin-separated operators that commute.
	\end{cor}

\subsection{Explicit spin-chain hamiltonians} \label{sec:explicit_hams}
Let us compute the general form of the spin-chain hamiltonians~\eqref{eq:spin chain hamiltonians}. 
Consider the expansion~\eqref{eq:Dtilde_n_expansion} of the spin-Ruijsenaars operators in $\hbar$.
To better understand the spin operators in the coefficient $\widetilde{D}_{n}^{(1)}$ of $\hbar^1$ we need to set up some notation. Recall that $P_I(\vect{x})$ was defined in \eqref{eq:P_I}. By \textsection\ref{app:grassmann} the permutation appearing in \eqref{eq:P_I} can be factorised as the product $w_I^{-1} = (s_n \, s_{n+1} \cdots s_{i_n-1}) \, \cdots \, (s_1 \, s_2 \cdots s_{i_1-1})$. Label the subscripts in this (reduced) decomposition from left to right as $j_1 = n$, \dots, $j_\ell = {i_1-1}$, where $\ell = \sum_{m=1}^n (i_m - m) = \sum_{i\in I} i \, - n\,(n+1)/2$. Then $w_k \coloneqq s_{j_{k+1}} \mspace{-2mu}\cdots s_{j_\ell}$ gives a set of $\ell+1$ permutations ranging from $w_0 = w_I^{-1}$, $w_1 = s_{n+1} \mspace{-2mu}\cdots s_{i_1 - 1}$ down to $w_{\ell-1} = s_{i_1 - 1}$, $w_\ell = e$. With this notation, we have
\begin{equation} \label{eq:P_I decomp}
	\begin{aligned}
		P_I(\vect{x}) = {}& \ordprod_{n\geqslant m \geqslant 1} \! \Biggl( \, \ordprodopp_{m\leqslant i' < i_m} \!\!\!\!\!\! P_{i',i'+1}\Bigl(x_{w_{\mspace{1mu}k_{m,i'}}(i')} - x_{i_m}\Bigr) \Biggr) \\
		= {}& \textcolor{NavyBlue}{P_{n,n+1}\Bigl(x_{w_1(n)} - x_{i_n} \Bigl) \cdots P_{i_n -1,i_n}\Bigl(x_{w_{i_n-n}(i_n - 1)} - x_{i_n} \Bigl)} \\
		{}& \quad\! \times \textcolor{Orange}{\cdots} \\[-.5ex]
		{}& \qquad\! \times \textcolor{Fuchsia}{P_{12}\bigl(x_{1} - x_{i_1}\bigr) \cdots P_{i_1 -1,i_1}\bigl(x_{i_1 -1} - x_{i_1}\bigr)} \\[1ex]
		={}&
		\tikz[baseline={([yshift=-.5*11pt*0.3]current bounding box.center)},xscale=-.5,yscale=0.25,font=\footnotesize]{
			\draw[->] (-6,0) node[below]{$x_N$} -- (-6,8) node[above]{$x_N$};
			\foreach \x in {-1,...,1} \draw (-5.25+.2*\x,4) node{$\cdot\mathstrut$};	
			\draw[->] (-4.5,0) -- (-4.5,8);
			\draw[rounded corners=2pt,->,NavyBlue] (-3.5,0) node[below]{$x_{i_n}$} -- (-3.5,1) -- (2.5,7) -- (2.5,8) node[above]{$\vphantom{x_i}\smash{x_{i_n}}$};
			\draw[rounded corners=2pt,->] (-2.5,0) -- (-2.5,1) -- (-3.5,2) -- (-3.5,8);
			\draw[rounded corners=2pt,->] (-1.5,0) -- (-1.5,2) -- (-2.5,3) -- (-2.5,8);
			\draw[rounded corners=2pt,->,Orange] (-.5,0) node[below]{$\vphantom{x}\cdots$} -- (-.5,1) -- (.5,2) -- (3.5,5) -- (3.5,8) node[above]{$\vphantom{x_i}\!\!\cdots$};
			\draw[rounded corners=2pt,->] (.5,0) -- (.5,1) -- (-.5,2) -- (-.5,3) -- (-1.5,4) -- (-1.5,8);
			\draw[rounded corners=2pt,->] (1.5,0) -- (1.5,2) -- (.5,3) -- (.5,4) -- (-.5,5) -- (-.5,8);
			\draw[rounded corners=2pt,->,Fuchsia] (2.5,0) node[below]{$x_{i_1}$} -- (2.5,1) -- (4.5,3) -- (4.5,8) node[above]{$\vphantom{x_i}\smash{x_{i_1}}$};
			\draw[rounded corners=2pt,->] (3.5,0) node[below]{$\vphantom{x}\!\!\cdots$} -- (3.5,1) -- (2.5,2) -- (2.5,3) -- (1.5,4) -- (1.5,5) -- (.5,6) -- (.5,8) node[above]{$\vphantom{x_i}\cdots$};
			\draw[rounded corners=2pt,->] (4.5,0) node[below]{$x_1$} -- (4.5,2) -- (3.5,3) -- (3.5,4) -- (2.5,5) -- (2.5,6) -- (1.5,7) -- (1.5,8) node[above]{$x_1$};
			\node at (5,2.5) {\smash{\textcolor{lightgray}{$\vec{a}$}}\vphantom{$a$}};
		} \, ,
	\end{aligned}
\end{equation}
where, in the first line, the arrows indicate the direction of increasing subscripts in the products, and $k_{m,i'} \coloneqq \sum_{m'(>m)}^n (i_{m'} - m') + i'-m+1$. 
Note that only some factors depend on the coordinates $x_i$ with respect to which we need to compute the derivative in \eqref{eq:Dtilde_n_expansion}. Namely, for any $i_m \in I$, $1\leqslant m \leqslant n$, the coordinate $x_{i_m}$ appears precisely in the $i_m - m$ factors $P_{i',i'+1}(x_{\bullet} - x_{i_m})$ with $m\leqslant i' < i_m$. 
	\begin{def*}
		We define the \emph{nearest-neighbour spin interaction} in $\text{End}(V^{\otimes N})$ as
		\begin{equation} \label{eq:h_i,i+1(x)}
			h_{i,i+1}(u) \coloneqq P_{i,i+1}(-u) \, P_{i,i+1}'(u) = \tikz[baseline={([yshift=-.5*11pt*0.4]current bounding box.center)},xscale=.5,yscale=0.25,font=\footnotesize]{
				\node at (-3,2.5) {\smash{\textcolor{lightgray}{$\vec{a}$}}\vphantom{$a$}};
				\draw[->] (-2.5,0) -- (-2.5,5);
				\foreach \x in {-1,...,1} \draw (-2.5+.75+.2*\x,2.5) node{$\cdot\mathstrut$};	
				\draw[->] (-1,0) -- (-1,5);
				\draw[rounded corners=2pt,->] (1,0) node[below]{$\,x\smash{''}$}-- (1,1) -- (0,2) -- (0,3) -- (1,4) -- (1,5) node[above]{$x\smash{''}$};
				\draw[rounded corners=2pt,->] (0,0) node[below]{$x\smash{'}$} -- (0,1) -- (1,2) -- (1,3) -- (0,4) -- (0,5) node[above]{$x\smash{'}$};
				\draw[->] (2,0) -- (2,5);
				\foreach \x in {-1,...,1} \draw (2+.75+.2*\x,2.5) node{$\cdot\mathstrut$};	
				\draw[->] (3.5,0) -- (3.5,5);
				\node at (0.5,1.5cm-.5pt) {\rotatebox{30}{$\circledast$}};
			} \ \ ,
			\qquad u = x'-x''\, ,
		\end{equation}
		where `$\raisebox{-1.5pt}{\rotatebox{30}{$\circledast$}}$' indicates the derivative of the $R$-matrix in our graphical notation. 
	\end{def*}
Then
\begin{align} \label{eq:P_I decomp cont'd}
	P_{I}(\vect{x})^{-1} \, \partial_{x_{i_m}} P_{I}(\vect{x}) = 
	\! \sum_{i'=m}^{i_m-1} & \textcolor{OliveGreen}{P_{i_1 -1,i_1}\pigl(x_{i_1} - x_{i_1 - 1}\pigr) \cdots P_{12}\pigl( x_{i_1} - x_{1} \pigr)} \nonumber \\[-.5ex]
	& \!\!\! \times \textcolor{Fuchsia}{\cdots} \nonumber \\[-.5ex]
	& \times \textcolor{Orange}{P_{i_m -1,i_m}\mspace{-1mu} \Bigl(x_{i_m} - x_{w_{\mspace{1mu}k_{\mspace{-1mu}m,i_m-1}}(i_m-1)} \Bigr) \cdots P_{i'+1,i'+2}\Bigl(x_{i_m}- x_{w_{\mspace{1mu}k_{\mspace{-1mu}m,i'+1}}(i'+1)} \Bigr)} \nonumber\\
	& \ \times -\textcolor{Orange}{h_{i',i'+1}\Bigl(x_{w_{\mspace{1mu}k_{\mspace{-1mu}m,i'}}(i')} - x_{i_m} \Bigr)} \\
	& \times \textcolor{Orange}{ P_{i'+1,i'+2}\Bigl(x_{w_{\mspace{1mu}k_{\mspace{-1mu}m,i'+1}}(i'+1)} - x_{i_m} \Bigr) \cdots P_{i_m -1,i_m}\mspace{-1mu} \Bigl(x_{w_{\mspace{1mu}k_{\mspace{-1mu}m,i_m-1}}(i_m - 1)} - x_{i_m} \Bigr) } \nonumber \\
	& \!\!\!\! \times \textcolor{Fuchsia}{\cdots} \nonumber \\[-1ex]
	\times \ & \textcolor{OliveGreen}{P_{12}\bigl(x_{1} - x_{i_1}\bigr) \cdots P_{i_1 -1,i_1}\bigl(x_{i_1 -1} - x_{i_1}\bigr)} \nonumber \\
	= -\!\! \sum_{i'=m}^{i_m-1} \!\!\!\! & \hphantom{-} 
	\tikz[baseline={([yshift=-.5*11pt*0.3]current bounding box.center)},xscale=-.5,yscale=0.25,font=\footnotesize]{
		\draw[->] (-6,0) node[below]{$x_N$} -- (-6,11) node[above]{$x_N$};
		\foreach \x in {-1,...,1} \draw (-5.25+.2*\x,5.5) node{$\cdot\mathstrut$};	
		\draw[->] (-4.5,0) -- (-4.5,11);
		\draw[rounded corners=2pt,->,Orange] (-3.5,0) node[below]{$x_{i_m}$} -- (-3.5,1) -- (-.5,4) -- (-.5,7) -- (-3.5,10) -- (-3.5,11) node[above]{$\vphantom{x_i}\smash{x_{i_m}}$};
		\draw[rounded corners=2pt,->] (-2.5,0) -- (-2.5,1) -- (-3.5,2) -- (-3.5,9) -- (-2.5,10) -- (-2.5,11);
		\draw[rounded corners=2pt,->] (-1.5,0) -- (-1.5,2) -- (-2.5,3) -- (-2.5,8) -- (-1.5,9) -- (-1.5,11);
		\draw[rounded corners=2pt,->,Fuchsia] (-.5,0) -- (-.5,1) -- (3.5,5) -- (3.5,6) -- (-.5,10) -- (-.5,11);
		\draw[rounded corners=2pt,->] (.5,0) node[below]{$x_{i'}$} -- (.5,1) -- (-.5,2) -- (-.5,3) -- (-1.5,4) -- (-1.5,7) -- (-.5,8) -- (-.5,9) -- (.5,10) -- (.5,11) node[above]{$x_{i'}$};
		\draw[rounded corners=2pt,->,OliveGreen] (2.5,0) -- (2.5,1) -- (4.5,3) -- (4.5,8) -- (2.5,10) -- (2.5,11);
		\draw[rounded corners=2pt,->] (1.5,0) -- (1.5,2) -- (.5,3) -- (.5,8) -- (1.5,9) -- (1.5,11);
		\draw[rounded corners=2pt,->] (3.5,0) node[below]{$\vphantom{x}\cdots$} -- (3.5,1) -- (2.5,2) -- (2.5,3) -- (1.5,4) -- (1.5,7) -- (2.5,8) -- (2.5,9) -- (3.5,10) -- (3.5,11) node[above]{$\vphantom{x_i}\cdots$};
		\draw[rounded corners=2pt,->] (4.5,0) node[below]{$x_1$} -- (4.5,2) -- (3.5,3) -- (3.5,4) -- (2.5,5) -- (2.5,6) -- (3.5,7) -- (3.5,8) -- (4.5,9) -- (4.5,11) node[above]{$x_1$};
		\node at (5,5.5) {\smash{\textcolor{lightgray}{$\vec{a}$}}\vphantom{$a$}};
		\node at (-1,3.5cm-.5pt) {\rotatebox{30}{$\circledast$}};
	} \, . \nonumber
\end{align}
Here we have used unitarity~\eqref{eq:unitarity} to remove some adjacent inverses. As the diagram shows, unitarity may allow for further simplifications: \textcolor{OliveGreen}{lines corresponding to $i \in I$ with $i<i'$ become trivial}, and \textcolor{Fuchsia}{lines with $i' < i < i_m$ can be reduced in part}. In this way we compute
\begin{equation} \label{eq:Dtilde_n^1}
	\begin{aligned}
		\widetilde{D}_{n}^{(1)} & = - \, \I \, \epsilon \sum_{I} A_{I}(\vect{x}) \sum_{m=1}^n P_{I}(\vect{x})^{-1} \, \partial_{x_{i_m}} P_{I}(\vect{x}) \ \Gamma_{I} \\
		& = \hphantom{-} \, \I \, \epsilon \sum_{I} A_{I}(\vect{x}) \sum_{m=1}^n \, \sum_{i'=m}^{i_m-1} 
		\tikz[baseline={([yshift=-.5*11pt*0.3]current bounding box.center)},xscale=-.5,yscale=0.25,font=\footnotesize]{
			\draw[->] (-6,0) node[below]{$x_N$} -- (-6,9) node[above]{$x_N$};
			\foreach \x in {-1,...,1} \draw (-5.25+.2*\x,4.5) node{$\cdot\mathstrut$};	
			\draw[->] (-4.5,0) -- (-4.5,9);
			\draw[rounded corners=2pt,->,Orange] (-3.5,0) node[below]{$x_{i_m}$} -- (-3.5,1) -- (-.5,4) -- (-.5,5) -- (-3.5,8) -- (-3.5,9) node[above]{$\vphantom{x_i}\smash{x_{i_m}}$};
			\draw[rounded corners=2pt,->] (-2.5,0) -- (-2.5,1) -- (-3.5,2) -- (-3.5,7) -- (-2.5,8) -- (-2.5,9);
			\draw[rounded corners=2pt,->] (-1.5,0) -- (-1.5,2) -- (-2.5,3) -- (-2.5,6) -- (-1.5,7) -- (-1.5,9);
			\draw[rounded corners=2pt,->,Fuchsia] (-.5,0) -- (-.5,1) -- (.5,2) -- (.5,7) -- (-.5,8) -- (-.5,9);
			\draw[rounded corners=2pt,->] (.5,0) node[below]{$x_{i'}$} -- (.5,1) -- (-.5,2) -- (-.5,3) -- (-1.5,4) -- (-1.5,5) -- (-.5,6) -- (-.5,7) -- (.5,8) -- (.5,9) node[above]{$x_{i'}$};
			\draw[rounded corners=2pt,->] (1.5,0) -- (1.5,9);
			\draw[rounded corners=2pt,->,OliveGreen] (2.5,0) -- (2.5,9);
			\draw[rounded corners=2pt,->] (3.5,0) node[below]{$\vphantom{x}\cdots$} -- (3.5,9) node[above]{$\vphantom{x_i}\cdots$};
			\draw[rounded corners=2pt,->] (4.5,0) node[below]{$x_1$} -- (4.5,9) node[above]{$x_1$};
			\node at (5,4.5) {\smash{\textcolor{lightgray}{$\vec{a}$}}\vphantom{$a$}};
			\node at (-1,3.5cm-.5pt) {\rotatebox{30}{$\circledast$}};
		} \!\! \times \Gamma_{I} \; .
	\end{aligned}
\end{equation}

Now apply the partial classical limit~\eqref{eq:cl_mapping_spin} and, for any $\elt \in \mathrm{PSL}(2,\mathbb{Z})$, the evaluation~\eqref{eq:evaluation} at the equilibrium $x_i^\star = x_i^\star{}^{(\elt)}$, $\gamma_j^\star = \E^{\I \mspace{2mu} p_j^\star{}^{(\elt)}}$. 
Define $a_{\{i_1,\dots,i_n\}}(\vect{x}^\star) \coloneqq \I \, \epsilon \, A_{\{i_1,\dots,i_n\}}(\vect{x}^\star) \, \gamma_{\{i_1,\dots,i_n\}}^\star$. Thanks to the definition \eqref{eq:A_coeffs} of $A_I$ and the identity \eqref{eq:v_1*}, it is independent of the equilibrium values of the momenta. Thus
we arrive at the following expression.
	\begin{prp} \label{prp:H_nB}
		The spin-chain hamiltonians~\eqref{eq:spin chain hamiltonians} admit the following explicit expression:
		\begin{subequations} \label{eq:H_n explicit}
			\begin{gather}
				H_{n,\elt} = \! \sum_{i_1 < \dots < i_n}^N \!\!\!\! a_{\{i_1,\dots,i_n\}}(\vect{x}^\star) \sum_{m=1}^n \, \sum_{i'=m}^{i_m-1} 
				\tikz[baseline={([yshift=-.5*11pt*0.3]current bounding box.center)},xscale=-.5,yscale=0.25,font=\footnotesize]{
					\draw[->] (-6,0) node[below]{$x_N^\star$} -- (-6,9) node[above]{$x_N^\star$};
					\foreach \x in {-1,...,1} \draw (-5.25+.2*\x,4.5) node{$\cdot\mathstrut$};	
					\draw[->] (-4.5,0) -- (-4.5,9);
					\draw[rounded corners=2pt,->] (-3.5,0) node[below]{$x_{i_m}^\star$} -- (-3.5,1) -- (-.5,4) -- (-.5,5) -- (-3.5,8) -- (-3.5,9) node[above]{$\vphantom{x_i}\smash{x_{i_m}^\star}$};
					\draw[rounded corners=2pt,->] (-2.5,0) -- (-2.5,1) -- (-3.5,2) -- (-3.5,7) -- (-2.5,8) -- (-2.5,9);
					\draw[rounded corners=2pt,->] (-1.5,0) -- (-1.5,2) -- (-2.5,3) -- (-2.5,6) -- (-1.5,7) -- (-1.5,9);
					\draw[rounded corners=2pt,->] (-.5,0) -- (-.5,1) -- (.5,2) -- (.5,7) -- (-.5,8) -- (-.5,9);
					\draw[rounded corners=2pt,->] (.5,0) node[below]{$x_{i'}^\star$} -- (.5,1) -- (-.5,2) -- (-.5,3) -- (-1.5,4) -- (-1.5,5) -- (-.5,6) -- (-.5,7) -- (.5,8) -- (.5,9) node[above]{$x_{i'}^\star$};
					\draw[rounded corners=2pt,->] (1.5,0) -- (1.5,9);
					\foreach \x in {-1,...,1} \draw (2.25+.2*\x,4.5) node{$\cdot\mathstrut$};	
					\draw[rounded corners=2pt,->] (3,0) node[below]{$x_1^\star$} -- (3,9) node[above]{$x_1^\star$};
					\node at (3.5,4.5) {\smash{\textcolor{lightgray}{$\vec{a}$}}\vphantom{$a$}};
					\node at (-1,3.5cm-.5pt) {\rotatebox{30}{$\circledast$}};
				} \, ,
				\intertext{with coefficients}
				\label{eq:ev class A Ga}
				a_{\{i_1,\dots,i_n\}}(\vect{x}^\star) 
				=  
				\I \, \epsilon \, \biggl(\frac{v_1^\star}{\epsilon}\biggr)^{\!\!n} \! \prod_{m<m'}^n \frac{\theta\bigl(x_{i_m}^\star - x_{i_{m'}}^\star\bigr)^2}{\theta\bigl(x_{i_m}^\star - x_{i_{m'}}^\star +\eta\bigr) \, \theta\bigl(x_{i_m}^\star - x_{i_{m'}}^\star - \eta\bigr)} \; .
			\end{gather}
		\end{subequations}
	\end{prp}
Besides the overall prefactor, we recognise an elliptic Vandermonde factor squared divided by the product of elliptic $q$- and $q^{-1}$-Vandermonde factors, for $q \sim \E^\eta$. \textcolor{OliveGreen}{For all $i \in I$ for which the $i$th line could be simplified, the only dependence on $x_i$ resides in the coefficient $a_{\{i_1,\dots,i_n\}}$.}
We stress that, while the classical equilibrium momenta have dropped out, by \textsection\ref{sec:equilibria_modular} both the classical equilibrium positions $\vect{x}^\star = \elt \cdot \vect{x}^{\star\,(1)}$ and the parameters $\eta,\epsilon,\tau$ depend on the choice of $\elt \in \mathrm{PSL}(2,\mathbb{Z})$.

To illustrate the general expressions \eqref{eq:H_n explicit}--\eqref{eq:H_-n explicit} we give their concrete form for the first few spin-chain hamiltonians.
\begin{ex*}
	For $n=1$, \eqref{eq:H_n explicit} with $(i',i) \rightsquigarrow (i,j)$ becomes the `chiral' spin-chain hamiltonian
	\begin{equation} \label{eq:Ham_left}
		\begin{aligned} 
			H_{1,\elt} = {}& \I \, v_1^\star \, \sum_{i<j}^N P_{(i+1~\dots~j)}(\vect{x}^\star)^{-1} \, h_{i,i+1}(x_{i}^\star - x_j^\star) \,  P_{(i+1~\dots~j)}(\vect{x}^\star) \\
			= {}& \I \, v_1^\star \, \sum_{i<j}^N
			\tikz[baseline={([yshift=-.5*11pt*0.2-1pt]current bounding box.center)},xscale=0.5,yscale=0.25,font=\footnotesize]{
				\def\b{0.3};
				\draw[->] (10.5,0) node[below]{$x_N^\star$} -- (10.5,10) node[above]{$x_N^\star$};
				\draw[->] (9,0) -- (9,10);
				\foreach \x in {-1,...,1} \draw (9.25+.2*\x,-1+\b) node{$\cdot\mathstrut$};
				\foreach \x in {-1,...,1} \draw (9.25+.2*\x,11-\b) node{$\cdot\mathstrut$};
				\draw[rounded corners=2pt,->] (8,0) node[below]{$x_j^\star$} -- (8,1.5) -- (5,4.5) -- (5,5.5) -- (8,8.5) -- (8,10) node[above]{$\smash{x_j^\star}$};
				\draw[rounded corners=2pt,->] (7,0) -- (7,1.5) -- (8,2.5) -- (8,7.5) -- (7,8.5) -- (7,10); 
				\draw[rounded corners=2pt,->] (6,0) -- (6,2.5) -- (7,3.5) -- (7,6.5) -- (6,7.5) -- (6,10);
				\foreach \x in {-1,...,1} \draw (6+.2*\x,-1+\b) node{$\cdot\mathstrut$};
				\foreach \x in {-1,...,1} \draw (6+.2*\x,11-\b) node{$\cdot\mathstrut$};
				\draw[rounded corners=2pt,->] (5,0) node[below]{$x_i^\star$} -- (5,3.5) -- (6,4.5) -- (6,5.5) -- (5,6.5) -- (5,10)  node[above]{$x_i^\star$}; 
				\draw[->] (4,0) -- (4,10);
				\foreach \x in {-1,...,1} \draw (3.75+.2*\x,-1+\b) node{$\cdot\mathstrut$};
				\foreach \x in {-1,...,1} \draw (3.75+.2*\x,11-\b) node{$\cdot\mathstrut$};
				\draw[->] (2.5,0) node[below]{$x_1^\star$} -- (2.5,10)  node[above]{$x_1^\star$};
				\foreach \x in {-1,...,1} \draw (3.25+.2*\x,5) node{$\cdot\mathstrut$};
				\foreach \x in {-1,...,1} \draw (9.75+.2*\x,5) node{$\cdot\mathstrut$};
				\node at (5.5,4cm-.5pt) {\rotatebox{30}{$\circledast$}};
				\node at (1.9,5) {\smash{\textcolor{lightgray}{$\vec{a}$}}\vphantom{$a$}};
		} .
	\end{aligned}
\end{equation}
In this expression we used the notations \eqref{eq:P_1...i} for the spin operator taking care of the `transport' to and from the nearest-neighbour spin interaction. The form~\eqref{eq:Ham_left} for a hamiltonian of an integrable $q$-deformed long-range spin chain was first found in \cite{Lam_18}, and was called $H^\textsc{l}$ in \cite{lamers2022spin}.
\end{ex*}

\begin{ex*}
The second spin-Ruijsenaars operator~\eqref{eq:Ruij_spin_pm n} is
\begin{equation} \label{eq:second_difference_op}
	\begin{aligned}
		\widetilde{D}_2 &= \sum_{j<j'}^N A_{\{j,j'\}}(\vect{x}) \, P_{\{j,j'\}}(\vect{x})^{-1} \; \Gamma_j \, \Gamma_{j'} \, P_{\{j,j'\}}(\vect{x}) \\
		&= \sum_{j<j'}^N A_{\{j,j'\}}(\vect{x}) \times \!
		\tikz[baseline={([yshift=-.5*11pt*0.4*.9-2pt]current bounding box.center)},xscale=0.5,yscale=0.25,font=\footnotesize]{
			\draw[->] (13.5,0) node[below]{$x_N$} -- (13.5,13) node[above]{$x_N\vphantom{x_i}$};
			\draw[->] (12,0) -- (12,13);
			\draw[very thick,rounded corners=3.5pt] (11,0) node[below]{$\mspace{1mu}x_{j'}$} -- (11,.5) -- (6,5.5) -- (6,6.5) node[yshift=-1pt]{$\mathllap{\epsilon\,}\tikz[baseline={([yshift=-.5*12pt*.35]current bounding box.center)},scale=.35]{\fill[black] (0,0) circle (.25)}$};
			\draw[rounded corners=3.5pt,->] (6,6.5) -- (6,7.5) -- (11,12.5) -- (11,13) node[above]{$x_{j'}$};
			\draw[rounded corners=3.5pt,->] (10,0) -- (10,.5) -- (11,1.5) -- (11,11.5) -- (10,12.5) -- (10,13);
			\node at (9.5,0) [below] {$\cdots$}; 
			\node at (9.5,13) [above] {$\cdots{\vphantom{x_j}}$};
			\draw[rounded corners=3.5pt,->] (9,0) -- (9,1.5) -- (10,2.5) -- (10,10.5) -- (9,11.5) -- (9,13);
			\draw[very thick,rounded corners=3.5pt] (8,0) node[below]{$x_j$} -- (8,1.5) -- (5,4.5) -- (5,6.5) node[yshift=-1pt]{$\mathllap{\epsilon \,}\tikz[baseline={([yshift=-.5*12pt*.35]current bounding box.center)},scale=.35]{\fill[black] (0,0) circle (.25)}$};
			\draw[rounded corners=3.5pt,->] (5,6.5) -- (5,8.5) -- (8,11.5) -- (8,13) node[above]{$x_j$};
			\draw[rounded corners=3.5pt,->] (7,0) -- (7,1.5) -- (9,3.5)	-- (9,9.5) -- (7,11.5) -- (7,13);
			\node at (6.5,0) [below] {$\cdots$}; 
			\node at (6.5,13) [above] {$\cdots{\vphantom{x_j}}$};
			\draw[rounded corners=3.5pt,->] (6,0) -- (6,2.5) -- (8,4.5) -- (8,8.5) -- (6,10.5) -- (6,13);
			\draw[rounded corners=3.5pt,->] (5,0) node[below]{$x_1$} -- (5,3.5) -- (7,5.5)	-- (7,7.5) -- (5,9.5) -- (5,13) node[above]{$x_1\vphantom{x_j}$};
			\foreach \x in {-1,...,1} \draw (12.75+.2*\x,6.5) node{$\cdot\mathstrut$};	
			\node at (4.4,3.5) {\smash{\textcolor{lightgray}{$\vec{a}$}}\vphantom{$a$}};
		} \\
		&= \sum_{j<j'}^N A_{\{j,j'\}}(\vect{x}) \, P_{\{j,j'\}}(\vect{x})^{-1} \, P_{\{j,j'\}}\bigl(\vect{x}-\I \, \epsilon\,\hbar\,(\vect{e}_{j} + \vect{e}_{j'})\bigr) \; \Gamma_j \, \Gamma_{j'} \, ,
	\end{aligned}
\end{equation}
where $\{\vect{e}_j\}_j$ is the standard orthonormal basis of $\C^N$. 
By \eqref{eq:P_ii'} we have
\begin{equation} \label{eq:P_j,j'}
	\begin{aligned} 
		P_{\{j,j'\}}\bigl(\vect{x}-\I \, \epsilon\,\hbar\,(\vect{e}_{j} + \vect{e}_{j'})\bigr) = {} & P_{(2~\dots~j')}(\textcolor{gray!40}{x_j-\I \, \epsilon\,\hbar,}\,x_1,\dots,x_{j-1},x_{j+1},\dots,x_{j'} -\I \, \epsilon\,\hbar\textcolor{gray!40}{,\dots,x_N}) \\
		& \times P_{(1~\dots~j)}(x_1,\dots,x_{j-1},x_j -\I \, \epsilon\,\hbar \textcolor{gray!40}{,x_{j+1},\dots, x_{j'}-\I \, \epsilon\,\hbar,\dots,x_N}) \, ,
	\end{aligned}
\end{equation}
where the arguments that do not actually appear in any $R$-matrix are printed in gray. 
Write
\begin{equation}
	(i~\dots~j) \cdot \vect{x} =  (
	\textcolor{lightgray}{x_1,\dots,x_{i-1},x_j,\,}x_i,
	\dots
	x_{j-1}, x_j,x_{j+1}
	\dots,x_{j'}\textcolor{lightgray}{,x_{j'+1},\dots,x_N}) \, , 
\end{equation}
with in gray all coordinates that do not actually appear as the  argument of any $R$-matrix contained in $P_{(i+2~\dots~j')}$. 
Using \eqref{eq:second_difference_op}--\eqref{eq:P_j,j'} we find the corresponding spin-chain hamiltonian:
\begin{equation} \label{eq:second_difference_op_after_evaluation}
	\begin{aligned}
		H_{2,\elt} = {} & \sum_{i<j}^N \, \Biggl(\, \sum_{k =1}^{i-1} a_{\{j,k\}}(\vect{x}^\star) + \! \sum_{k =j+1}^N \!\! a_{\{j,k\}}(\vect{x}^\star) \Biggr) 
		\textcolor{Orange}{ P_{(i+1~\dots~j)}(\vect{x}^\star)^{-1} \, h_{i,i+1}(x_i^\star - x_j^\star) \,  P_{(i+1~\dots~j)}(\vect{x}^\star) } \\
		& + \! \sum_{i<j<k}^N \! a_{\{j,k\}}(\vect{x}^\star) \, \textcolor{Fuchsia}{ P_{(i~\dots~j)}(\vect{x}^\star)^{-1} } \, 
		\textcolor{Orange}{ P_{(i+2~\dots~k)}\pigl(\bigl((i~\dots~j) \cdot \vect{x}\bigr)^\star\pigr){}^{-1} \, h_{i+1,i+2}(x_i^\star-x_k^\star) } \\[-1ex]
		& \qquad \qquad \qquad \qquad \qquad \qquad \quad \textcolor{Orange}{ \times \, P_{(i+2~\dots~k)}\pigl(\bigl((i~\dots~j) \cdot \vect{x}\bigr)^\star\pigr) } \, \textcolor{Fuchsia}{ P_{(i~\dots~j)}(\vect{x}^\star) } \, .
	\end{aligned}
\end{equation}
The first line contains the terms in \eqref{eq:H_n explicit} at $m=1$ (with $(i',i_1,i_2) \rightsquigarrow (i,j,k)$, so $k>j$) along with those terms at $m=2$ for which \textcolor{OliveGreen}{all spin operators depending on $x_{i_1}$ dropped out by unitarity} ($i_1<i'$, with $(i',i_1,i_2) \rightsquigarrow (i,k,j)$). Here, the spin interactions are identical to those in \eqref{eq:Ham_left} but with different coefficients. The second line comprises the remaining terms with $m=2$ whose spin interactions do not simplify much ($i'< i_1$, with $(i',i_1,i_2) \rightsquigarrow (i,j,k)$). 
Once again, $H_{-2,\elt}$ arises from \eqref{eq:second_difference_op_after_evaluation} by a vertical reflection of its diagrammatic representation. 
The form of \eqref{eq:second_difference_op_after_evaluation} is compatible with that of the second hamiltonians from \cite{lamers2022spin}, and appeared more explicitly in \cite{MZ_23b}.
\end{ex*} 

For $n>N/2$ it is slightly easier to work with the spin-Ruijsenaars operators 
$\widetilde{D}_{-n}$ `beyond the equator'. Recall the identity \eqref{eq:D_N-n^k}. Proceeding as above, the resulting spin-chain hamiltonians are found to have the same form as \eqref{eq:H_n explicit}, but with parity-reversed spin interactions. Set $a_{-\{i_1,\dots,i_n\}}(\vect{x}^\star) \coloneqq \I \, \epsilon \, A_{-\{i_1,\dots,i_n\}}(\vect{x}^\star) \, \gamma_{-\{i_1,\dots,i_n\}}^\star$. Then we obtain
\begin{prp} \label{prp:H_-nB}
	The spin-chain hamiltonians `beyond the equator' take the explicit form
	\begin{subequations} \label{eq:H_-n explicit}
		\begin{gather}
			H_{-n,\elt} 
			= 
			\sum_{i_1 < \dots < i_n}^N \!\!\!\! a_{-\{i_1,\dots,i_n\}}(\vect{x}^\star) \sum_{m=1}^n \, \sum_{i'=i_m+1}^{N-m+1} 
			\tikz[baseline={([yshift=-.5*11pt*0.3]current bounding box.center)},xscale=.5,yscale=0.25,font=\footnotesize]{
				\draw[->] (-6,0) node[below]{$x_1^\star$} -- (-6,9) node[above]{$x_1^\star$};
				\foreach \x in {-1,...,1} \draw (-5.25+.2*\x,4.5) node{$\cdot\mathstrut$};	
				\draw[->] (-4.5,0) -- (-4.5,9);
				\draw[rounded corners=2pt,->] (-3.5,0) node[below]{$x_{i_m}^\star$} -- (-3.5,1) -- (-.5,4) -- (-.5,5) -- (-3.5,8) -- (-3.5,9) node[above]{$\vphantom{x_i}\smash{x_{i_m}^\star}$};
				\draw[rounded corners=2pt,->] (-2.5,0) -- (-2.5,1) -- (-3.5,2) -- (-3.5,7) -- (-2.5,8) -- (-2.5,9);
				\draw[rounded corners=2pt,->] (-1.5,0) -- (-1.5,2) -- (-2.5,3) -- (-2.5,6) -- (-1.5,7) -- (-1.5,9);
				\draw[rounded corners=2pt,->] (-.5,0) -- (-.5,1) -- (.5,2) -- (.5,7) -- (-.5,8) -- (-.5,9);
				\draw[rounded corners=2pt,->] (.5,0) node[below]{$x_{i'}^\star$} -- (.5,1) -- (-.5,2) -- (-.5,3) -- (-1.5,4) -- (-1.5,5) -- (-.5,6) -- (-.5,7) -- (.5,8) -- (.5,9) node[above]{$x_{i'}^\star$};
				\draw[rounded corners=2pt,->] (1.5,0) -- (1.5,9);
				\foreach \x in {-1,...,1} \draw (2.25+.2*\x,4.5) node{$\cdot\mathstrut$};	
				\draw[rounded corners=2pt,->] (3,0) node[below]{$x_N^\star$} -- (3,9) node[above]{$x_N^\star$};
				\node at (-6.5,4.5) {\smash{\textcolor{lightgray}{$\vec{a}$}}\vphantom{$a$}};
				\node at (-1,3.5cm-.5pt) {\rotatebox{30}{$\circledast$}};
			} \, ,
			\intertext{with coefficients that can be expressed in terms of the constants $v_{-1}^{\star} = v_{N-1}^{\star}/v_N^\star$ as}
			\label{eq:ev class A- Ga}
			\begin{aligned} 
				a_{-\{i_1,\dots,i_n\}}(\vect{x}^\star) 
				= 
				\I \, \epsilon \, \biggl(\frac{v_{-1}^\star}{\epsilon}\biggr)^{\!\!n} \! \prod_{m<m'}^n \frac{\theta\bigl(x_{i_m}^\star - x_{i_{m'}}^\star\bigr)^2}{\theta\bigl(x_{i_m}^\star - x_{i_{m'}}^\star +\eta\bigr) \, \theta\bigl(x_{i_m}^\star - x_{i_{m'}}^\star - \eta\bigr)} \; .
			\end{aligned}
		\end{gather}
	\end{subequations}
\end{prp}
\begin{ex*}
	From \eqref{eq:H_-n explicit} with $i\rightsquigarrow j$ we obtain the counterpart of \eqref{eq:Ham_left} with opposite `chirality',
	\begin{equation} \label{eq:Ham_right}
		\begin{aligned} 
			H_{-1,\elt} = {} & \I \, v_{-1}^\star \sum_{i<j}^{N}  P_{(j-1~\dots~i)}(\vect{x}^\star)^{-1} \, h_{j-1,j}(x_i^\star - x_{j}^\star) \, P_{(j-1~\dots~i)}(\vect{x}^\star) \\ 
			= {} & \I \, v_{-1}^\star \sum_{i<j}^N 
			\tikz[baseline={([yshift=-.5*11pt*0.2-1pt]current bounding box.center)},xscale=-0.5,yscale=0.25,font=\footnotesize]{
				\def\b{0.3};
				\draw[->] (10.5,0) node[below]{$x_1^\star$} -- (10.5,10) node[above]{$x_1^\star$};
				\draw[->] (9,0) -- (9,10);
				\foreach \x in {-1,...,1} \draw (9.25+.2*\x,-1+\b) node{$\cdot\mathstrut$};
				\foreach \x in {-1,...,1} \draw (9.25+.2*\x,11-\b) node{$\cdot\mathstrut$};
				\draw[rounded corners=2pt,->] (8,0) node[below]{$x_i^\star$} -- (8,1.5) -- (5,4.5) -- (5,5.5) -- (8,8.5) -- (8,10) node[above]{$\smash{x_i^\star}$};
				\draw[rounded corners=2pt,->] (7,0) -- (7,1.5) -- (8,2.5) -- (8,7.5) -- (7,8.5) -- (7,10); 
				\draw[rounded corners=2pt,->] (6,0) -- (6,2.5) -- (7,3.5) -- (7,6.5) -- (6,7.5) -- (6,10);
				\foreach \x in {-1,...,1} \draw (6+.2*\x,-1+\b) node{$\cdot\mathstrut$};
				\foreach \x in {-1,...,1} \draw (6+.2*\x,11-\b) node{$\cdot\mathstrut$};
				\draw[rounded corners=2pt,->] (5,0) node[below]{$x_j^\star$} -- (5,3.5) -- (6,4.5) -- (6,5.5) -- (5,6.5) -- (5,10)  node[above]{$x_j^\star$}; 
				\draw[->] (4,0) -- (4,10);
				\foreach \x in {-1,...,1} \draw (3.75+.2*\x,-1+\b) node{$\cdot\mathstrut$};
				\foreach \x in {-1,...,1} \draw (3.75+.2*\x,11-\b) node{$\cdot\mathstrut$};
				\draw[->] (2.5,0) node[below]{$x_N^\star$} -- (2.5,10)  node[above]{$x_N^\star$};
				\foreach \x in {-1,...,1} \draw (3.25+.2*\x,5) node{$\cdot\mathstrut$};
				\foreach \x in {-1,...,1} \draw (9.75+.2*\x,5) node{$\cdot\mathstrut$};
				\node at (5.5,4cm-.5pt) {\rotatebox{30}{$\circledast$}};
				\node at (11.1,5) {\smash{\textcolor{lightgray}{$\vec{a}$}}\vphantom{$a$}};
			} ,
		\end{aligned}
	\end{equation}
	using the notation \eqref{eq:P_N...i} for the spin operators. This expression first appeared in \cite{lamers2022spin}, where it was called $H^\textsc{r}$.
\end{ex*}

In conclusion, at any fixed classical equilibrium configuration $(\vect{x}^\star,\vect{p}^\star) = \elt \cdot (\vect{x}^{\star\,(1)},\vect{p}^{\star\,(1)}) \in M^\star_\mathbb{C}$, parametrised by $\elt \in \mathit{PSL}(2,\mathbb{Z})$, we obtain a hierarchy of commuting spin-chain hamiltonians with long-range spin interactions. This hierarchy contains one spin-chain operator $H_{n,B}$ for each spin-Ruijsenaars operator $\widetilde{D}_n$. The exception is $n=N$. Indeed, since the total shift operator $\widetilde{D}_N = \Gamma_1 \cdots \Gamma_N \, \mathrm{id}$ acts trivially on spins, its classical limit lies in the centre $Z(\widetilde{\alg}_\hbar) = \mathrm{Fun}(\vect{x}) \, \mathrm{id}$, and commutes trivially with all other hamiltonians. Instead, it gives rise to the lattice translation operator, at least in the trigonometric case~\cite{lamers2022spin}. Although we do not yet have the proof in the elliptic case, such translation operators are known in the face and vertex examples \cite{klabbers2025landscapes}.

\subsection{Application: examples of spin chains} \label{sec:examples} 
For specific choices of the input data, freezing yields various known spin chains\,---\,or rather, modular families thereof, indexed by $\elt \in \mathit{PSL}(2,\mathbb{Z})$.

\begin{itemize}
\item \textbf{Vertex-type.} If, as in \eqref{eq:P^v}, $P_{i,i+1}(u) = P^\textsc{v}_{i,i+1}(u)$ is the Baxter--Belavin $R$-matrix \eqref{eq:Vertex-type-R-matrices} of type $\mathfrak{gl}_r$,
the elliptic spin-Ruijsenaars system in \textsection\ref{sec:ell_sRuij} is 
that of \cite{MZ_23a}. 
In the eight-vertex case ($r=2$), there exist (explicit) functions $V_\alpha(u;\eta \,|\, \tau)$ for $\alpha \in \{0,x,y,z\}$ such that
the nearest-neighbour interaction from \eqref{eq:h_i,i+1(x)} can be recast via Pauli matrices as \cite{klabbers2025landscapes}
\begin{equation} \label{eq:h=sum V spin}
	r=2: \quad 
	h^{\mspace{-1mu}\textsc{v}}(u) = \!\!
	\tikz[baseline={([yshift=-.5*11pt*0.4]current bounding box.center)},xscale=.5,yscale=0.25,font=\footnotesize]{
		\draw[rounded corners=2pt,->] (1,0) node[below]{$\,x\smash{''}$}-- (1,1) -- (0,2) -- (0,3) -- (1,4) -- (1,5) node[above]{$x\smash{''}$};
		\draw[rounded corners=2pt,->] (0,0) node[below]{$x\smash{'}$} -- (0,1) -- (1,2) -- (1,3) -- (0,4) -- (0,5) node[above]{$x\smash{'}$};
		\node at (0.5,1.5cm-.5pt) {\rotatebox{30}{$\circledast$}};
	} 
	\!\!\! = \, \sum_{\alpha=0}^z \frac{1}{4} \, V_\alpha(u) \!\!
	\tikz[baseline={([yshift=-.5*11pt*0.4]current bounding box.center)},xscale=.5,yscale=0.25,font=\footnotesize]{
		\draw[->] (1,0) node[below]{$\,x\smash{''}$} -- (1,5) node[above]{$x\smash{''}$};
		\draw[->] (0,0) node[below]{$x\smash{'}$} -- (0,5) node[above]{$x\smash{'}$};
		\draw[style={decorate, decoration={zigzag,amplitude=.5mm,segment length=1mm}}] (0,2.5) -- node[above]{$\alpha$} (1,2.5);
	} \!\!\! = \, \sum_{\alpha=0}^z \frac{1}{4} \, V_\alpha(u) \bigl( 1 - P \, \sigma^\alpha \otimes \sigma^\alpha\bigr) \, , \quad u = x' - x'' \, .
\end{equation}	
Freezing at $\elt=\SLid \in \text{PSL}(2,\Z)$ gives the MZ chain \cite{MZ_23b}. Instead taking $\elt=S$ yields a variant that we call the MZ$'$ chain. For $r=2$ the latter was introduced in \cite{klabbers2025landscapes}, and various limits were evaluated; remarkably, the difference $H_{\pm1,\SLid} - H_{\pm1,S}$ is a multiple of the identity \cite{klabbers2025landscapes}, in such a way that (only) the MZ$'$ chain admits a short-range limit. \\[-1.5ex] 

\item \textbf{Face-type.} When, like in \eqref{eq:P^f}, $P_{i,i+1}(u) = P^{\mspace{1mu}\textsc{f}}_{i,i+1}(u)$ is Felder's dynamical $R$-matrix of type $\mathfrak{gl}_r$, we instead obtain a dynamical elliptic spin-Ruijsenaars system which for $r=2$ appeared in \cite{klabbers2024deformed}. In that case, there is a physically meaningful factorisation
\begin{equation} \label{eq:h=V spin factorised}
	r=2: \quad 
	h^{\textsc{f}}(u,a) = \!\!
	\tikz[baseline={([yshift=-.5*11pt*0.4]current bounding box.center)},xscale=.5,yscale=0.25,font=\footnotesize]{
		\draw[rounded corners=2pt,->] (1,0) node[below]{$\,x\smash{''}$}-- (1,1) -- (0,2) -- (0,3) -- (1,4) -- (1,5) node[above]{$x\smash{''}$};
		\draw[rounded corners=2pt,->] (0,0) node[below]{$x\smash{'}$} -- (0,1) -- (1,2) -- (1,3) -- (0,4) -- (0,5) node[above]{$x\smash{'}$};
		\node at (-.4,1.5) {$a$};
		\node at (0.5,1.5cm-.5pt) {\rotatebox{30}{$\circledast$}};
	} 
	\!\!\! = \theta(\eta) \, V(u,a) \!
	\tikz[baseline={([yshift=-.5*11pt*0.4]current bounding box.center)},xscale=.5,yscale=0.25,font=\footnotesize]{
		\draw[->] (1,0) node[below]{$\,x\smash{''}$} -- (1,5) node[above]{$x\smash{''}$};
		\draw[->] (0,0) node[below]{$x\smash{'}$} -- (0,5) node[above]{$x\smash{'}$};
		\draw[style={decorate, decoration={zigzag,amplitude=.5mm,segment length=1mm}}] (0,2.5) -- (1,2.5);
		\node at (-.4,1.5) {$a$};
	} \!\!\! = \theta(\eta) \, V(u,a) \, E(u,a) \, , \quad 
	\begin{aligned}
		u & = x'-x'' \, , \\
		a & = a_1 - a_2 \, ,
	\end{aligned}
\end{equation}	
for an (explicit) function $V(u,a;\eta\,|\,\tau)$, i.e.\ the potential, and operator $E(u,a;\eta \,|\, \tau)$, i.e.\ the nearest neighbour spin interaction \cite{klabbers2024deformed}. Freezing at $\elt=S$ results in the $q$-deformed Inozemtsev chain, introduced for $r=2$ in \cite{klabbers2024deformed}. While its cousin with $\elt=\SLid$ also generalises the (long-range) $q$-deformed Haldane--Shastry chain \cite{Ugl_95u,Lam_18}, it does not have a short-range limit.
\end{itemize}
Either example spans a `landscape' of integrable spin chains. The limiting spin chains were explicitly evaluated in \cite{klabbers2025landscapes} for $r=2$, see also below. The vertex- and face-type landscapes only overlap in a single point, namely the rational limit ($\tau \to \I\,\infty$, $N\to\infty$). At higher rank ($r>2$) the spin chains with $\elt=S$ have not yet been subject of detailed investigation.

The freezing procedure of \textsection\ref{sec:partial class lim}--\ref{sec:explicit_hams} can be executed in various limits. 

\subsubsection{Trigonometric limit} \label{sec:lim_trig}
When the modular parameter $\tau \to \I\,\infty$ is removed, the elliptic functions degenerate into trigonometric ones. Physically, this limit corresponds to long(est)-range interactions. The (appropriate) function space $\mathrm{Fun}(\vect{x})$ from $\alg_\hbar$ and $\widetilde{\alg}_\hbar$ may be replaced by the space $\mathbb{C}[z_1^{\pm1},\dots,z_N^{\pm1}]$ of Laurent polynomials in multiplicative coordinates $z_j = \E^{\I\,x_j}$, cf.~just above \eqref{eq:A_0}. The trigonometric spin-Ruijsenaars operators still have the form \eqref{eq:Ruij_spin_pm n}. 

Freezing works the same, but there is no more modular action. 
In particular, there is a single classical equilibrium configuration, with equispaced (real) positions and $p^\star_j = c$ all equal.\,%
\footnote{\ Here the summands in \eqref{eq:v_n*} all coincide for any given $n$: the coefficients of the trigonometric Ruijsenaars (--Schneider) functions equal $A_I^\text{tri}(\vect{x}^\star) = \binom{N}{n}_q/\binom{N}{n}$ for $n=|I|$, with $\binom{N}{n}_q$ a $q$-binomial coefficient with $q \sim \E^\eta$.}
From the elliptic perspective, the coordinates in the $\elt=S$ solution \eqref{eq:freezing_type2} are all sent to $0$ in the trigonometric limit, 
so this solution cannot be used due to divergences. Appropriate regularisation amounts to computing limits of the operators evaluated at equilibrium, i.e.\ of the spin-chain hamiltonians, see \cite{klabbers2025landscapes} for examples of this when $r=2$. The spin-chain hamiltonians still have the form \eqref{eq:H_n explicit} and \eqref{eq:H_-n explicit}.

\begin{itemize}
\item \textbf{Vertex-type.} We obtain the trigonometric $R$-matrix in the principal grading, which for $r=2$ is the (symmetric) six-vertex $R$-matrix. This gives the trigonometric MZ chain. For $r=2$ the nearest-neighbour spin interaction can be factorised like in \eqref{eq:h=V spin factorised} \cite{klabbers2025landscapes}.
\item \textbf{Face-type.} After an intermediate level with dynamical $R$-matrix, sending $a\to -\I \, \infty$ yields the trigonometric $R$-matrix in the homogeneous grading, related to the Hecke generators by `Baxterisation'. The resulting spin chain is the $q$-deformed Haldane--Shastry chain~\cite{Ugl_95u,Lam_18,lamers2022spin}.\,%
\footnote{\ Note that \cite{Ugl_95u,lamers2022spin} used an expansion in $\epsilon$ rather than $\hbar$. Removing the derivatives coming from expanding $\Gamma$ then requires a shift by the part linear in $\epsilon$ of $\widetilde{D}_N = D_N \, \mathrm{id}$. Whilst the resulting spin-chain hamiltonians are the same, this shift is visible in the eigenvalues, which are explicitly known: see (1.77)--(1.78) in \cite{lamers2022spin}. It would be interesting to understand the origin of this shift in the eigenvalues of the trigonometric `chiral' hamiltonians in the more precise setting of the current paper.}
The connection to Hecke algebras enables a detailed understanding of the spectrum, with eigenvectors featuring Macdonald polynomials~\cite{lamers2022spin}, and quantum-affine invariance~\cite{drinfel1986degenerate,bernard1993yang,Ugl_95u,lamers2022spin}.
\end{itemize}

\subsubsection{Undeformed limit} \label{sec:lim_undeformed}
The `undeformed' limit arises by to setting $\epsilon = \eta/g$ and sending $\eta \to 0$, which may again be achieved using deformation quantisation. Then $\alg_\hbar$ reduces to the Weyl algebra, with $\mathbb{C}[\Gamma_1^{\pm1},\dots,\Gamma_N^{\pm1}]$ replaced by $\mathbb{C}[\hat{p}_1,\dots,\hat{p}_N]$, and similarly for $\widetilde{\alg}_\hbar$. The elliptic spin-Ruijsenaars operators reduce to (nonrelativistic) elliptic spin-Calogero--Sutherland operators, depending on $(\vect{x},\vect{p};g \, | \, \tau)$. 
\begin{cor}
In the trigonometric limit, the modular action \eqref{eq:S_T_extended_with_shift} becomes
\begin{equation} \label{eq:S_T_extended_with_shift_CS_case}
	\begin{aligned}
		S & \colon (\vect{x},\vect{p} ; g \,|\, \tau) 
		\longmapsto \biggl(-\frac{\vect{x}}{\tau} , -\tau\,\vect{p} - \frac{2\pi \mspace{2mu}\I}{g} \, \sum_{j=1}^N \bigl( |\vect{x}| - N\mspace{1mu}x_{j} \bigr) \, \vect{e}_j \, ; -\tau \, g \,\bigg|\, {-}\frac{1}{\tau} \biggr) \, , \\
		T & \colon (\vect{x},\vect{p}; g\,|\, \tau) 
		\longmapsto (\vect{x},\vect{p}; g\,|\, \tau + 1) \, .
	\end{aligned}
\end{equation}
\end{cor}
Precisely as in \textsection\ref{sec:equilibria_modular}, this generates a modular family of equilibria of the (classical) elliptic Calogero--Moser--Sutherland system, which were previously found in \cite{doreyEllipticSuperpotentialSoftly1999,bourgetDualityModularityElliptic2015,bourgetN1gauge2016}.
In the trigonometric limit, the only equilibrium has equispaced positions, see also \cite{ruijsenaars1995action}.

\begin{itemize}
\item \textbf{Vertex-type.} Here $\elt=1$ gives the Sechin--Zotov chain~\cite{sechin2018r}. For $r=2$ the interactions remain of the form \eqref{eq:h=sum V spin}, and the spin chain is fully anisotropic even at $\eta=0$, except in the trigonometric limit where it becomes the antiperiodic Fukui--Kawakami chain~\cite{fukui1996exact}. Instead taking $\elt=S$ gives what we call the SZ$'$ chain, which admits a short-range limit as well: the antiperiodic \textsc{xx} chain \cite{klabbers2025landscapes}.
\item \textbf{Face-type.} Here the limit $\eta\to0$ is the isotropic limit, in which the hamiltonians are $\mathfrak{gl}_r$-invariant. This limit results in the Inozemtsev chain~\cite{Inozemtsev:1989yq}, whose integrability was recently proven using freezing in \cite{chalykh2024integrability}. The trigonometric limit gives the Haldane--Shastry chain. The choice $\elt=S$ gives the Weierstra{\ss} pair potential used in \cite{Inozemtsev_1995}, which directly admits a short-range limit: the Heisenberg \textsc{xxx} chain; see also \textsection2 in \cite{klabbers2022coordinate}. By rescaling the dynamical parameter $a = a'/\eta$ before taking $\eta\to0$ one obtains a one-parameter deformation of the Inozemtsev chain~\cite{klabbers2024deformed}. 
\end{itemize}

\section{Hybrid systems}
\label{sec:hybrid_systems}

\noindent 
As a by-product of freezing, a class of hybrid systems naturally arises along the way. They can be interpreted as integrable systems in their own right, including a notion of hamiltonian dynamics, following \cite{mikhailovCommutativePoissonAlgebras2024,liashyk2024classical}. In this section we review how this works, and elaborate on the interpretation of freezing in this setting.

\subsection{Hybrid dynamics}
\label{sec:hybrid_dynamics}
As argued in \cite{mikhailovCommutativePoissonAlgebras2024}, the first step is the identification of a (noncommutative) Poisson subalgebra of the (noncommutative) associative algebra $\widetilde{\alg}_\hbar$, which will contain all suitable hamiltonians. It will be more transparent to use the identification $\tilde{c}_\hbar$ from \eqref{eq:ctilde_hbar} to work in $\widetilde{\alg}_0[\mspace{-2mu}[\hbar]\mspace{-2mu}]$ instead of $\widetilde{\alg}_\hbar$. In particular, inside $\widetilde{\alg}_0[\mspace{-2mu}[\hbar]\mspace{-2mu}]$ the spin-Ruijsenaars operators \eqref{eq:Ruij_spin_pm n} take the form
\begin{equation} \label{eq:Ruij_spin_n_in_A0hbar}
\tilde{c}_\hbar\big(\widetilde{D}_{\pm n}\big) \eqqcolon \widetilde{\DA}_{\pm n} = \widetilde{\DA}_{\pm n}^{(0)} + \widetilde{\DA}_{\pm n}^{(1)} \,  \hbar + O(\hbar^2)\, , 
\end{equation}
for some  $\widetilde{\DA}_{\pm n}^{(k)} \in \widetilde{\alg}_0$. Note the difference in fonts! The precise form of these elements depends on the choice of $\tilde{c}_\hbar$, cf.~the proof of \eqref{eq:class integrab}. In the case of \eqref{eq:Ruij_spin_n_in_A0hbar} we have
\begin{equation}
\tilde{c}_\hbar\bigl(\widetilde{D}_{\pm n} \bigr) = D_{\pm n}^\mathrm{cl} \, \mathrm{id} + \hbar \, \biggl(B_{\pm n} + \sum_I \widetilde{A}_{\pm I}^{(1)} \, \gamma_I^{\pm1} \biggr) + O(\hbar^2)
\end{equation}
for some $B_{\pm n} \in \widetilde{\alg}_0$. Observe that $\sum_I \widetilde{A}_{\pm I}^{(1)} \, \gamma_I^{\pm1} = \tilde{c}_0\bigl(\widetilde{D}_{\pm n}^{(1)}\bigr)$. We will require $\tilde{c}_\hbar$ to be such that $B_{\pm n} \in Z(\widetilde{A}_0)$.
This leads us to consider the following subalgebra of $\widetilde{\alg}_0[\mspace{-2mu}[\hbar]\mspace{-2mu}]$.
\begin{prp*}[Mikhailov--Vanhaecke \cite{mikhailovCommutativePoissonAlgebras2024}]
	The subalgebra
	\begin{equation} \label{eq:H_hbar}
		\mathcal{H}_\hbar \coloneqq Z\bigl(\widetilde{\alg}_0\bigr) + \hbar \, \widetilde{\alg}_0[\mspace{-2mu}[\hbar]\mspace{-2mu}] \,\subset\, \widetilde{\alg}_0[\mspace{-2mu}[\hbar]\mspace{-2mu}]
	\end{equation}
	equipped with the rescaled bracket $[\,\cdot\,,\cdot\,]_\hbar \coloneqq \hbar^{-1}\,[\,\cdot\,,\cdot\,]$ forms a (noncommutative) Poisson algebra.
	Moreover, $\mathcal{H}_\hbar$ acts on on $\widetilde{\alg}_0[\mspace{-2mu}[\hbar]\mspace{-2mu}]$ by derivations,
	\begin{equation} \label{eq:action_on_Ahbar}
		(z+ \hbar \, H) \cdot A = [z+\hbar \, H,A]_\hbar\, , \qquad 
		z \in Z\bigl(\widetilde{\alg}_0\bigr)
		,\quad H,A \in \widetilde{\alg}_0[\mspace{-2mu}[\hbar]\mspace{-2mu}] \, . 
	\end{equation}
	Together with the (left and right) action induced by the product on $\widetilde{\alg}_0[\mspace{-2mu}[\hbar]\mspace{-2mu}]$, i.e.
	\begin{equation}
		\label{eq:quantum_product_action}
		(z+ \hbar \, H) A = z A + \hbar \, H A \quad \text{ and } \quad A(z+ \hbar \, H) = A z + \hbar\, A H
	\end{equation}
	this gives $\widetilde{\alg}_0[\mspace{-2mu}[\hbar]\mspace{-2mu}]$ the structure of a $\mathcal{H}_\hbar$-Poisson module. 
\end{prp*}
\begin{proof}[Proof.]
	Since commutators in $\mathcal{H}_\hbar$ are necessarily at least of order $\hbar$, the rescaled bracket is well defined. All other checks are straightforward.
\end{proof}

By our assumption \eqref{eq:Atilde=A id+Ohbar}, the spin-Ruijsenaars operators $\widetilde{\DA}_{n}$ are elements of $\mathcal{H}_\hbar$, and, more precisely, by \eqref{eq:Dtilde commute} their span is a Poisson-commuting subalgebra $\mathcal{B}_\hbar \subset \mathcal{H}_\hbar$, with
\begin{equation}
\pigl[\widetilde{\DA} ,\widetilde{\DA}' \mspace{1mu} \pigr]_\hbar =0 \qquad \text{for all} \quad \widetilde{\DA},\widetilde{\DA}' \in \mathcal{B}_\hbar\, .  
\end{equation}
Together with \eqref{eq:action_on_Ahbar}, this implies that, in the Heisenberg picture of quantum mechanics, for an operator $A \in \widetilde{\alg}_0[\mspace{-2mu}[\hbar]\mspace{-2mu}]$ the system of evolution equations
\begin{equation} \label{eq:time ev}
\frac{\partial\mspace{-1mu}A}{\partial\mspace{1mu}t_n} = 	\frac{\I}{\hbar} \, \pigl[\widetilde{\DA}_{n},A\pigr]  \, , \quad 1\leqslant n \leqslant N \, , 
\end{equation}
is compatible. 

In this context, the partial classical limit $\tilde{c}_0$ should make part of the action \eqref{eq:action_on_Ahbar} defining the time evolution \eqref{eq:time ev} classical. The partial classical limit of $\widetilde{\alg}_0[\mspace{-2mu}[\hbar]\mspace{-2mu}]$ is $\widetilde{\alg}_0[\mspace{-2mu}[\hbar]\mspace{-2mu}]/\hbar \,\widetilde{\alg}_0[\mspace{-2mu}[\hbar]\mspace{-2mu}] \cong \widetilde{\alg}_0$, whose centre is $Z(\widetilde{\alg}_0) = \mathrm{Fun}(\vect{x})[\gamma_1^{\pm1},\ldots,\gamma_N^{\pm1}] \, \mathrm{id}$. For noncommutative algebras such as $\widetilde{\alg}_0$, there is no good notion of a Poisson structure, so one cannot immediately define a compatible system of partially classical time evolutions. 

Like for $\widetilde{\alg}_0[\mspace{-2mu}[\hbar]\mspace{-2mu}] \cong \widetilde{\alg}_\hbar$, the `classical limit' of $\mathcal{H}_\hbar$ is a quotient, $\mathcal{H}_\hbar/\hbar \, \mathcal{H}_\hbar$. However, elements in the latter quotient still contain $\hbar$: as a vector space,
\begin{subequations} \label{eq:MV Poisson alg}
\begin{gather}
	\label{eq:MV space}
	\mathcal{H}_\hbar/\hbar \, \mathcal{H}_\hbar = Z\bigl(\widetilde{\alg}_0\bigr) \oplus \widetilde{\alg}_0[\mspace{-2mu}[\hbar]\mspace{-2mu}]\big/Z\bigl(\widetilde{\alg}_0\bigr) \, ,
	\intertext{with elements being of the (\emph{partially} classical) form}
	\label{eq:MV elements}
	z + \hbar \, \,\overline{\!H}\, , \qquad z\in Z\bigl(\widetilde{\alg}_0\bigr)\, ,
	\quad \,\overline{\!H} \coloneqq H \text{ mod } Z\bigl(\widetilde{\alg}_0\bigr) 	\, , 
	\quad H\in \widetilde{\alg}_0[\mspace{-2mu}[\hbar]\mspace{-2mu}]
	\, .
\end{gather}
\end{subequations}
	
\begin{prp*}[Mikhailov--Vanhaecke \cite{mikhailovCommutativePoissonAlgebras2024}]
	The subalgebra $\hbar \, \mathcal{H}_\hbar$ is a Poisson ideal of $\bigl(\mathcal{H}_\hbar,[\,\cdot\,,\cdot\,]_\hbar\bigr)$. Moreover, the quotient $\mathcal{H}_\hbar/\hbar \, \mathcal{H}_\hbar$ inherits a product from $\mathcal{H}_\hbar$,
	\begin{equation}
		(z+\hbar\, \overline{\!H})(z'+\hbar\, \overline{\!H}') = z z' + \hbar \, \overline{ \bigl(z H' + z' H+m_1(z,z') \bigr) }\, , 
	\end{equation}
	as well as a Poisson structure defined by the bracket
	\begin{equation} \label{eq:MV_bracket}
		\bigl[\mspace{-7.5mu}\bigl\{ z + \hbar \, \,\overline{\!H}, z' + \hbar \, \,\overline{\!H}' \bigr\}\mspace{-7.5mu}\bigr] = \I \, \{z,z'\} + \hbar \, \overline{ \bigl( - \{z,z'\}_2 + \I \, \{z,H'\} + \I \, \{H,z'\} + [H,H'] \bigr) }\, , 
	\end{equation}
	where the right-hand side features the second-order bracket defined in \eqref{eq:m_2}, the Poisson bracket \eqref{eq:PB_spin}, and the commutator. 
	Finally, $\widetilde{\alg}_0$ is an $\mathcal{H}_\hbar/\hbar \, \mathcal{H}_\hbar$-Poisson-module under the classical limit of the action in \eqref{eq:action_on_Ahbar}\footnote{As a direct computation shows $\tilde{c}_0\bigl( [z+\hbar\, H,A]_\hbar \bigr) = \I \, \{z,A\}+ [\hbar \, H,A]_\hbar$.}, 
	\begin{equation}
		\label{eq:hybrid_action}
		(z+ \hbar\, \,\overline{\!H})\cdot A  = \I \, \{z,A\}+ [\hbar\, \,\overline{\!H},A]_\hbar\, , 
	\end{equation}
	and the (left- and right-) product actions 
	\begin{equation}
		\label{eq:hybrid_product_action}
		(z+ \hbar \, \, \overline{\!H}) A  = z A \quad \text{ and } \quad A (z+ \hbar \, \, \overline{\!H}) = z A = Az\, .
	\end{equation}  
\end{prp*}

In other words, on these quotients,
the algebra homomorphism $\mathcal{H}_\hbar \longrightarrow \mathrm{End}(\widetilde{\alg}_0[\mspace{-2mu}[\hbar]\mspace{-2mu}])$ defined by the action \eqref{eq:action_on_Ahbar} descends to a new algebra homomorphism 
$\mathcal{H}_\hbar/\hbar \, \mathcal{H}_\hbar \longrightarrow \mathrm{End}(\widetilde{\alg}_0)$ defined by \eqref{eq:hybrid_action}

The image $\mathcal{B}_0 \subset \mathcal{H}_\hbar/\hbar \, \mathcal{H}_\hbar$ of $\mathcal{B}_\hbar$ in this quotient is still a Poisson-commuting subalgebra. Moreover, the action \eqref{eq:hybrid_action} is a  derivation, so in full analogy with the above it defines a compatible set of partially classical evolution equations for $A \in \widetilde{\alg}_0$,
\begin{equation} \label{eq:hybrid_evolution}
\frac{\partial \mspace{-1mu} A}{\partial \mspace{1mu} t_{n}} = -\bigl\{\widetilde{\DA}_n^{(0)},A\bigr\}+ \frac{\I}{\hbar} \, \bigl[\hbar\, \,\overline{\widetilde{\DA}_n^{(1)}} ,A\bigr]  \, , \qquad 1\leqslant n \leqslant N \, ,
\end{equation}
where $\widetilde{\DA}_n^{(0)} + \hbar \, \,\overline{\widetilde{\DA}_n^{(1)}}$ denotes the image of $\widetilde{\DA}_n$ in the quotient. Dynamical systems described by such equations were dubbed integrable \emph{hybrid} dynamical systems in \cite{liashyk2024classical}. 

\subsection{Poisson-algebraic interpretation of freezing} \label{sec:freezing_hybrid}

We now investigate what happens to the $\DA$s under the partial classical limit, and first re-establish that the spin-chain hamiltonians~\eqref{eq:spin chain hamiltonians} commute.
\begin{proof}[Alternative proof of Theorem~\ref{thm:HnB commute} (Chalykh \cite{chalykh2024integrability}).]
	Since $[\widetilde{\DA}_n,\widetilde{\DA}_m]_\hbar =0$, it holds in $\mathcal{B}_0$ that
	\begin{equation}
		\label{eq:freezing_condition_in_A0}
		-\{\widetilde{\DA}_n^{(0)},\widetilde{\DA}_m^{(0)}\}_2+\I \{\widetilde{\DA}_n^{(0)},\overline{\widetilde{\DA}_m^{(1)}}\}+ \I \{\overline{\widetilde{\DA}_n^{(1)}},\widetilde{\DA}_m^{(0)}\} + \pigl[\overline{\widetilde{\DA}_n^{(1)}},\overline{\widetilde{\DA}_m^{(1)}}\pigr] \equiv 0 \ \mathrm{mod}\ Z\bigl(\widetilde{\mathcal{A}}_0\bigr)\, . 
	\end{equation}
	Plugging in the explicit expressions for the $\DA$s as prescribed by \eqref{eq:Ruij_spin_n_in_A0hbar} we obtain 
	\begin{equation}
		\begin{aligned}
			-\pigl\{\tilde{c}_0\bigl(D_n^{(0)}\bigr),\tilde{c}_0\bigl(D_m^{(0)}\bigr) \pigr\}_2 & + \I \, \pigl\{ \tilde{c}_0\bigl(D_n^{(0)}\bigr),\overline{\tilde{c}_0\bigl(D_m^{(1)}\bigr)} \pigr\} \\ 
			& + \I \, \pigl\{ \overline{\tilde{c}_0\bigl(D_n^{(1)}\bigr)},\tilde{c}_0\bigl(D_m^{(0)}\bigr) \pigr\} + \pigl[ \overline{\tilde{c}_0\bigl(D_n^{(1)}\bigr)},\overline{\tilde{c}_0\bigl(D_m^{(1)}}\bigr) \pigr] \equiv 0 \ \mathrm{mod}\ Z\bigl(\widetilde{\mathcal{A}}_0\bigr)\, ,
		\end{aligned}
	\end{equation}
	where we note that the ambiguity in $\widetilde{\DA}_{n}^{(1)}$ and $\widetilde{\DA}_{m}^{(1)}$ due to the choice of $\tilde{c}_\hbar$ is precisely taken care of by the quotient. The first term vanishes, as shown in \eqref{eq:m_2}, so
	\begin{equation} \label{eq:zero curv}
		\I \, \pigl\{ \tilde{c}_0\bigl(D_n^{(0)}\bigr), \overline{\tilde{c}_0\bigl(D_m^{(1)}\bigr)} \pigr\} + \I \, \pigl\{\overline{\tilde{c}_0\bigl(D_n^{(1)}\bigr)},\tilde{c}_0\bigl(D_m^{(0)}\bigr)\pigr\} + \pigl[\overline{\tilde{c}_0\bigl(D_n^{(1)}\bigr)},\overline{\tilde{c}_0\bigl(D_m^{(1)}\bigr)}\pigr] \equiv 0  \ \mathrm{mod}\ Z\bigl(\tilde{\mathcal{A}}_0\bigr)\, .
	\end{equation}
	This is the analogue of \eqref{eq:after_c0} in $\mathcal{B}_0$. Proceeding as in \textsection\ref{sec:proof}, evaluating at equilibria shows that the first two terms also vanish, leaving us with
	\begin{equation}
		\pigl[\mathrm{ev}_{\mspace{-2mu}\elt}\bigl(\overline{\tilde{c}_0(D_n^{(1)})}\bigr), \mathrm{ev}_{\mspace{-2mu}\elt}\bigl(\overline{\tilde{c}_0(D_m^{(1)})}\bigr)\pigr] \equiv 0 \ \mathrm{mod}\ Z\bigl(\tilde{\mathcal{A}}_0\bigr)\, .
	\end{equation}
	Since the commutator of finite-size square matrices on the left-hand side is traceless, and $Z\bigl(\widetilde{\mathcal{A}}_0\bigr)$ is spanned by $\alg_0 \, \mathrm{id}$, we find that in fact
	\begin{equation} \label{eq:commuting_elms_B0}
		\pigl[ \mathrm{ev}_{\mspace{-2mu}\elt}\bigl(\overline{\tilde{c}_0(D_n^{(1)})}\bigr), \mathrm{ev}_{\mspace{-2mu}\elt}\bigl(\overline{\tilde{c}_0(D_m^{(1)})}\bigr)\pigr] =0\, .
	\end{equation}
	Finally, as any central terms drop from the commutator, we may omit the bars.
	In this way \eqref{eq:spin chain integrable} is recovered from the hybrid point of view. 
\end{proof}
\begin{rmk*}
	From a physics perspective, discarding terms proportional to the identity matrix is consistent with the fact that such additive corrections only shift the energy, and play no physical role. 
\end{rmk*}
With this in place, we turn to the dynamical systems defined by the hamiltonians in $\mathcal{B}_0$ and how to freeze these.

\begin{def*}
	For any $\elt \in \text{PSL}(2,\mathbb{Z})$ we define the evaluation map: $\evq \colon \mathcal{B}_0 
	\longrightarrow \mathrm{Mat}(r^N, \C) \ \mathrm{mod} \ Z(\widetilde{\alg}_0)$ on the quotient by $z+\hbar \, \,\overline{\!H} \longmapsto \hbar \ \mathrm{ev}_{\mspace{-2mu}\elt}(\,\overline{\!H})$.
\end{def*}
Freezing the hamiltonians amounts to the application of $\evq$ to $\mathcal{B}_0$:
\begin{prp}
	The map $\evq$ is a morphism of Poisson algebras from the subalgebra $\mathcal{B}_0 \subset \mathcal{H}_\hbar/\hbar \, \mathcal{H}_\hbar$ into $\bigl(\mathrm{Mat}(r^N, \C) \ \mathrm{mod} \ Z(\widetilde{\alg}_0),[\,\cdot\,,\,\cdot\,]\bigr)$. 
\end{prp} 

The image $\evq(\mathcal{B}_0)$ is spanned by $\hbar\; \mathrm{ev}_{\mspace{-2mu}\elt} \pigl(\tilde{c}_0\bigl(\overline{\widetilde{D}_n^{(1)}}\bigr)\pigr)$ for $1 \leqslant n \leqslant N$, which notably all come with a factor $\hbar$, 
and can be interpreted as the algebra of spin-chain hamiltonians.
\begin{proof}[Proof.]
	All Poisson brackets in $\mathcal{B}_0$ vanish, so for all $z+\hbar \, \,\overline{\!H}$, $z'+\hbar \, \,\overline{\!H}' \in \mathcal{B}_0$,
	\begin{equation}
		\evq \pigl( \bigl[\mspace{-7.5mu}\bigl\{  z+\hbar \, \,\overline{\!H} , z' + \hbar \,  \,\overline{\!H}' \bigr\}\mspace{-7.5mu}\bigr] \pigr)= 
		0 = 
		\pigl[ \mathrm{ev}_{\mspace{-2mu}\elt}\bigl(\,\overline{\!H}\bigr) , \mathrm{ev}_{\mspace{-2mu}\elt}\bigl(\,\overline{\!H}'\bigr)\pigr] \, , 
	\end{equation}
	where the second equality follows from \eqref{eq:commuting_elms_B0}.
\end{proof}

In the same way we can reinterpret the role of the evaluation map on $\widetilde{\alg}_0$, in order to freeze the latter: 
\begin{lem}
	The map $\mathrm{ev}_{\mspace{-2mu}\elt}$ defines an (associative-)algebra homomorphism $A\longmapsto \mathrm{ev}_{\mspace{-2mu}\elt}(A)$ from $\widetilde{\alg}_0$ into $\bigl(\text{Mat}(r^N, \C),\,\cdot\,\bigr)$.
\end{lem}	
\begin{proof}[Proof.]
	The property $\mathrm{ev}_{\mspace{-2mu}\elt}(A \, A') = \mathrm{ev}_{\mspace{-2mu}\elt}(A) \, \mathrm{ev}_{\mspace{-2mu}\elt}(A')$ is evident.
\end{proof}
We can once more define a set of actions\,---\,now of $\evq(\mathcal{B}_0) \subset (\mathrm{Mat}(r^N, \C) \ \mathrm{mod} \ Z(\widetilde{\alg}_0))$ on $\mathrm{Mat}(r^N, \C)$. 
\begin{prp*} $\mathrm{Mat}(r^N, \C)$ is a $\evq(\mathcal{B}_0)$-Poisson module with Lie algebra action
	\begin{equation}
		\label{eq:action_quantum}
		H\cdot A = [H,A]_\hbar\, , \qquad H \in \evq(\mathcal{B}_0)\, , \quad A \in \text{Mat}(r^N, \C) 
	\end{equation}
	and the product actions given by left- and right- matrix multiplication. 
\end{prp*}
\begin{proof}[Proof.]
	Since $[\cdot,\cdot]_\hbar$ is a matrix commutator, all properties follow from matrix multiplication. 
\end{proof}

The action \eqref{eq:action_quantum} is also a derivation for the (matrix) product in $\mathrm{Mat}(r^N, \C)$, and thus defines a now fully quantum-mechanical (!)\ evolution equation for operators $A\in \text{Mat}(r^N, \C)$,
\begin{equation}
\frac{\partial \mspace{-1mu} A}{\partial \mspace{1mu} t_n} = 
\frac{\I}{\hbar} \, \pigl[\hbar\, H_{n,\elt} ,A\pigr] \, , \qquad H_{n,\elt} = \mathrm{ev}_{\mspace{-2mu}\elt}\pigl(\tilde{c}_0\bigl(\widetilde{D}_n^{(1)}\bigr)\pigr) \, .
\end{equation}
This is the Heisenberg picture for operators acting on a (spin-chain) Hilbert space $(\C^r)^{\otimes N}$, cf.\ the last step in \eqref{eq:2-step-process}. Comparing with \eqref{eq:hybrid_evolution} we see that the evaluation at the classical equilibrium generated by $\elt \in \text{PSL}(2,\Z)$ has effectively removed the classical contribution in the hybrid evolution. Interestingly, the above suggests that the converse is also true: to obtain a quantum-mechanical system from a hybrid system we need to evaluate at a simultaneous equilibrium of all hamiltonians $\tilde{c}_0\big(\widetilde{D}_n^{(0)}\big)$, $1\leqslant n \leqslant N$.

By construction, this action is compatible with the projections to $\text{Mat}(r^N, \C)$, in the same way as the action \eqref{eq:hybrid_action} is compatible with the projections to the hybrid setup. More precisely, we have
\begin{thm} \label{thm:comm diagram}
The two-step freezing process in \eqref{eq:2-step-process} can be summarised by the following commutative diagram:
\begin{equation} \label{eq:commutative diagram}
	\tikz[baseline={([yshift=-.5*11pt*0.3]current bounding box.center)},xscale=.5,yscale=0.25]{
		\def\r{15};
		\def\s{10};
		\node at (0,5) {\textup{Poisson algebra}};
		\node at (\r,5) {\textup{Poisson module}};
		\node (A1) at (0,1) {$\bigl(\mathcal{H}_\hbar,[\cdot,\cdot]_\hbar\bigr)$};
		\node (A2) at (\r,1) {$ \displaystyle \mathrm{End}\bigl(\widetilde{\alg}_0[\mspace{-2mu}[\hbar]\mspace{-2mu}]\bigr)$};
		\draw[->] (A1) -- (A2) node[midway,above] {\small $z+\hbar \, H \longmapsto [z+\hbar \, H, \,\cdot\, ]_\hbar$};
		\node (B1) at (0,1-\s) {
			\begin{tikzpicture}
				\node at (1,0) {$\bigl(\mathcal{H}_\hbar/\hbar \, \mathcal{H}_\hbar,[\mspace{-7.5mu}\{\cdot,\cdot\}\mspace{-7.5mu}]\bigr)$};
				\node at (1,-1) {$\bigl(\mathcal{B}_0,[\mspace{-7.5mu}\{\cdot,\cdot\}\mspace{-7.5mu}]\bigr)$};
				\node at (1,-0.5) {$\bigcup$};
			\end{tikzpicture}
		};
		\node (B2) at (\r,1-\s) {$ \displaystyle \mathrm{End}\bigl(\widetilde{\alg}_0\bigr)$};
		\draw[shorten <= -.2cm,->] (B1) -- (B2) node[midway,above] {\small\!\!\!$z+\hbar \, \,\overline{\!H} \longmapsto \I \, \{z,\,\cdot\,\} + \bigl[\hbar\, \,\overline{\!H}, \,\cdot\, \bigr]_\hbar$};
		\node (C1) at (0,1-2*\s) {$\bigl(\evq(\mathcal{B}_0) ,[\cdot,\cdot]\bigr)$};
		\node (C2) at (\r,1-2*\s) {$\displaystyle \mathrm{End}\bigl(\mathrm{Mat}(r^N, \C)\bigr)$};
		\draw[->] (C1) -- (C2) node[midway,above] {\, \small $\hbar\,\evq \,\overline{\!H} \longmapsto \bigl[\hbar\,\mathrm{ev}_{\mspace{-2mu}\elt} \,\overline{\!H},\,\cdot\,\bigr]_\hbar$};
		\draw[->>] (A1) -- (B1);
		\draw[->>] (A2) -- (B2); 
		\draw[->>] (B1) -- (C1) node[midway,left] {\small$\evq$};
		\draw[->>] (B2) -- (C2) node[midway,right] {\small $\mathrm{ev}_{\mspace{-2mu}\elt}$};
		
		\def\u{6}
		\node[align=center] (A3) at (\r+\u,1) {\textup{\textsc{qmbs}} \\
			\textup{with spins}};
		\node[align=center] (B3) at (\r+\u,1-\s) {\textup{hybrid} \\ \textup{system}};
		\node[align=center] (C3) at (\r+\u,1-2*\s) {\textup{quantum} \\ \textup{spin chain}};
		\draw[->] (A3) -- (B3);
		\draw[->] (B3) -- (C3);
	}
\end{equation}
The horizontal arrows are the (Lie algebra) actions in \eqref{eq:action_on_Ahbar}, \eqref{eq:hybrid_action}, and \eqref{eq:action_quantum}, which together with the (left- and right-) product actions (in \eqref{eq:quantum_product_action}, \eqref{eq:hybrid_product_action} and by matrix multiplication) give rise to Poisson modules. The left vertical arrows are morphisms of Poisson algebras, the right vertical arrows are algebra homomorphisms.
\end{thm}
\begin{proof}[Proof.]
	It only remains to check that
	\begin{equation}
		\begin{aligned}
			\mathrm{ev}_{\mspace{-2mu}\elt}\bigl( (z+\hbar \, \,\overline{\!H} \mspace{2mu}) \cdot A \bigr) & = \mathrm{ev}_{\mspace{-2mu}\elt} \pigl(\I\,\{z,A\}+ [\hbar \, \,\overline{\!H},A]_\hbar \pigr) \\
			& = \mathrm{ev}_{\mspace{-2mu}\elt} \pigl( [\hbar \, \,\overline{\!H},A]_\hbar\pigr) \\
			& = \pigl[\mathrm{ev}_{\mspace{-2mu}\elt}( \hbar \, \,\overline{\!H} \mspace{1mu}),\mathrm{ev}_{\mspace{-2mu}\elt}(A) \pigr]_\hbar \\
			&=  \evq(z+\hbar \, \,\overline{\!H} \mspace{1mu}) \cdot \mathrm{ev}_{\mspace{-2mu}\elt}(A)\, . 
		\end{aligned}
	\end{equation}
	Here we once more used that Poisson brackets involving $z+0\,\hbar\in \mathcal{B}_0$ vanish upon evaluation.
\end{proof}
The top square of \eqref{eq:commutative diagram}, discussed in \textsection\ref{sec:hybrid_dynamics} and following \cite{mikhailovCommutativePoissonAlgebras2024,liashyk2024classical}, shows how a hybrid system can be obtained as a partial classical limit of a quantum mechanical system. This requires the hamiltonians of the latter to 
semiclassically spin separated as in \eqref{eq:H_hbar}, cf.~\eqref{eq:Dtilde=Dcl id+Ohbar}, and utilises a pair of projections compatible with the hamiltonian actions driving the time evolution at each level. 
The bottom square in \eqref{eq:commutative diagram}, discussed in the present subsection, shows how the same approach can be used to further understand the role of the evaluation map. It essentially projects both the (commuting) hamiltonians and the operators down to the setting of a quantum spin chain, in a way compatible with the actions at each level.

\section{Discussion}
\label{sec:discussion}

\noindent
In this paper we demonstrated how 
matrix-valued (quantum) elliptic Ruijsenaars operators give rise to an elliptic long-range spin chain for each equilibrium configuration of the (classical) elliptic Ruijsenaars--Schneider system by freezing, and showed that this procedure preserves quantum integrability. 

In more detail, in 
\textsection\ref{sec:scalar} we first reviewed the scalar (spinless) quantum elliptic Ruijsenaars system and its classical limit in the formalism of deformation quantisation.
In \textsection\ref{sec:modularity}--\ref{sec:equilibria_modular} we 
defined an action of the modular group $\mathrm{PSL}(2,\mathbb{Z})$ on the (spinless) classical elliptic Ruijsenaars--Schneider system (Theorem \ref{thm:RS_modular}), and used it (via Theorem~\ref{thm:equilibria_modular}) to exhibit a modular family of (discrete) equilibrium configurations \eqref{eq:modular_family_equilib}, cf.~Fig.~\ref{fg:modular_action}. 
This generalises the known equilibria of the (spinless) classical elliptic Calogero--Sutherland--Moser system, which also appears in context of supersymmetric gauge theory \cite{doreyEllipticSuperpotentialSoftly1999}. The construction requires a shift of the momenta, see \eqref{eq:S_T_extended_with_shift}, which survives in the Calogero--Sutherland--Moser limit. 
In \textsection\ref{sec:ell_sRuij} we recalled the structure of the matrix-valued difference operators $\widetilde{D}_n$ defining elliptic spin-Ruijsenaars systems, using a formulation that covers both the (vertex-type) systems of \cite{MZ_23a} and the (face-type) systems of \cite{klabbers2024deformed}.
For any classical equilibrium, in \textsection\ref{sec:freezing} we used the framework of deformation quantisation, building on \cite{mikhailovCommutativePoissonAlgebras2024,liashyk2024classical,chalykh2024integrability}, to derive a quantum spin chain with long-range interactions by freezing. We proved that quantum integrability is preserved (Theorem~\ref{thm:HnB commute}) in the sense that if the $\widetilde{D}_n$ commute among each other, then so do the spin-chain hamiltonians $H_{n,\elt}$ defined in \eqref{eq:spin chain hamiltonians}, frozen at the equilibrium labelled by $\elt \in \mathrm{PSL}(2,\mathbb{Z})$. We obtained the explicit expression 
from Propositions~\ref{prp:H_nB}--\ref{prp:H_-nB} for these hamiltonians for arbitrary $n$. 
In \textsection\ref{sec:hybrid_systems} we connected freezing to the setting of \cite{mikhailovCommutativePoissonAlgebras2024}, generalising \textsection9 of \cite{liashyk2024classical}  to the difference case, and adding a Poisson-algebraic interpretation of the `evaluation' at the classical equilibrium configuration (Theorem~\ref{thm:comm diagram}).

As described in \textsection\ref{sec:examples}, the vertex- and face-type cases from \cite{MZ_23b} and \cite{klabbers2024deformed} give rise to two separate landscapes of long-range spin chains, see Figures 2--3 in \cite{klabbers2025landscapes}. 
In particular, our results allow one to freeze at $\elt = S$, which (unlike $\elt = \SLid$) provides spin chains admitting a short-range limit. As outlined in \textsection\ref{sec:lim_trig}--\ref{sec:lim_undeformed}, our results also carry over to the (trigonometric) long-range limit, and, in the face case, for $\eta\to0$ agree with \cite{chalykh2024integrability}, and \textsection9 of \cite{liashyk2024classical}. Besides proving integrability, we believe that freezing holds the key to a complete understanding of these $q$-deformed long-range spin chains\,---\,including exact descriptions of their spectra, like for the (face-type) Haldane--Shastry chain \cite{bernard1993yang,lamers2024fermionic} and its $q$-deformation \cite{Ugl_95u,lamers2022spin}. It would be interesting to see whether any of the quantum many-body systems with spins discussed in this paper, or any of their limits, can be obtained from the double elliptic (`\textsc{dell}') system with spins that was proposed in \cite{koroteev2020quantum}, or if they are related to the new spin-Calogero--Sutherland models recently introduced in \cite{bourgine2024calogero,hu2024hilbert}.

Another intriguing question is whether $\mathcal{B}_\hbar \subset \mathcal{H}_\hbar$ is maximal commutative, or whether it is possible to construct additional hamiltonians akin to \cite{chalykh2024integrability} in at least the face-type example.

A final interesting avenue that we mention is  computation of various different classical limits of the quantum elliptic Ruijsenaars operators. In the vertex and face examples, in addition to $\hbar = \hbar_1$ one can introduce another deformation parameter $\hbar_2$ in the deformed spin permutations such that $P(x) = P \, \bigl(1 + \hbar_2 \, r(x) + O(\hbar_2)\bigr)$ where $r(x)$ obeys the (in the face case: modified \cite{felder1995conformal}) classical Yang--Baxter equation. One can then consider the opposite partial classical limit in which only $\hbar_2$ vanishes, or the fully classical limit in which $\hbar_1 \propto \hbar_2$ both vanish. In this setting, the freezing limit $\hbar_1\to 0$ is a Nekrasov--Shatashvili-type limit, cf.~\cite{nekrasovQuantizationIntegrableSystems2010}. We expect the fully classical limit and its degenerations to be closely related to \cite{kricheverSpinGeneralizationRuijsenaarsSchneider1995,gibbonsGeneralisationCalogeroMoserSystem1984,billey1994exact,arutyunovHamiltonianStructureSpin1998,chalykh2020hamiltonian,feherReductionBiHamiltonianHierarchy2020}. We plan to return to this question in a future publication.

\subsection*{Acknowledgements}
\noindent We thank M.~Mikhailov, J.~Pulmann and M.~Volk for discussions, and A.~Bourget for correspondence. We are particularly grateful to O.~Chalykh for numerous discussions and feedback that led to significant improvements. We further thank the referees for constructive feedback.

Rob Klabbers's research is partially funded by the Deutsche Forschungsgemeinschaft (DFG, German Research Foundation)\,--\,Projektnummer 417533893/GRK2575 ``Rethinking Quantum Field Theory''.

\appendix 

\section{Elliptic preliminaries}
\label{app:elliptics}

\subsection{Elliptic functions}

\noindent We summarise our conventions for the elliptic functions we use and list their most important properties. Standard references are \cite{DLMF,abramowitz1948handbook,whittaker1904course}.

The odd Jacobi theta function with lattice parameter $\tau \in \C$ ($\mathrm{Im}\,\tau>0$) is defined as
\begin{equation}\label{eq:vartheta}
\begin{aligned}
	\vartheta(x | \tau ) & = 2 \sum_{n=0}^{\infty} (-1)^n \, p^{(n+1/2)^2} \sin[(2n+1)x] \\
	& =  2 \, p^{1/4} \sin(x) \prod_{n=1}^\infty (1-p^{2n}) (1 - \mathrm{e}^{2\I x}  p^{2n})(1 - \mathrm{e}^{-2\I x} \, p^{2n})\, , 
\end{aligned}
\qquad p = \E^{\I \pi \tau} \, , 
\end{equation}
where $p$ is called the `nome'. The normalised theta function $\theta(x) = \vartheta(\pi x)/(\pi \vartheta'(x))$  is odd and entire as well as is doubly quasiperiodic, $\theta(x + 1)  = -\theta(x)$ and $\theta(x + \,\tau) = - p^{-1} \, \E^{-2\pi\I x} \, \theta(x)$, with a simple zero at the origin, and satisfies $\theta'(0)=1$. Its trigonometric and rational degenerations are $\theta(x) = \sin( \pi x)/\pi+ O(p)$ and $\sin(\pi x)/\pi = x + O(N^{-2})$. It obeys the addition formula 
\begin{equation} \label{eq:theta_addition}
\begin{aligned}
	\theta(x+y) \, \theta(x-y)\,\theta(z+w) \, \theta(z-w) =
	{} & \theta(x+z) \, \theta(x-z) \, \theta(y+w) \, \theta(y-w) \\
	& \!+\, \theta(x+w) \, \theta(x-w) \, \theta(z+y) \, \theta(z-y) \, .
\end{aligned}
\end{equation}

Other standard Jacobi theta functions are defined in terms of \eqref{eq:vartheta} by
\begin{equation}
\begin{aligned}
	& \vartheta_1(z \,|\, \tau) = \vartheta(z) \, ,  
	&& \vartheta_2(z \,|\, \tau) = \vartheta(z+1/2 \,|\, \tau)\, , \\
	& \vartheta_3(z \,|\, \tau) = \E^{\I \pi \tau/4} \, \E^{\I \pi x} \, \vartheta(z+(1+\tau)/2 \,|\, \tau) \, , \quad
	&& \vartheta_4(z \,|\,\tau) = -\I \, \E^{\I \pi \tau/4} \, \E^{\I \pi x} \, \vartheta(z+\tau/2 \,|\, \tau)\, , 
\end{aligned}
\end{equation}
hence one can think of these as (rescaled) versions of the odd Jacobi theta function \eqref{eq:vartheta} shifted by half-periods of the lattice, generalising the well-known identity $\cos(x) = \sin(x+\pi/2)$ between the elementary trigonometric functions. Expanding in the nome $p$ we find 
\begin{equation}
\begin{aligned}
	& p^{-1/4} \, \vartheta_1(z \,|\, \tau) = 2 \sin \pi x + O(p^2) \, , 
	&& p^{-1/4} \, \vartheta_2(z \,|\, \tau) = 2 \cos \pi x+ O(p^2) \, , \\
	& \hphantom{p^{-1/4} \, } \vartheta_3(z \,|\, \tau) = 1 + 2\,p \, \cos \pi x + O(p^2) \, , \quad
	&& \hphantom{p^{-1/4} \, } \vartheta_4(z \,|\, \tau) = 1 - 2\, p \cos \pi x+ O(p^2) \, . 
\end{aligned} 
\end{equation}

There are many more relations between the $\vartheta_a$, for our purposes the Jacobi imaginary transformation will be particularly important. For a given $\tau$ let $\tau' = -1/\tau$, then the following relations hold 
\begin{equation} \label{eq:jac_imag_arb_tau}
\begin{aligned}
	\I \, (-\I \tau)^{1/2} \, \vartheta_1(z \,|\, \tau) & =  \E^{\I \tau' z^2/\pi} \, \vartheta_1(\tau' z \,|\, \tau')\, , \\
	(-\I \tau)^{1/2} \, \vartheta_2(z \,|\, \tau) &=  \E^{\I \tau' z^2/\pi} \, \vartheta_4(\tau' z \,|\, \tau') \, , \\
	(-\I \tau)^{1/2} \, \vartheta_3(z \,|\, \tau) &=  \E^{\I \tau' z^2/\pi} \, \vartheta_3(\tau' z \,|\, \tau')\, , \\
	(-\I \tau)^{1/2} \, \vartheta_4(z \,|\, \tau) &= \E^{\I \tau' z^2/\pi} \, \vartheta_2(\tau' z \,|\, \tau')\, , 
\end{aligned}
\end{equation}
effectively relating the two regimes in which Im$(\tau)$ is either large or small. Note that $\vartheta_2$ and $\vartheta_4$ switch places under this transformation, whereas $\vartheta_1$ and $\vartheta_3$ transform into themselves. 

The (normalised) Kronecker elliptic function is 
\begin{equation}
\begin{aligned}
	\label{eq:Kronecker_def}
	\phi(u,v \,|\, \tau) \coloneqq \frac{\theta(u+v \,|\, \tau)}{\theta(u \,|\, \tau) \, \theta(v \,|\, \tau)} & = \frac{\pi \sin \bigl(\pi (u+v)\bigr)}{\sin(\pi u) \, \sin(\pi v)} + O\bigl(p\bigr) \\
	& = \pi \bigl(\cot (\pi u) + \cot(\pi v)\bigr) + O\big(p\bigr) \, .
\end{aligned}
\end{equation}
If no confusion can arise we will suppress its dependence on $\tau$, simply writing $\phi(u,v)$. This function is symmetric and doubly (quasi)periodic, $\phi(u+1,v) = \phi(u,v)$ and $\phi(u+\tau,v ) = \E^{-2\pi \I v} \, \phi(u,v)$. 

\subsection{Elliptic \textit{R}-matrices}
In the following, for $1\leqslant \alpha,\beta \leqslant r$ let $E_{\alpha \beta} \in \text{Mat}(r\times r,\C)$ denote the $r\times r$ matrix units, with entries $(E_{\alpha \beta})_{\gamma \delta} = \delta_{\alpha \gamma}\, \delta_{\beta \delta}$. 

\subsubsection{Vertex-type} 
\label{app:R-mat_vx}
For $r\geqslant 2$ the Baxter--Belavin \textit{R}-matrix for $\mathfrak{gl}_r$ can be given as 
\begin{subequations} \label{eq:Vertex-type-R-matrices}
\begin{gather}
	R(x ;\eta\,|\, \tau) \coloneqq  \sideset{}{'}{\sum}_{\alpha,\beta,\gamma,\delta=1}^{r} \!\!\!\! R_{\alpha\gamma, \beta\delta}(x ; \eta \,|\, \tau) \, E_{\alpha \beta} \otimes E_{\gamma \delta}\, ,
	\intertext{where the prime indicates that sum is restricted to the `weakened ice rule' $\alpha + \gamma \equiv \beta+\delta \text{ mod } r$, and the nonzero entries read}
	\begin{aligned}
		R_{\alpha\gamma, \beta\delta}(x ; \eta \,|\, \tau) \coloneqq \frac{1}{\phi(x,\eta \, | \,\tau)} & \exp\biggl(\frac{2\pi \I}{r} \pigl( (\beta-\alpha)\,x + (\gamma-\beta) \, \eta +   (\gamma-\beta)(\gamma-\alpha)\,\tau \pigr)\biggr) \\
		& \times \phi(x+\gamma-\beta \, \tau,  \eta + (\beta-\alpha) \,\tau \, | \, r \, \tau )
	\end{aligned}
\end{gather}
\end{subequations}
Then $\check{R}(x) = P \, R(x)$ again satisfies the relations \eqref{eq:unitarity}--\eqref{eq:commutativity} as well as \eqref{eq:initial}. 
The Baxter--Belavin \textit{R}-matrix is also commonly written in terms of a representation of the Heisenberg group \cite{belavinDynamicalSymmetryIntegrable1981}; for this and many further relations see e.g.\ the appendix of \cite{zabrodinFieldAnalogueRuijsenaarsSchneider2022}.\,%
\footnote{\label{fn:eta} Our $\eta$ is related to that of (B.14) in \cite{zabrodinFieldAnalogueRuijsenaarsSchneider2022} via $\eta_{\text{ZZ}} \coloneqq \eta/r$.}

For $r=2$ the weakened ice rule allows for $R_{11,22} ,  R_{22,11}\neq 0$ in addition to nonzero entries in the same six positions as for the dynamical $R$-matrix and this definition yields the eight-vertex \textit{R}-matrix in the conventions of \cite{klabbers2025landscapes}, see Section 2.1 therein. The trigonometric limit gives the usual (symmetric) six-vertex $R$-matrix in the `principal grading' (possibly up to a global spin rotation), see \textsection{}B.2 in \cite{klabbers2025landscapes}.

\subsubsection{Face-type} \label{app:R-mat_face}
The elliptic dynamical \textit{R}-matrix of type $\mathfrak{gl}_r$ with $r \geqslant 2 $ reads \cite{felder1994elliptic,felder1997elliptic}
\begin{equation} \label{eq:dynamical_R_matrices}
\begin{aligned}
	R(x,a;\eta \, | \, \tau ) \coloneqq {}& \sum_{\alpha=1}^r  E_{\alpha\alpha} \otimes E_{\alpha\alpha}+ \frac{1}{\phi(x,\eta\,|\, \tau)} \sum_{\alpha\neq \beta}^r \phi(a_\beta-a_\alpha,\eta\,|\, \tau) \, E_{\alpha\alpha} \otimes E_{\beta \beta} \\
	& + \frac{1}{\phi(x,\eta\,|\, \tau)}\sum_{\alpha \neq \beta}^r \phi(x,a_\beta- a_\alpha \,|\, \tau ) \, E_{\alpha \beta} \otimes E_{\beta \alpha}\, .
\end{aligned}
\end{equation}
Then $\check{R}(x,a) \coloneqq P \, R(x,a)$ satisfies the unitarity relation~\eqref{eq:unitarity}, the (dynamical) Yang--Baxter equation~\eqref{eq:braided_YBe}, `commutativity at a distance' \eqref{eq:commutativity}, as well as the initial condition~\eqref{eq:initial}. For a graphical interpretation, including the connection to the `interaction (a)round the face' (\textsc{irf}) picture, see \textsection{}B in \cite{klabbers2024deformed}.
This \textit{R}-matrix is associated to the elliptic quantum group $E_{\tau, \eta}(\mathfrak{gl}_r)$ \cite{felder1997elliptic}. 

For $r=2$, \eqref{eq:dynamical_R_matrices} coincides with the dynamical \textit{R}-matrix in \cite{klabbers2025landscapes} and \cite{klabbers2024deformed} after identifying $a = a_1 - a_2$.%
\footnote{\ One obtains the dynamical \textit{R}-matrix of  \cite{zabrodinFieldAnalogueRuijsenaarsSchneider2022} after a transposition and passing to $\eta_{\text{ZZ}} \coloneqq \eta/r$.} 
Taking the trigonometric, and then non-dynamical limit, on obtains the trigonometric $R$-matrix in the `homogeneous grading', related to the Hecke algebra, see e.g.\ \textsection5.1 of \cite{klabbers2024deformed}.

\subsubsection{Face-vertex transformation} 
The face- and vertex-type elliptic \textit{R}-matrices are related by a `face-vertex transformation' \cite{baxter1973eight,felderAlgebraicBetheAnsatz1996}, which can be interpreted as a Drinfeld twist \cite{jimbo1999quasi}. As we emphasised in \textsection5.2 of \cite{klabbers2025landscapes}, this transformation does \emph{not} extend from the level of the $R$-matrix to a simple conjugation of the corresponding spin-Ruijsenaars models, yielding very different spectral and physical properties.

\section{Deformed spin permutations}
\label{app:deformed_spin_perm}

\noindent
In this appendix we provide more details for the deformed spin permutations defined in \textsection\ref{sec:P_w(x)}.

\subsection{Generalities}
\label{app:deformed_spin_perm gen}

\noindent
Let $w\longmapsto s_w$ denote the natural action of $S_N$ on $\mathrm{Fun}(\vect{x})$ by permuting variables. According to \eqref{eq:unitarity}--\eqref{eq:commutativity}, $\tilde{s}_i \coloneqq s_{i,i+1} \, P_{i,i+1}(x_i - x_{i+1})$ gives an $S_N$-action on $\mathrm{Fun}(\vect{x}) \otimes V^{\otimes N}$ \cite{felder1995conformal}, see also \cite{finch2016theta}. By \eqref{eq:initial} it deforms the diagonal action. Write $w \longmapsto \widetilde{w}$ for this representation. Define $P_w(\vect{x})$ by $\widetilde{w} = s_w \, P_w(\vect{x})$. Then considering $\widetilde{w\,w\smash{'}} = \widetilde{w} \, \widetilde{w'}$ yields the general cocycle condition\,%
\footnote{\ Alternatively, one can set $\widetilde{w} = P'_w(\vect{x}) \, s_w$ and work with $P'_w(\vect{x}) = s_w \, P_w(\vect{x}) \, s_w^{-1} = P_w(s_w \cdot \vect{x})$; in particular, $P'_{(i~i+1)}(\vect{x}) = P_{i,i+1}(x_{i+1} - x_i)$. }
\begin{equation}
P_{w\,w'}(\vect{x}) = s_{w'}^{-1} \, P_{w}(\vect{x}) \, s_{w'} \; P_{w'}(\vect{x}) = P_{w}\bigl(s_{w'}^{-1} \cdot \vect{x}\bigr) \; P_{w'}(\vect{x}) \, . 
\end{equation}
This includes the cocycle condition~\eqref{eq:cocycle} as the special case $w' = (i~i+1)$, which allows one to recursively construct any $P_w(\vect{x})$ starting from the identity operator~\eqref{eq:id}. 

In more detail, given $w\in S_N$, pick a(ny) reduced decomposition $w = s_{j_1} \mspace{-2mu}\cdots s_{j_\ell}$. 
Put $w_k \coloneqq s_{j_{k-1}} \mspace{-2mu}\cdots s_{j_\ell}$ to get a set of permutations $w_0 = s_{j_1} \mspace{-2mu}\cdots s_{j_\ell} = w$, $w_1 = s_{j_2} \mspace{-2mu}\cdots s_{j_\ell}$ down to $w_{\ell-1} = s_{j_\ell}$, $w_\ell^{-1} = e$. Then the recursive description \eqref{eq:id}--\eqref{eq:cocycle} implies that
\begin{equation} \label{eq:P_w decomp}
\begin{aligned}
	P_w(\vect{x}) & = \ordprodopp_{1\leqslant k\leqslant \ell} \!\! P_{j_k,j_k+1}\pigl(x_{w_{I\mspace{-2mu},\mspace{2mu}k}^{-1}(j_k)}  - x_{w_{I\mspace{-2mu},\mspace{2mu}k}^{-1}(j_k +1)} \pigr) \\
	& = P_{j_1,j_1+1}\pigl(x_{w_{I\mspace{-2mu},\mspace{1mu}1}^{-1}(j_1)} - x_{w_{I\mspace{-2mu},\mspace{1mu}1}^{-1}(j_1 +1)} \pigr) \cdots P_{j_\ell,j_\ell+1}\bigl(x_{j_\ell} - x_{j_{\ell}+1}\bigr) \, ,
\end{aligned}
\end{equation}
where the arrow indicates the direction of increasing subscripts in the product. Each factor depends on the difference of only two coordinates, whose subscripts are permuted by the $s_j$ to its left due to the cocycle condition~\eqref{eq:cocycle}, accounting for how the coordinates follow lines in diagrams. Let us give some explicit examples for the particular permutations that appear in the spin-Ruijsenaars operators.

\subsection{Cycles} \label{app:cycles}
The spin-Ruijsenaars operators $\widetilde{D}_n$ from 
\eqref{eq:Ruij_spin_pm n} are all built from deformed cycles. Here are some examples.
Clearly,
\begin{equation}
P_{(i,i+1)}(\vect{x}) = P_{i,i+1}(x_i - x_{i+1}) = \tikz[baseline={([yshift=-.5*11pt*0.3]current bounding box.center)},xscale=.5,yscale=0.25,font=\footnotesize]{
	\draw[->] (0,0) node[below]{$x_1$} -- (0,3) node[above]{$\vphantom{x}\smash{x_1}$};
	\foreach \x in {-1,...,1} \draw (.75+.2*\x,1.5) node{$\cdot\mathstrut$};	
	\draw[->] (1.5,0) -- (1.5,3);
	\draw[rounded corners=2pt,->] (3.5,0) node[below, yshift=.05cm]{$\,\vphantom{t}\smash{x_{i+1}}$} -- (3.5,1) -- (2.5,2) -- (2.5,3) node[above]{$\vphantom{x}\smash{x_{i+1}}\,$};
	\draw[rounded corners=2pt,->] (2.5,0) node[below, yshift=.05cm]{$\vphantom{t}x_{i}$} -- (2.5,1) -- (3.5,2) -- (3.5,3) node[above]{$\vphantom{x}\smash{x_{i}}$};
	\draw[->] (4.5,0) -- (4.5,3);
	\foreach \x in {-1,...,1} \draw (5.25+.2*\x,1.5) node{$\cdot\mathstrut$};	
	\draw[->] (6,0) node[below]{$x_N$} -- (6,3) node[above]{$\vphantom{x}\smash{x_N}$};
	\node at (-.5,1.5) {\smash{\textcolor{lightgray}{$\vec{a}$}}\vphantom{$a$}};
} .
\end{equation}
Next, for $(i~i+1~i+2) = (i~i+1)\,(i+1~i+2)$ we get
\begin{gather} 
\\[.75ex]
P_{(i~i+1~i+2)}(\vect{x}) = P_{i,i+1}(x_i - x_{i+2}) \, P_{i+1,i+2}(x_{i+1} - x_{i+2}) = 
\smash{ 
	\tikz[baseline={([yshift=-.5*11pt*0.3]current bounding box.center)},xscale=-.5,yscale=0.25,font=\footnotesize]{
		\draw[->] (0,0) node[below]{$x_N$} -- (0,4) node[above]{$x_N$};
		\foreach \x in {-1,...,1} \draw (.75+.2*\x,2) node{$\cdot\mathstrut$};	
		\draw[->] (1.5,0) -- (1.5,4);
		\draw[rounded corners=2pt,->] (2.5,0) node[below]{$\ \, x_{i+2}$} -- (2.5,1) -- (4.5,3) -- (4.5,4) node[above]{$x_{i+2}\;$};
		\draw[rounded corners=2pt,->] (3.5,0) node[below]{$x_{i+1}\;$} -- (3.5,1) -- (2.5,2) -- (2.5,4) node[above]{$\,x_{i+1}$};
		\draw[rounded corners=2pt,->] (4.5,0) node[below]{$x_{i}\ $} -- (4.5,2) -- (3.5,3) -- (3.5,4) node[above]{$x_{i}$};
		\draw[->] (5.5,0) -- (5.5,4);
		\foreach \x in {-1,...,1} \draw (6.25+.2*\x,2) node{$\cdot\mathstrut$};	
		\draw[->] (7,0) node[below]{$x_1$} -- (7,4) node[above]{$x_1$};
		\node at (7.5,2) {\smash{\textcolor{lightgray}{$\vec{a}$}}\vphantom{$a$}};
} } .
\nonumber \\[1.5ex] \nonumber 
\end{gather}
More generally, for $k<l$ the (uninterrupted) cycle $(k~k+1~\dots~l) = (k~k+1) \cdots (l-1~l)$ yields
\begin{gather} \label{eq:P_i...j}
\\[1.5ex] 
P_{(k~k+1~\dots~l)}(\vect{x}) = P_{k,k+1}(x_k - x_{l}) \cdots P_{l-1,l}(x_{l-1} - x_l) =
\smash{ 
	\tikz[baseline={([yshift=-.5*11pt*0.3]current bounding box.center)},xscale=-.5,yscale=0.25,font=\footnotesize]{
		\draw[->] (-1,0) node[below]{$x_N$} -- (-1,5) node[above]{$x_N$};
		\foreach \x in {-1,...,1} \draw (-.25+.2*\x,2.5) node{$\cdot\mathstrut$};	
		\draw[->] (.5,0) -- (.5,5);
		\draw[rounded corners=2pt,->] (1.5,0) node[below]{$x_l$} -- (1.5,1) -- (4.5,4) -- (4.5,5) node[above]{$x_l$};
		\draw[rounded corners=2pt,->] (2.5,0) node[below]{$x_{l-1}\;$} -- (2.5,1) -- (1.5,2) -- (1.5,5) node[above]{$\,x_{l-1}$};
		\draw[rounded corners=2pt,->] (3.5,0) node[below]{$\!\!\!\!\vphantom{x_i}\dots$} -- (3.5,2) -- (2.5,3) -- (2.5,5) node[above]{$\!\dots$};
		\draw[rounded corners=2pt,->] (4.5,0) node[below]{$x_k$} -- (4.5,3) -- (3.5,4) -- (3.5,5) node[above]{$x_k$};
		\draw[->] (5.5,0) -- (5.5,5);
		\foreach \x in {-1,...,1} \draw (6.25+.2*\x,2.5) node{$\cdot\mathstrut$};	
		\draw[->] (7,0) node[below]{$x_1$} -- (7,5) node[above]{$x_1$};
		\node at (7.5,2.5) {\smash{\textcolor{lightgray}{$\vec{a}$}}\vphantom{$a$}};
} } .
\nonumber \\[2ex] \nonumber 
\end{gather}
The case $k=1$, $l=i$ gives \eqref{eq:P_1...i}.

Similarly,
\begin{gather} \label{eq:P_j...i}
\\[1.5ex] 
P_{(l~l-1~\dots~k)}(\vect{x}) = P_{l-1,l}(x_k - x_{l}) \cdots P_{k,k+1}(x_k - x_{k+1}) =
\smash{ 
	\tikz[baseline={([yshift=-.5*11pt*0.3]current bounding box.center)},xscale=.5,yscale=0.25,font=\footnotesize]{
		\draw[->] (-1,0) node[below]{$x_1$} -- (-1,5) node[above]{$x_1$};
		\foreach \x in {-1,...,1} \draw (-.25+.2*\x,2.5) node{$\cdot\mathstrut$};	
		\draw[->] (.5,0) -- (.5,5);
		\draw[rounded corners=2pt,->] (1.5,0) node[below]{$x_k$} -- (1.5,1) -- (4.5,4) -- (4.5,5) node[above]{$x_k$};
		\draw[rounded corners=2pt,->] (2.5,0) node[below]{$\;x_{k+1}$} -- (2.5,1) -- (1.5,2) -- (1.5,5) node[above]{$x_{k+1}\,$};
		\draw[rounded corners=2pt,->] (3.5,0) node[below]{$\vphantom{x_i}\dots\!\!\!$} -- (3.5,2) -- (2.5,3) -- (2.5,5) node[above]{$\dots\!$};
		\draw[rounded corners=2pt,->] (4.5,0) node[below]{$x_l$} -- (4.5,3) -- (3.5,4) -- (3.5,5) node[above]{$x_l$};
		\draw[->] (5.5,0) -- (5.5,5);
		\foreach \x in {-1,...,1} \draw (6.25+.2*\x,2.5) node{$\cdot\mathstrut$};	
		\draw[->] (7,0) node[below]{$x_N$} -- (7,5) node[above]{$x_N$};
		\node at (-1.5,2.5) {\smash{\textcolor{lightgray}{$\vec{a}$}}\vphantom{$a$}};
} } .
\nonumber \\[2ex] \nonumber 
\end{gather}
The case $k=i$, $l=N$ yields \eqref{eq:P_N...i}.\,%
\footnote{\ Observe that \eqref{eq:P_j...i} is not quite the inverse of \eqref{eq:P_i...j} since the mismatch in inhomogeneities does not allow for composition. Rather, the inverse $P_w(\vect{x})^{-1} = P_{w^{-1}}(x_{w(1)},\dots,x_{w(N)})$ is graphically obtained by flipping the diagram of $P_w(\vect{x})$ upside down and reversing the orientations of all lines back from bottom to top.}

\subsection{Grassmannian permutations} \label{app:grassmann}
We are particularly interested in \emph{Grassmannian permutations}, which are minimal-length representatives of the coset $S_N/(S_n \times S_{N-n})$ for some $1\leqslant n\leqslant N$, i.e.\ permutations with (at most) one descent $w(n)>w(n+1)$.
More explicitly, given an $n$-element subset $I=\{i_1 < \dots < i_n\} \subseteq \{1,\dots,N\}$, the corresponding Grassmannian permutation, which we denote by $w_I \in S_N$, sends $k \longmapsto i_k$ for all $1\leqslant k \leqslant n$ without permuting either $\{1,\dots,n\}$ or $\{n+1,\dots,N\}$ amongst each other. It can be defined recursively, starting from $w_\varnothing = e$ the identity, through the recursion relation 
\begin{equation}
w_{\mspace{-1mu}J \cup \{ j\}} = w_J \; (j ~ j-1 \dots |J|+1) \quad \text{if} \quad j > \max(J) \, . 
\end{equation}
For instance, $n=1$ gives the cycle $w_{\{i\}} = (i ~ i-1 \dots 1)$, at $n=2$ we get the product of cycles $w_{\{i,i'\}} = (i ~ i-1 \dots 1) \, (i' ~ i'-1 \dots 2)$, and so on. In general, $w_I = (i_1 ~ i_1-1 \dots 1) \cdots (i_n ~ i_n -1 \dots n)$. Note that at $n=N$ we simply retrieve $w_{\{1,\dots,N\}} =e$ the identity. This motivates further defining $w_{-I} \coloneqq w_{\{1,\dots,N\} \setminus I}$. For example, $w_{-\{i\}} = w_{\{1,\dots,i-1,i+1,\dots,N\}} = (i~i+1 \dots N)$ is again a cycle, $w_{-\{i,i'\}} = (i~i + 1 \dots N-1) \, (i'~i' + 1 \dots N)$ a product of two cycles, and in general $w_{-I} = (i_1~i_1 + 1 \dots N-n+1) \cdots (i_n~i_n + 1 \dots N)$. 

From these $w_I$ we 
construct the operators $P_I(\vect{x}) \coloneqq P_{w_I^{-1}}(\vect{x})$ as in \eqref{eq:P_I}. Note the inverse! 

Let us again give a few examples. $P_\varnothing(\vect{x})$ is just \eqref{eq:id}. 
For $n=1$ we get \eqref{eq:P_1...i}. Next,
\begin{equation} \label{eq:P_ii'}
\begin{aligned}
	P_{\{i,i'\}}(\vect{x}) & = P_{(2~\dots~i'-1 ~ i')}(x_i,x_1,\dots,x_{i-1},x_{i+1},\dots,x_N) \, P_{(1~\dots~i-1 ~ i)}(\vect{x}) \\
	& = 
	\tikz[baseline={([yshift=-.5*11pt*0.3]current bounding box.center)},xscale=-.5,yscale=0.25,font=\footnotesize]{
		\draw[->] (-3,0) node[below]{$x_N$} -- (-3,6) node[above]{$x_N$};
		\foreach \x in {-1,...,1} \draw (-2.25+.2*\x,3) node{$\cdot\mathstrut$};	
		\draw[->] (-1.5,0) -- (-1.5,6);
		\draw[rounded corners=2pt,->] (-.5,0) node[below]{$x_{i'}$} -- (-.5,1) -- (.5,2) -- (3.5,5) -- (3.5,6) node[above]{$x_{i'}$};
		\draw[rounded corners=2pt,->] (.5,0) -- (.5,1) -- (-.5,2) -- (-.5,6);
		\draw[rounded corners=2pt,->] (1.5,0) node[below]{$x_i$} -- (1.5,1) -- (4.5,4) -- (4.5,6) node[above]{$x_i$};
		\draw[rounded corners=2pt,->] (2.5,0) node[below]{$x_{i-1}\;$} -- (2.5,1) -- (.5,3) -- (.5,6) node[above]{$\,x_{i-1}$};
		\draw[rounded corners=2pt,->] (3.5,0) node[below]{$\!\!\!\!\vphantom{x_i}\dots$} -- (3.5,2) -- (1.5,4) -- (1.5,6) node[above]{$\!\dots$};
		\draw[rounded corners=2pt,->] (4.5,0) node[below]{$x_1$} -- (4.5,3) -- (2.5,5) -- (2.5,6) node[above]{$x_1$};
		\node at (5,3) {\smash{\textcolor{lightgray}{$\vec{a}$}}\vphantom{$a$}};
	} .
\end{aligned} 
\end{equation}
Finally, since $w_{\{1,\dots,N\}} = e$ the first nontrivial example of $P_{-I}(\vect{x}) \coloneqq P_{\{1,\dots,N\} \setminus I}(\vect{x})$ is \eqref{eq:P_N...i}.

As these examples illustrate, one can compute $P_w(\vect{x})$ graphically. First draw $w \in S_N$: start with a row of numbers $1~2~\dots~N$, a little above it $w(1)~w(2)~\dots~w(N)$, and connect equal numbers, drawing so that no more than two lines cross at any point. Remove unnecessary double crossings, anticipating \eqref{eq:unitarity_diagram}, to get a reduced decomposition of $w$. Now replace all~$i$ by~$x_i$ in both rows and reinterpret the crossings as deformed permutations via~\eqref{eq:deformed_permutation_diagram}.

\bibliography{bibliography}

\end{document}